\documentclass{aa}
\usepackage{graphicx}
\usepackage{txfonts}
\usepackage{xcolor}
\usepackage{amsmath}    
\usepackage{amssymb}    
\usepackage{multicol}        
\usepackage{bm}         
\usepackage{pdflscape}  

\begin{document}

  \title{Tracing obscured galaxy build-up at high redshift using deep radio surveys}

  \author{Stergios Amarantidis\inst{1,2,3}\thanks{samarant@iram.es},
         Jose Afonso\inst{2,3},
         Israel Matute\inst{2,3},
         Duncan Farrah\inst{4,5},
         A. M. Hopkins\inst{6},
         Hugo Messias\inst{7,8},
         Ciro Pappalardo\inst{2,3},
         N. Seymour\inst{9}
         }

  \institute{Instituto de Radioastronomía Milimétrica, Av. Divina Pastora 7, Núcleo Central E-18012, Granada, Spain
        \and
            Instituto de Astrofísica e Ciências do Espaço, Universidade de Lisboa, OAL, Tapada da Ajuda, 1349-018 Lisbon, Portugal
        \and
        Departamento de Física, Faculdade de Ciências, Universidade de Lisboa, Edifício C8, Campo Grande, 1749-016 Lisbon, Portugal
        \and
        Department of Physics and Astronomy, University of Hawai‘i at Mānoa, 2505 Correa Rd., Honolulu, HI, 96822, USA
        \and
        Institute for Astronomy, University of Hawai‘i, 2680 Woodlawn Dr., Honolulu, HI, 96822, USA
        \and
        School of Mathematical and Physical Sciences, 12 Wally’s Walk, Macquarie University, NSW 2109, Australia
        \and
        Joint ALMA Observatory, Alonso de Córdova 3107, Vitacura 763-0355, Santiago de Chile, Chile
        \and
        European Southern Observatory, Alonso de Córdova 3107, Vitacura, Casilla 19001, Santiago de Chile, Chile
        \and
        International Centre for Radio Astronomy Research, Curtin University, 1 Turner Avenue, Bentley, WA, 6102, Australia
        }

  \date{Received March 14, 2023; accepted July 25, 2023}

 \abstract
  {A fundamental question of extra-galactic astronomy that is yet to be fully understood, concerns the evolution of the star formation rate (SFR) and supermassive black hole (SMBH) activity with cosmic time, as well as their interplay and how it impacts galaxy evolution. A primary focus that could shed more light on these questions is the study of merging systems, comprising highly star-forming galaxies (SFGs) and active galactic nuclei (AGN) at the earliest stages of galactic formation. However, considering the challenges associated with identifying these objects, it is essential to explore complementary selection methods across multiple wavelengths.}
  {The primary objective of this study is to conduct a comprehensive analysis of a sample of high-redshift ($z>3$) far-infrared (far-IR) and radio-emitting galaxies in the highest possible spatial resolution. The aim is to study the properties of this population, such as their morphological characteristics, and to explore the interplay of SFR and SMBH activity at this epoch.}
  {In order to select the galactic population of our interest, we employed two selection criteria that have frequently been used as separate methods in the literature. In more detail, we selected galaxies that present relatively compact radio morphologies at 1.4 GHz (i.e. an angular size smaller than 10 arcsec) as well as a far-IR spectrum that peaks in flux at $\lambda \geq 350 \, \mu m$ (i.e. $flux_{350\mu m}>flux_{250\mu m}$). For these selection criteria, we used the COSMOS and ECDF-S fields, two of the most extensively observed astronomical fields currently available, which provide high spectral and spatial resolution at a multi-wavelength scale. By accepting only galaxies that satisfied these selection criteria, we derived a sample of eight galaxies that were identified either photometrically or spectroscopically at $z>3$  from literature studies and by our team.}
  {A thorough investigation of available optical, near-IR, and millimetre (mm) imaging reveals a possible merging scenario in five out of eight cases in our sample. Additionally, available multi-wavelength photometry strongly suggests active star formation at the $10^3 \, M_{\odot}/yr$ level in massive systems (stellar masses of $M_{\star}\sim 10^{11} \, M_{\odot}$) co-hosting an active SMBH.}
  {Comparison of these results with previous studies, suggests that our selection method preferentially identifies galaxies hosting an active SMBH, as well as a strong SFG component, resulting in high SFR and IR luminosity. An additional examination of the efficacy of the radio and far-IR selection criteria provides further support for their combined application in selecting co-evolving AGN and star formation activity at high redshift. In this regard, future use of these selection criteria on radio and far-IR/mm observations of statistically larger galaxy samples is of high interest.}


  \titlerunning{Tracing obscured galaxy buildup at high redshift using deep radio surveys}
  \authorrunning{S. Amarantidis}
  \maketitle

\section{Introduction}\label{introduction}
One of the main challenges in the field of extra-galactic astronomy is understanding how the star formation rate density (SFRD - $M_{\odot}/yr/Mpc^3$) evolves with redshift \citep[see][for a detailed review]{2017ApJ...840...39M,2018ApJ...855..105O,2020ARA&A..58..157T}. While we have accurate estimates of the SFRD at intermediate redshifts ($1<z<3$) using various selection methods from deep multi-wavelength data \citep[e.g.][for infrared (IR), radio, ultraviolet-UV and millimetre (mm) criteria, respectively]{2020A&A...643A...8G,2017A&A...602A...5N,2018ApJ...854...73I,2017MNRAS.466..861D} that agree with results from hydrodynamical simulations \citep[e.g.][]{2018MNRAS.473.4077P,2018IAUS..333..228L}, uncertainties regarding higher redshifts ($z>3$) still remain \citep[e.g.][]{2019ApJ...877...45M,2020A&A...643A...8G}. For instance, studies of UV-selected galaxy samples \citep[e.g.][]{2016MNRAS.459.3812M,2018ApJ...854...73I} estimate a steep decline in the SFRD at $z>3$, while other works employing far-IR radio selection criteria \citep[e.g.][]{2013ApJ...779...32V,2016MNRAS.461.1100R,2017A&A...602A...5N,2022MNRAS.509.4291M} or gamma-ray-burst-selected samples \citep[e.g.][]{2008ApJ...683L...5Y,2009ApJ...705L.104K} suggest a flatter behaviour.

These uncertainties may originate from our current instrumentational limitations, including low sensitivity, coarse spatial resolution, and coverage of small parts of the sky, which hinder our ability to properly characterise distant objects and obtain a statistically unbiased sample at higher redshifts. For instance, observations based on telescopes such as the \textit{Herschel} Space Observatory \citep[][]{2010A&A...518L...1P} or the Submillimetre Common-User Bolometer Array \citep[SCUBA;][]{2013MNRAS.430.2513H} are bound to limit the redshift detection to $z_{\rm max} \sim 4$ \citep[e.g.][]{2018A&A...614A..33D}, and \textcolor{black}{to} uncover only the very bright end of the far-IR luminosity function \citep[e.g.][]{2013MNRAS.432...23G,2021A&A...648A...8W}. On the other hand, more sensitive telescopes, such as the Atacama Large Millimeter/submillimeter Array \citep[ALMA;][]{2019clrp.2020...19W} or the Northern Extended Millimeter Array \citep[NOEMA;][]{2016ITTST...6..223C} are limited to smaller areas. Thereby, these telescopes provide fewer objects \citep[e.g.][]{2017MNRAS.466..861D,2020ApJS..247...61F,2021ApJ...923..215C} and reveal a fainter star-forming galaxy (SFG) population \citep[e.g.][]{2021MNRAS.507.3998P}. Nevertheless, recent studies using the capabilities of both single telescopes and interferometers have pushed the limits of our current understanding regarding the star formation (SF) history, exposing a mm dusty galaxy population at $z>3$ \citep[see e.g. ][]{2018A&A...614A..33D,2019ApJ...878...73Y,2020A&A...643A...8G}.

Another aspect of our current limitations regarding the accurate estimation of the SFRD at high redshifts might be hidden in the selection criteria that are currently used to detect SFGs. Various methods have been employed over the years in order to detect this galaxy population \citep[e.g.][]{2013MNRAS.432...23G,2015ApJ...811..140B}, however, the bulk of the research is focused on the detection of Lyman-break galaxies in the rest frame UV and the application of dust-correction low-redshift empirical relations for the derivation of their star formation rates \citep[e.g.][]{2015ApJ...811..140B,2018ApJ...855..105O}. These relations might not be necessarily valid at $z>3,$ where the SFG population remains uncertain. Alternatively, recent studies using sensitive far-IR or mm telescopes \citep[e.g. the ALPINE-ALMA survey;][]{2020A&A...643A...8G} or radio surveys \citep[e.g.][]{2021ApJ...909...23T,2022ApJ...927..204E} aim to reveal a \textcolor{black}{an SFG population that has so far remained undetected} in the optical/UV bands, which is estimated to significantly contribute to the SFRD at $z>3$ \citep[e.g.][]{2020A&A...643A...8G}. In order to obtain accurate SFR estimates, these studies separate the SFG from the AGN population \citep[e.g.][]{2017A&A...602A.123L,2020A&A...643A...8G} by removing the AGN contribution from the IR luminosity. However, this approach can introduce a bias in the sample. Interestingly, results from the recent surveys of the \textit{James Webb} Space Telescope (\textit{JWST)} \citep[e.g.][]{2023ApJS..265....5H,2023arXiv230112825N,2023ApJ...946L..16P,2023MNRAS.518L..19R} reveal a dusty-galaxy population \textcolor{black}{like this} at even higher redshifts (i.e. $z>6$) \textcolor{black}{that have not been explored before}.

Previous studies, such as the work by \citet{2019ApJ...876..110D}, indicated that the galaxy pair and merging ratio increases with redshift (from $\sim 5 \, \%$ at $z<1$ to $\sim 40 \,\%$ at $z>3$ and for $M_{\star}>10^{10.3}\, M_{\odot}$). Therefore, in this work, we question the strategy of removing AGN-contaminated radio sources from selected samples \textcolor{black}{as} this approach may overlook merging systems in which the AGN emission conceals an additional SFG population that could significantly contribute to the SFRD at high redshifts. Such galaxies could be detected from the identification of far-IR bright \textit{Herschel} galaxies along with a radio emission. Even though these selection criteria might favour sources with cold gas and negative k-correction, their application to fields with deep coverage \citep[e.g. the COSMOS field;][]{cosmos_paper} along with photometric data and high spatial resolution information from telescopes such as the \textit{Hubble} Space Telescope (HST) and ALMA could potentially reveal a population of SFGs that would be omitted by the current selection criteria because of the AGN contribution of a neighbour galaxy.

In an attempt to explore new techniques to assist radio surveys reaching $z>3$, we present a complementary study combining the radio selection and far-IR peak spectrum criteria \citep[e.g.][]{2022MNRAS.516.5471Y}. Furthermore, we suggest that at $z \sim 3-4$, a substantial amount of star formation occurs that is hardly revealed from optical or near-IR surveys, and which is consequently excluded by the current selection methods \textcolor{black}{due to} the AGN contribution in the system. In more detail, we explore a sample of $z>3$ far-IR radio-emitting galaxies, detected in the COSMOS and ECDF-S fields, \textcolor{black}{that might uncover} interacting systems and merging galaxies with a substantial contribution of an AGN component, \textcolor{black}{along with the presence of} high star formation activity. If radio emission can help to pinpoint this important `missing' SF component, upcoming radio surveys, such as the Evolutionary Map of the Universe (EMU) survey \citep[e.g.][]{norris,2021PASA...38...46N} as well as novel selection criteria in the radio \citep[e.g.][]{2021ApJ...909...23T,2022ApJ...927..204E} could become even more important to our understanding of the SFRD at $z>3$.

The paper is structured as follows: in Sect. 2 we present the method and criteria that were employed in the selection of the original sample and the subsequent reduction of the sample to the high-redshift far-IR radio candidates. Section 3 briefly highlights the photometric and spectroscopic properties of the sample, along with the results obtained from various properties of the entire sample. In Sect. 4 we compare our work with previous studies and provide a brief discussion of the results, emphasising the importance of the merging population for SFRD estimations. Finally, we provide the conclusions of this study in Sect. 5 and discuss future prospects regarding our method. Throughout this paper, we assume a $\Lambda$-CDM cosmology using $H_{\rm 0}=70 \, kms^{-1}Mpc^{-1}$, $\Omega_{\rm m} = 0.3$, and $\Omega_{\rm \Lambda} = 0.7$.

\section{Sample selection}\label{sample_selection}
\begin{figure*}
\centering
\includegraphics[width=2\columnwidth]{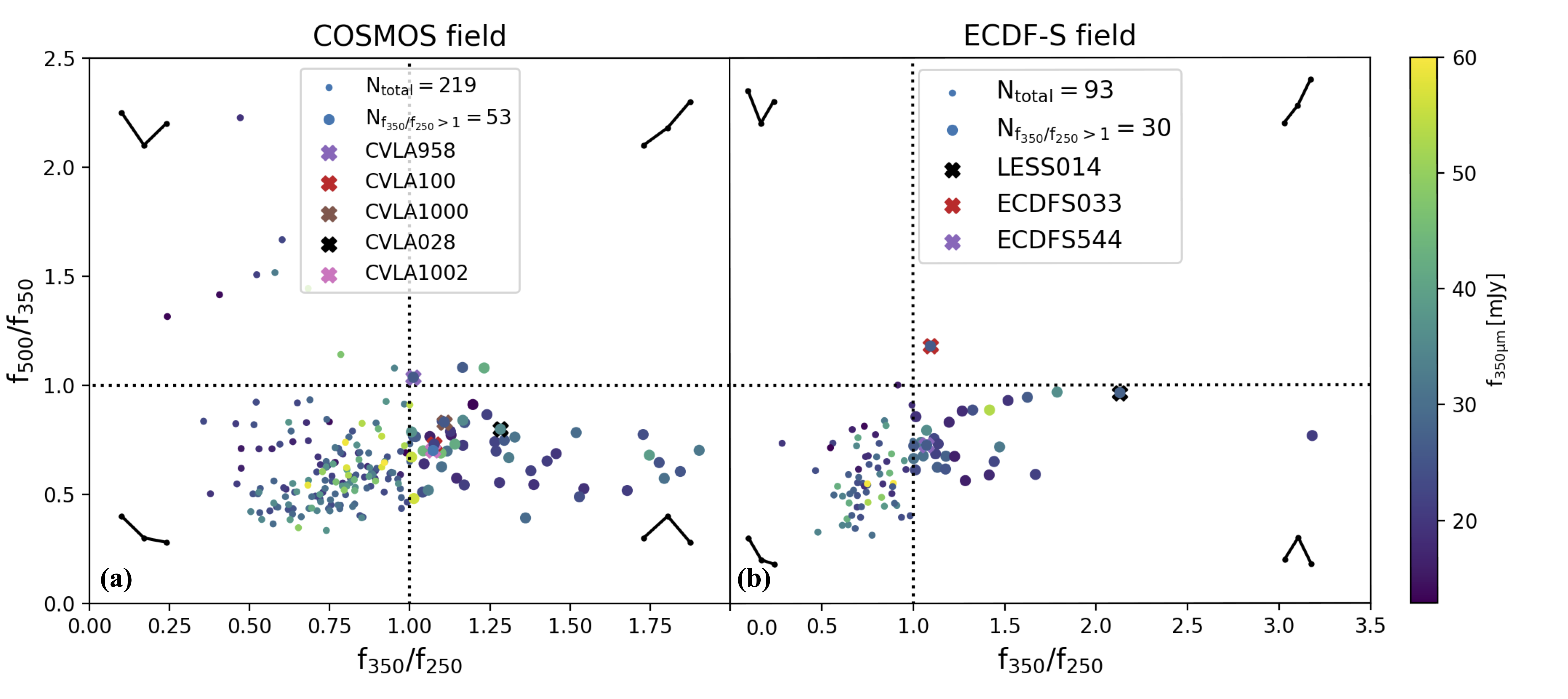}
\caption{Flux density ratios of 350 and 250 $\mu m$ vs the ratios of 500 and 350 $\mu m$ of HerMES in COSMOS (panel a) and ECDF-S (panel b). The sample was selected by cross-matching the HerMES catalogue with the VLA-COSMOS and VLA-ECDFS data. The dashed black lines indicate $f_{350}/f_{250}$=1 and $f_{500}/f_{350}$=1, while the points surpassing the limit of $f_{350}/f_{250}$>1 are presented with larger markers. The colour bar corresponds to the 350 $\mu m$ flux of each source, and the legend shows the number of sources for each panel and the selected high-redshift candidates (crosses). The black lines in the subregion of each panel provide information regarding the shape of the far-IR (FIR) emission. For instance, sources with $f_{500}/f_{350}$<1 and $f_{350}/f_{250}$>1 have a peak in flux between 250 and 500 $\mu m$. The photometric properties of the sources presented in the legend are provided in Table \ref{radio_sources_table_2}.}
\label{cosmos_ecdfs}
\end{figure*}
For our selection, we examined two of the most deeply explored extra-galactic fields that were observed under the projects Extended \textit{Chandra} Deep Field-South \citep[ECDF-S;][]{2005ApJS..161...21L} and Cosmic Evolution Survey \citep[COSMOS;][]{cosmos_paper}. In past years, these two fields have been extensively observed in a wide range of electromagnetic frequencies, ranging from X-rays to radio waves. Radio data were obtained from both fields at 1.4 GHz with the VLA \citep[VLA-COSMOS; the flux density limit is 7 $\mu Jy/beam,$ and VLA-ECDF-S; the flux density limit is 7.4 $\mu Jy/beam$;][]{2010ApJS..188..384S,smolcic17,2013ApJS..205...13M} at an angular resolution of $\sim 2.5''$.\\
\indent The far-IR wavelength regime, which is necessary for our selection, is available through the \textit{Herschel} Multi-tiered Extragalactic Survey \citep[HerMES;][]{hermes}. This survey covered a total of 380 $deg^2$ using the two main instruments on board \textit{Herschel} , namely the Spectral and Photometric Imaging Receiver \citep[SPIRE;][; at 250, 350 and 500 $\mu m$; the confusion noise level is $\sim 5.8$, $\sim 6.3,$ and $\sim 6.8 \, mJy$, respectively]{2010A&A...518L...3G,2010A&A...518L...5N} and the Photodetector Array Camera and Spectrometer \citep[PACS;][; at 100 and 160 $\mu m$; the detection limits are $\sim 7.7$ and $\sim 14.5 \, mJy$, respectively]{2010A&A...518L...2P}. The ECDFS and COSMOS fields are also included in the covered area of HerMES, comprising a total of $\sim 2 \, deg^2$. Previous studies using HerMES data have demonstrated the efficient use of \textit{Herschel}'s far-IR fluxes for detecting high-redshift ($z>3$) dusty galaxies. For instance, \citet{2014ApJ...780...75D} employed simple criteria for the increasing fluxes for the 250, 350, and 500 $\mu m$ bands, which is indicative of a far-IR peak at longer (redshifted) $> 350 \, \mu m$ wavelengths, revealing a dusty SF galaxy population with a mean redshift distribution of $<z>\sim 4.7$.\\
\indent To identify potential high-redshift hybrid AGN-SFG systems, we conducted a cross-match analysis between the VLA COSMOS and ECDF-S 1.4 GHz catalogues with those of HerMES, applying a maximum separation of 18 arcsec (motivated by the beam size of \textit{Herschel} at 250 $\mu m$ of 18.1 arcsec). The resulting cross-matched sample contained 2292 sources for COSMOS, and for the ECDF-S field is provided 769 sources. This sample was further reduced by considering robust radio sources (signal-to-noise ratio; S/N>5) without an indication for extended or multicomponent radio emission \citep[according to each catalogue; for more details, see][]{2005ApJS..161...21L,cosmos_paper}. To ensure the correspondence between radio and far-IR emission, \textcolor{black}{we excluded from the sample sources that present multiple radio detections} within one \textit{Herschel} beam. Furthermore, only sources with 250, 350, and 500 $\mu m$ flux errors smaller than 50 $\%$ were accepted, which ensured that the shape of the far-IR emission was properly determined. This selection method resulted in 219 and 93 sources for COSMOS and ECDF-S, respectively.\\
\indent Furthermore, motivated by the efficacy of studies such as \citet{2014ApJ...780...75D} and \citet{2018A&A...614A..33D} in selecting high-$z$ dusty SF galaxies, we adopted a similar method to identify sources with $f_{350 \rm \mu m}/f_{250 \rm \mu m}>1$, which would translate into a potential $z > 3$ redshift. This selection method resulted in a sample of 53 and 30 galaxies (see Fig. \ref{cosmos_ecdfs}) for the COSMOS and ECDF-S fields, respectively. The radio 1.4 GHz and far-IR ($250 \, \mu m$) images of each of these sources were visually inspected in order to secure the robustness of the matching process and identify potentially uncertain or confused radio/far-IR detections. We confirmed that all 83 selected sources display compact morphologies both in the radio and in the FIR, and that the detected counterparts are robust, with average separations between the radio and far-IR detections below $\sim 2$ arcsec.

\subsection*{High-redshift candidates}
Application of these far-IR criteria resulted in a sample of $53+30=83$ sources from both fields for which we retrieved possible photometric or spectroscopic identifications from the literature. For this exploration, we consulted Simbad \citep{2000A&AS..143....9W}, the NED\footnote{\href{https://ned.ipac.caltech.edu}{https://ned.ipac.caltech.edu}} online search tools, and the Cosmos2015 \citep[][]{Laigle16} online catalogue\footnote{\href{https://irsa.ipac.caltech.edu/cgi-bin/Gator/nph-scan?projshort=COSMOS}{https://irsa.ipac.caltech.edu}}, retrieving redshift values from sources that are within 2 arcsec of our candidates.\\
\indent From this analysis, we found 18 sources in the ECDF-S and 19 sources in the COSMOS field with spectroscopic redshift determination. Five of these radio sources present a spectroscopic redshift of $z_{\rm spec}>3$, namely J033140.1-275631 \citep[ECDF-S field; $z_{\rm spec}$=5.14;][]{2015ApJ...815..129S}, J100028.71+023203.7 \citep[COSMOS field; $z_{\rm spec}$=4.76;][]{2018ApJ...858...77H}, [LBX2017] 214 \citep[ECDF-S field; $z_{\rm spec}$=3.74;][]{2012A&A...546A..84I}, COSMOSVLA J100256.53+021158.4 \citep[COSMOS field; $z_{\rm spec}$=3.503;][] {2014A&A...563A..54P}, and [DSS2017] 554 \citep[ECDF-S field; $z_{\rm spec}$=3.1988;][]{2017ApJ...840...78D}. The reported spectroscopic redshifts of the two highest- $z$ sources, however, are not reliable. For instance, the candidate with the highest redshift, found in the ECDF-S, reveals contradicting spectroscopic measurements. More specifically, the $z_{\rm spec}=5.137$ value \citep[derived from near-IR spectroscopy;][]{2015ApJ...815..129S} contradicts \citet{2012MNRAS.427..520C} and \citet{2017ApJ...840...78D}, who estimated a lower spectroscopic value of $z_{\rm spec}= 1.617$ (with spectra from optical and far-IR wavelengths) for this candidate. \textcolor{black}{Because} a high-redshift nature of this source cannot be excluded, we further explored its available spectra. This investigation indicated that the emission line responsible for the $z_{\rm spec}$=5.137 solution can be also interpreted as an emission line at $z_{\rm spec}$=1.617, which further supports the notion for the intermediate-redshift solution derived in the literature. Therefore, we considered this source as an intermediate-redshift galaxy and did not explore it further.\\
\begin{figure}
\centering
\includegraphics[width=\columnwidth]{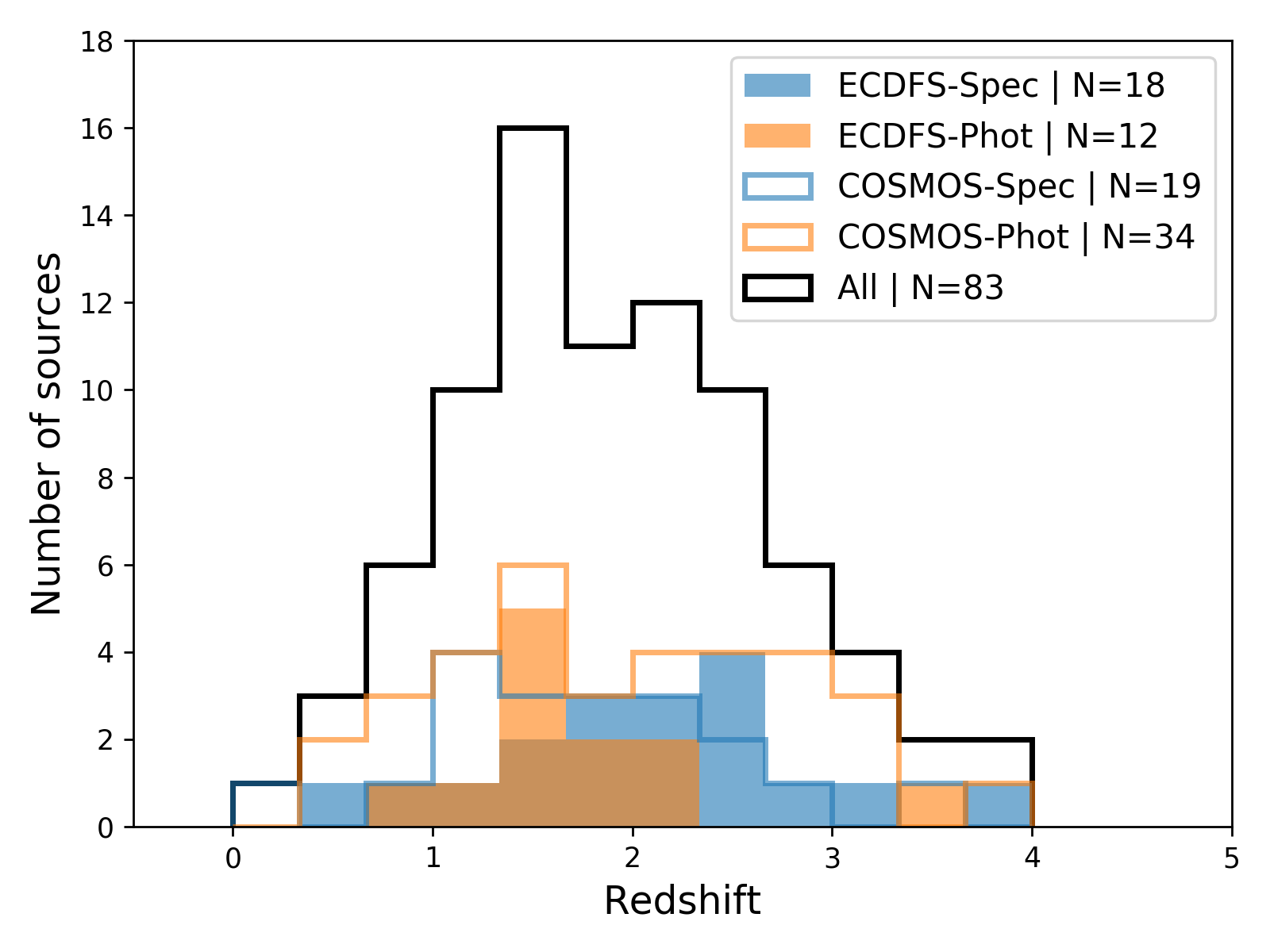}
\caption{Histograms presenting the number of sources for each redshift bin selected by the criteria for the COSMOS (empty histograms) and ECDF-S (filled histograms). The blue histograms correspond to spectroscopic values derived from the literature, and the orange histograms show the photometric values. The combined sample is depicted with an empty black histogram. The number of sources for each category is presented in the legend.}
\label{hist_z}
\end{figure}
\setlength{\tabcolsep}{4pt}
\begin{table*}
\centering
\caption[The names (COSMOSVLA is abbreviated as CVLA for visualisation purposes), abbreviations, equatorial coordinates, and flux at 1.4 GHz, 250 $\mu m$, 350 $\mu m$ and 500 $\mu m$ for our radio candidates]{Names, abbreviations, equatorial coordinates, flux densities at 1.4 GHz, 250 $\mu m$, 350 $\mu m$ and 500 $\mu m$ (along with their total error - instrumental and confusion), and angular diameter (derived from the 1.4 GHz emission) for our high-redshift ($z_{\rm spec,phot}>3$) radio candidates, ordered by $f_{\rm 1.4GHz}$ (descending order). The second row of the table corresponds to the units of each parameter.}
\label{radio_sources_table_2}
\begin{tabular}{cccccccc}
Source name & Abbreviation & RA & DEC & $f_{\rm 1.4GHz}$  & $f_{\rm 250}$ & $f_{\rm 350}$ & $f_{\rm 500}$  \\
-- & -- & [\textit{deg}] & [\textit{deg}] & [\textit{mJy}] & [\textit{mJy}] & [\textit{mJy}] & [\textit{mJy}] \\
\hline
CVLA J100256.53+021158.4  &CVLA1002  & 150.73553&2.19963  & $0.104 \pm 0.020$  & $37.0 \pm 3.4$  & $39.7 \pm 6.6$ & $25.1 \pm 5.1$ \\
ALESS J033152.49-280319.1 &ALESS14   & 52.96871&-28.05531 & $0.103 \pm 0.016$ & $13.0 \pm 6.2$  & $27.7 \pm 3.9$ & $26.7 \pm 4.5$\\
CVLA J100004.81+023045.2 &CVLA1000  & 150.0200&2.51255  & $0.096 \pm 0.015$ & $25.1 \pm 6.3$  & $27.8 \pm 6.6$ & $23.1 \pm 5.3$\\
CVLA J095845.94+024329.2 &CVLA958 &149.69144&2.72479  & $0.095 \pm 0.018$ & $27.4 \pm 6.3$  & $27.7 \pm 6.6$ & $28.7 \pm 5.4$ \\
CVLA J100233.16+020626.5 &CVLA100  &150.63817&2.10736  & $0.088 \pm 0.015$ & $22.3 \pm 6.3$  & $24.0 \pm 6.6$ & $17.5 \pm 5.3$  \\
CVLA J100028.75+023203.5  &CVLA028 & 150.11958&2.53438  & $0.057 \pm 0.015$ & $28.6 \pm 6.3$ & $36.7 \pm 6.6$ & $29.3 \pm 5.2$  \\
ALESS J033211.34-275211.9  & ALESS33  & 53.04738&-27.87003  & $0.043 \pm 0.012$ & $18.6 \pm 6.2$ & $25.6 \pm 3.9$ & $25.4 \pm 4.7$ \\
$[$DSS2017$]$ 554  &ECDFS544 & 53.42108&-27.92731  & $0.042 \pm 0.014$  &$24.5 \pm 3.4$ & $27.0 \pm 3.9$ & $23.8 \pm 3.9$ \\
\hline
\end{tabular}
\end{table*}
\indent Likewise, for J100028.71, inspection of its DEIMOS 10k survey spectrum \citep[][]{2018ApJ...858...77H} revealed a very low S/N spectrum without an obvious line detection that could explain the $z=4.76$ indication. We assumed an erroneous redshift estimate caused by a possible false-positive of the automated feature recognition from the DEIMOS survey. This source is however kept in this work as a high-redshift source, \textcolor{black}{prompted by a} photometric redshift determination of $z_{\rm phot}\sim 3.2$ by \citet{2015A&A...577A..29M}. Inspection of the remaining three sources with a spectroscopic redshift higher than 3 revealed that the emission lines used to determine the redshifts of these sources have either a high S/N (i.e. S/N$>10$) or are denoted with a secure spectroscopic quality flag. This means that these measurements are accurate.\\
\indent Our search identifies 12 sources in the ECDF-S field and 34 sources in the COSMOS field based on photometric redshifts from the literature. These results are depicted in Fig. \ref{hist_z}, where the histograms for the redshifts of each survey and the total values are plotted with different colours. This figure reveals that our selection favours radio sources around $z\sim 2$ with a distribution that extends to higher redshifts. Four sources of this sample in the COSMOS field have photometric redshifts (identified in the Cosmos2015 catalogue) of $z>3$ (i.e. COSMOSVLA J100233.16+020626.5 with $z_{\rm phot}$=3.016, COSMOSVLA J095845.94+024329.2 with $z_{\rm phot}$=3.172, COSMOSVLA J100028.75+023203.5 with $z_{\rm phot}$=3.175, and COSMOSVLA J100004.81+023045.2 with $z_{\rm phot}$=3.76). Similarly, in the ECDF-S sample, we identified one source with $z_{\rm phot}$>3, namely ALESS J033152.49-280319.1 \citep[$z_{\rm phot}$=3.56;][]{2011MNRAS.415.1479W}. These five high-redshift candidates, along with the three sources with a spectroscopic redshift above 3, are presented along with their 1.4 GHz and far-IR HerMES flux densities in Table \ref{radio_sources_table_2}.\\
\indent \textcolor{black}{We are particularly interested in sources with the highest 
redshifts. We therefore decided to further investigate the five sources} for which the photo-$z$ indicates a $z>3$ nature. Follow-up observations of these sources aimed to improve the redshift determination (for the photometric redshift estimates), and a more detailed analysis of their properties was initiated, using mm observatories from the northern and southern hemispheres (for the COSMOS and ECDF-S fields, respectively), the first results of which are described in the following section.\\
\indent \textcolor{black}{The comparison of} our acquired spectroscopic/photometric redshift values for our sample with the corresponding values from alternative photometric catalogues in the literature is a valuable exercise. This comparison is presented in more detail in Appendix \ref{appA} and indicates that the COSMOS and the ECDF-S fields do not contain additional candidates at $z>3$ that our analysis could have miss-identified at lower $z$, and the $z>3$ nature of the five sources with photo-$z$ considered here is confirmed. Therefore, we are more confident that the sources that are presented in the following sections are the only ones with $z_{\rm phot,spec}>3$ that could be selected with our selection criteria from the two fields of study.

\section{Results}\label{results}
To characterise the morphological and physical properties of the eight high-redshift galaxies under investigation, a detailed examination of the available photometric and spectroscopic data from the literature was conducted. Simultaneously, follow-up observations were initiated, in particular, using millimetre facilities. The results are described below.

During this analysis, a spectral energy distribution (SED) fitting analysis was performed for each galaxy in the sample using CIGALE \citep[][]{cigale1,cigale2}. The resulting fits yield two values for each modelled physical property of a galaxy (e.g. stellar mass and redshift), one corresponding to the best-fit solution and another that is derived through a Bayesian statistical analysis. In the following text, we refer to the best-fit values when presenting the sample, while the Bayesian-derived estimates along with their corresponding uncertainties are introduced in Table \ref{table2}. The images of the eight sources for four different photometric bands (optical, near-IR, radio, and mm wavelengths) are presented in Fig. \ref{figure3}. For each wavelength regime, the observations with the highest available spatial resolution were selected.

\subsection{SED fitting results}
\begin{figure*}
\centering
\includegraphics[width=0.88\columnwidth]{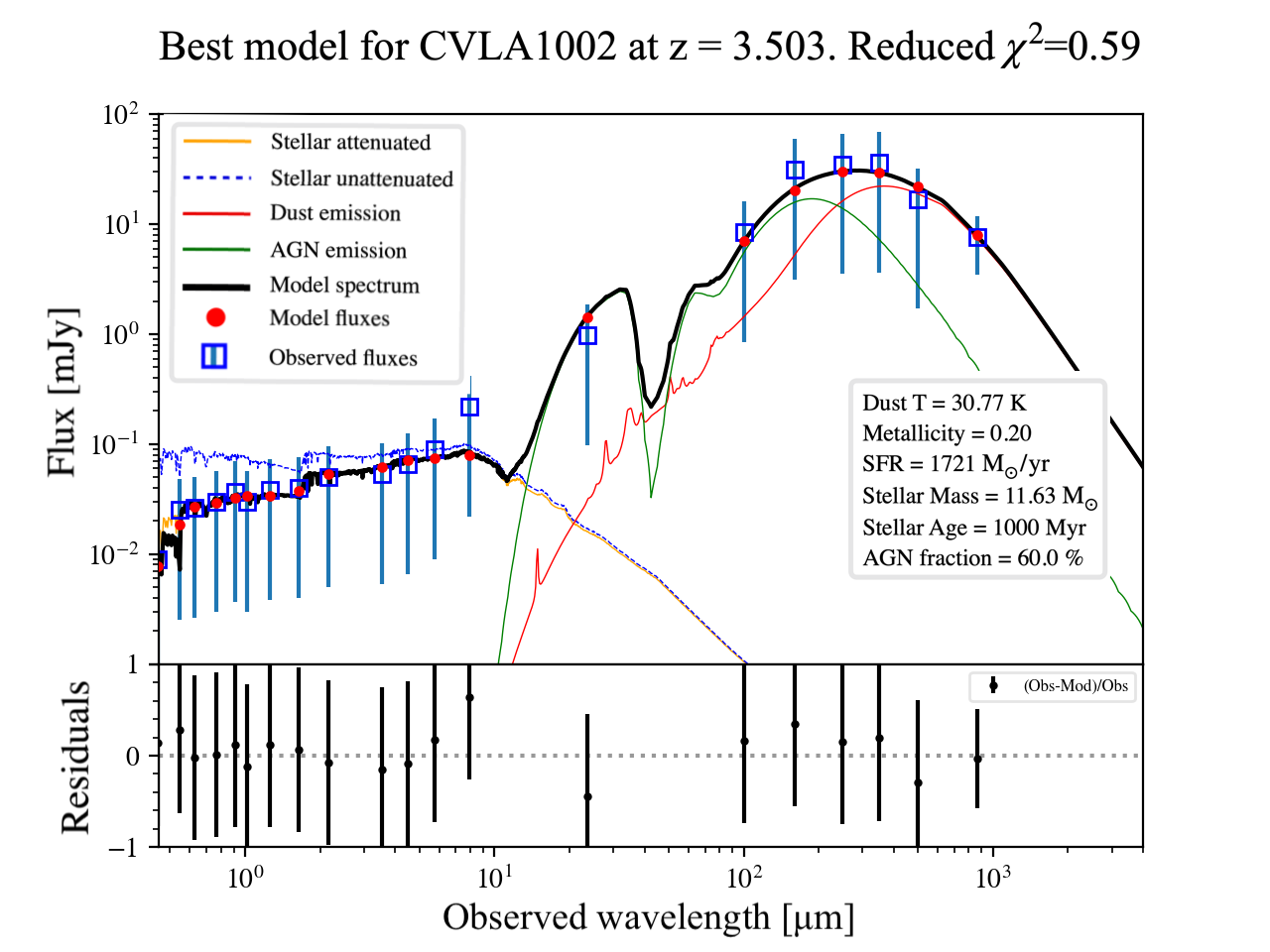}
\includegraphics[width=0.88\columnwidth]{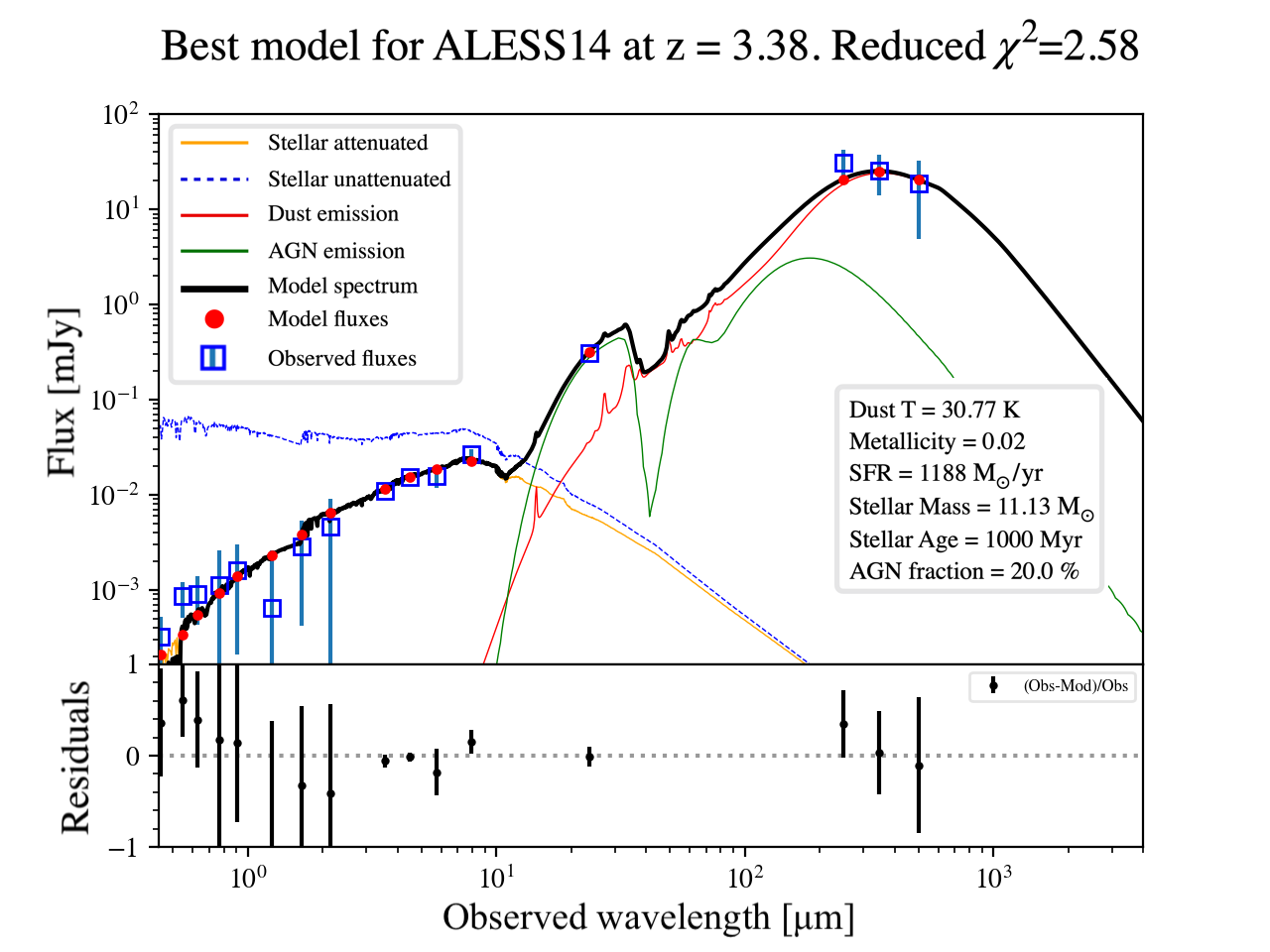}

\includegraphics[width=0.88\columnwidth]{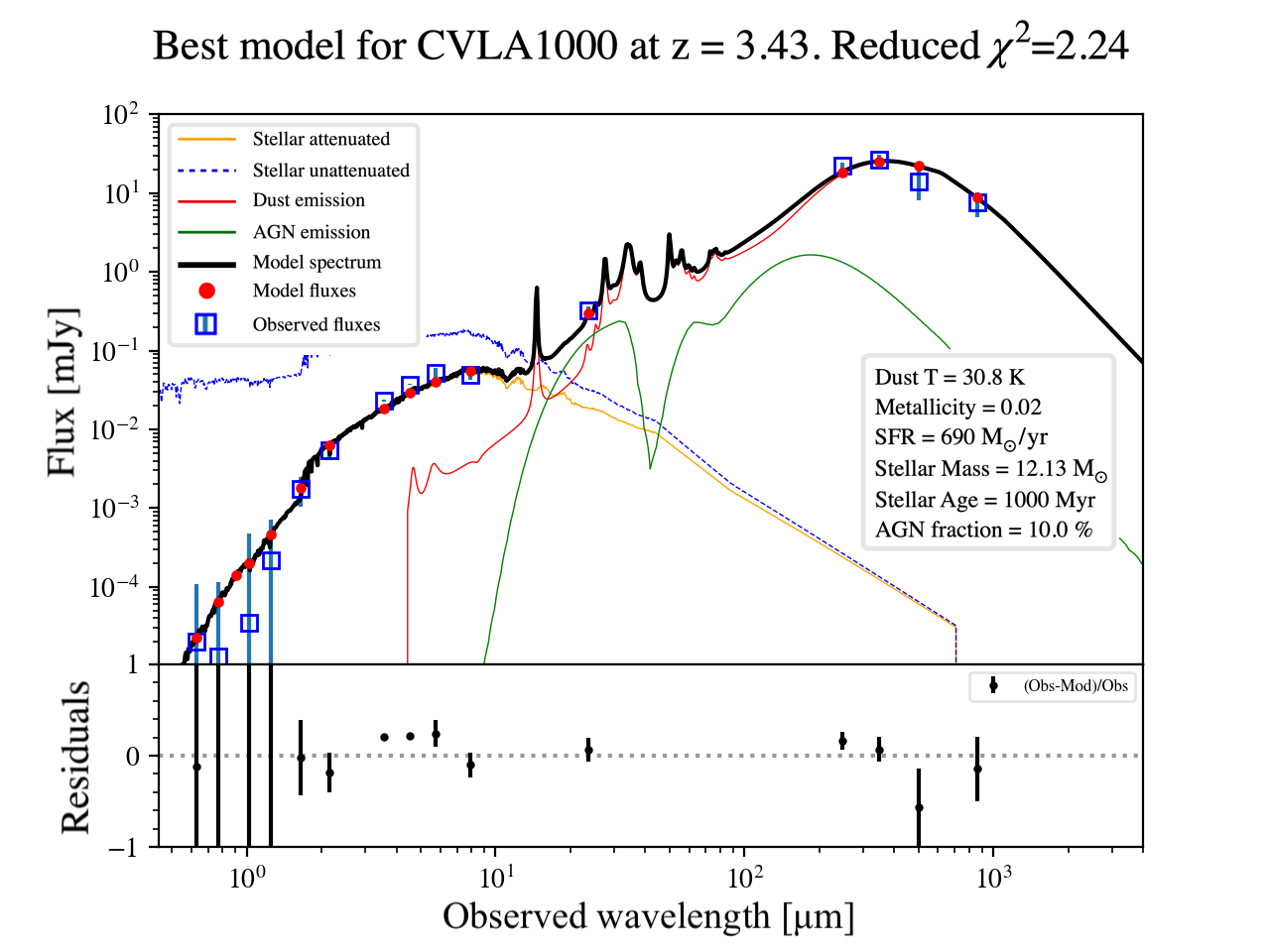}
\includegraphics[width=0.88\columnwidth]{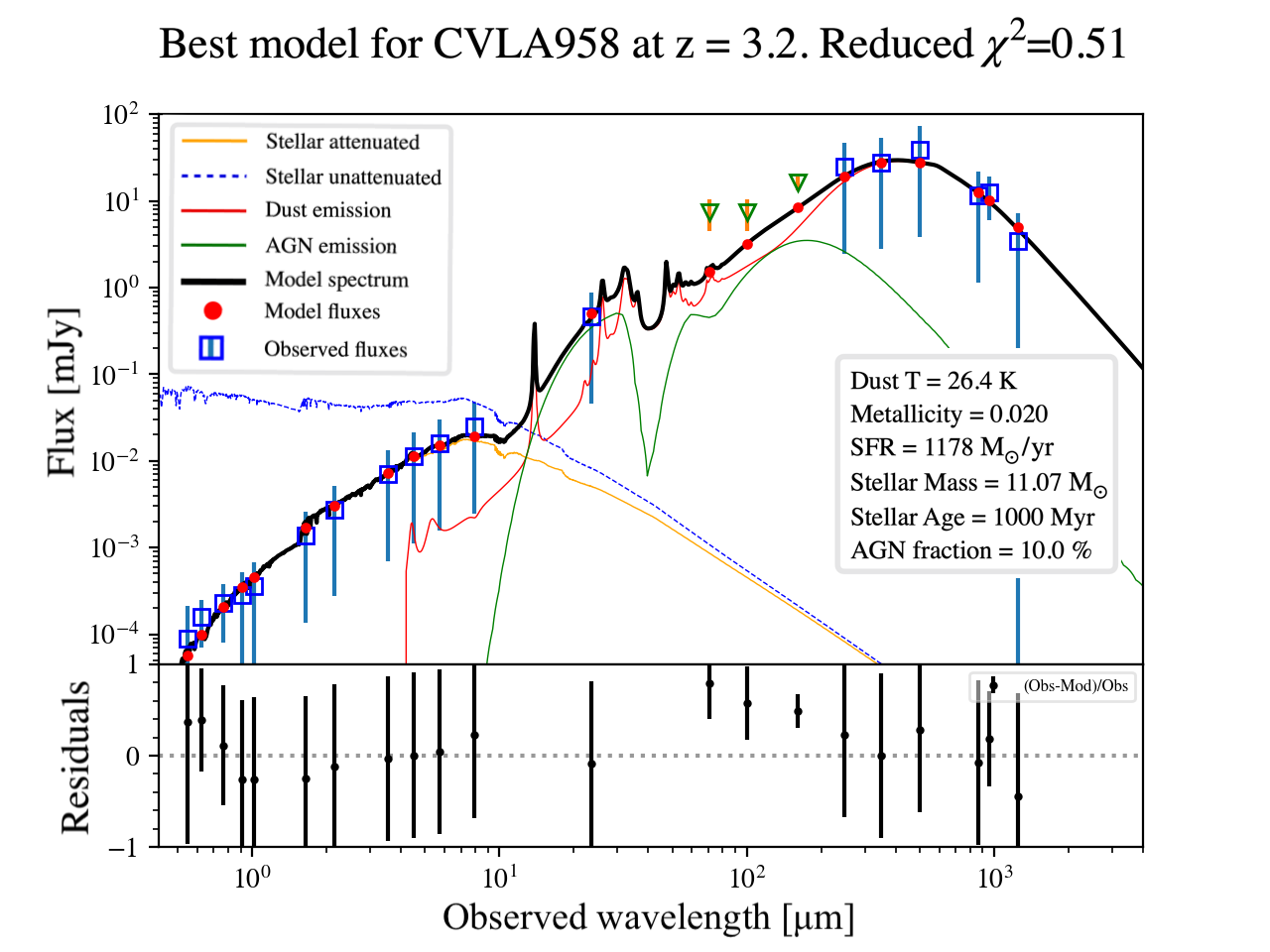}

\includegraphics[width=0.88\columnwidth]{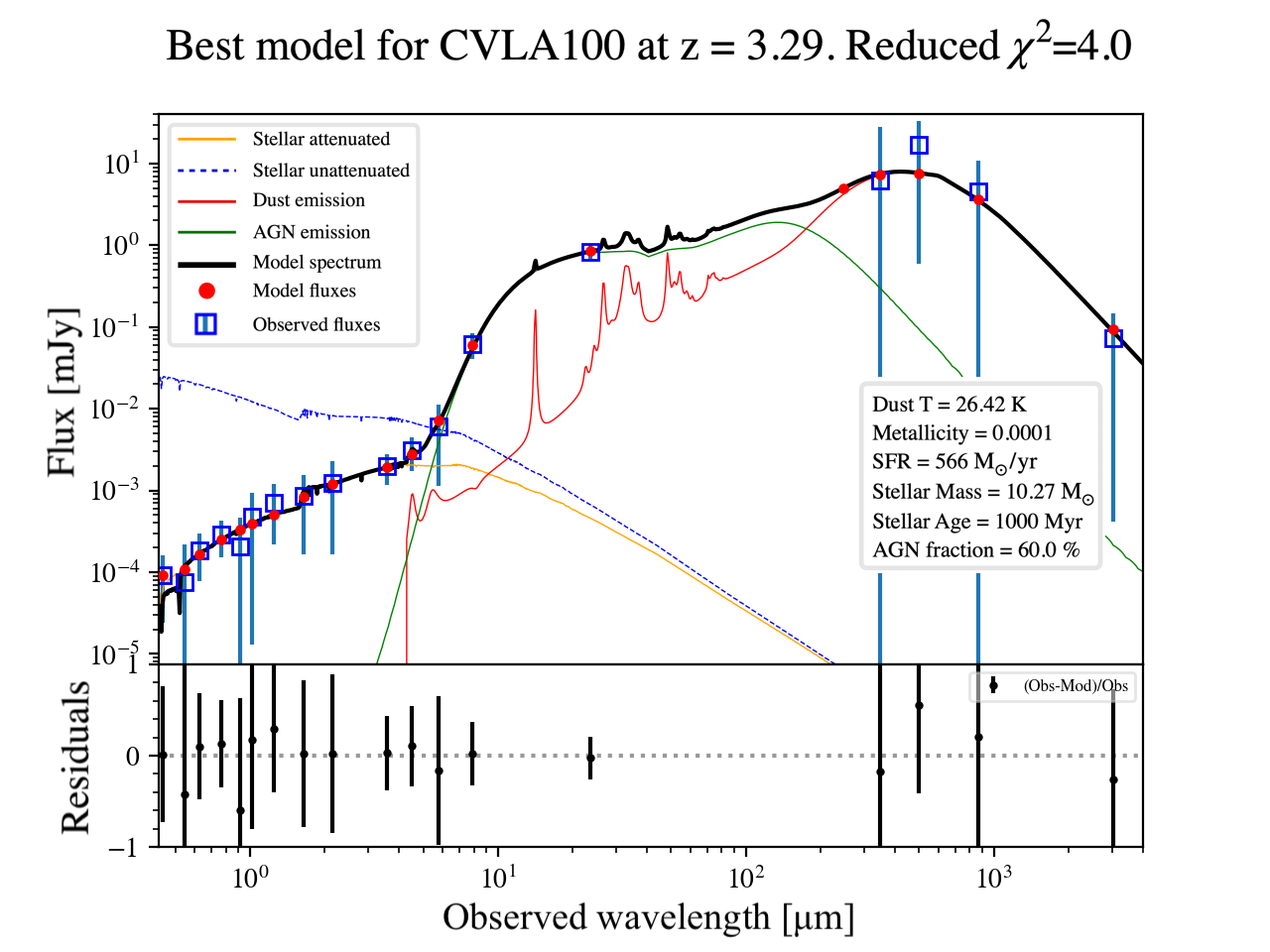}
\includegraphics[width=0.88\columnwidth]{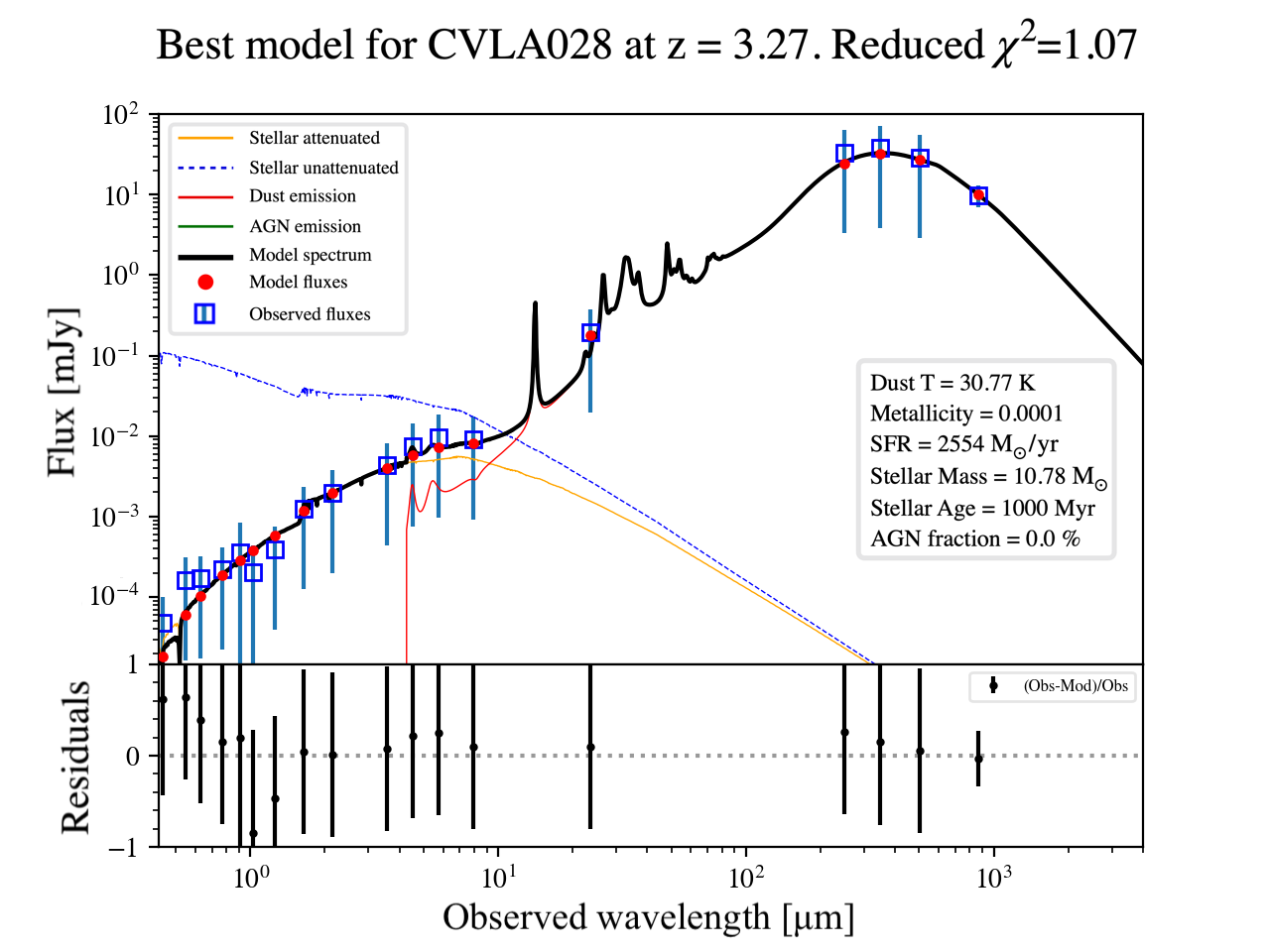}

\includegraphics[width=0.88\columnwidth]{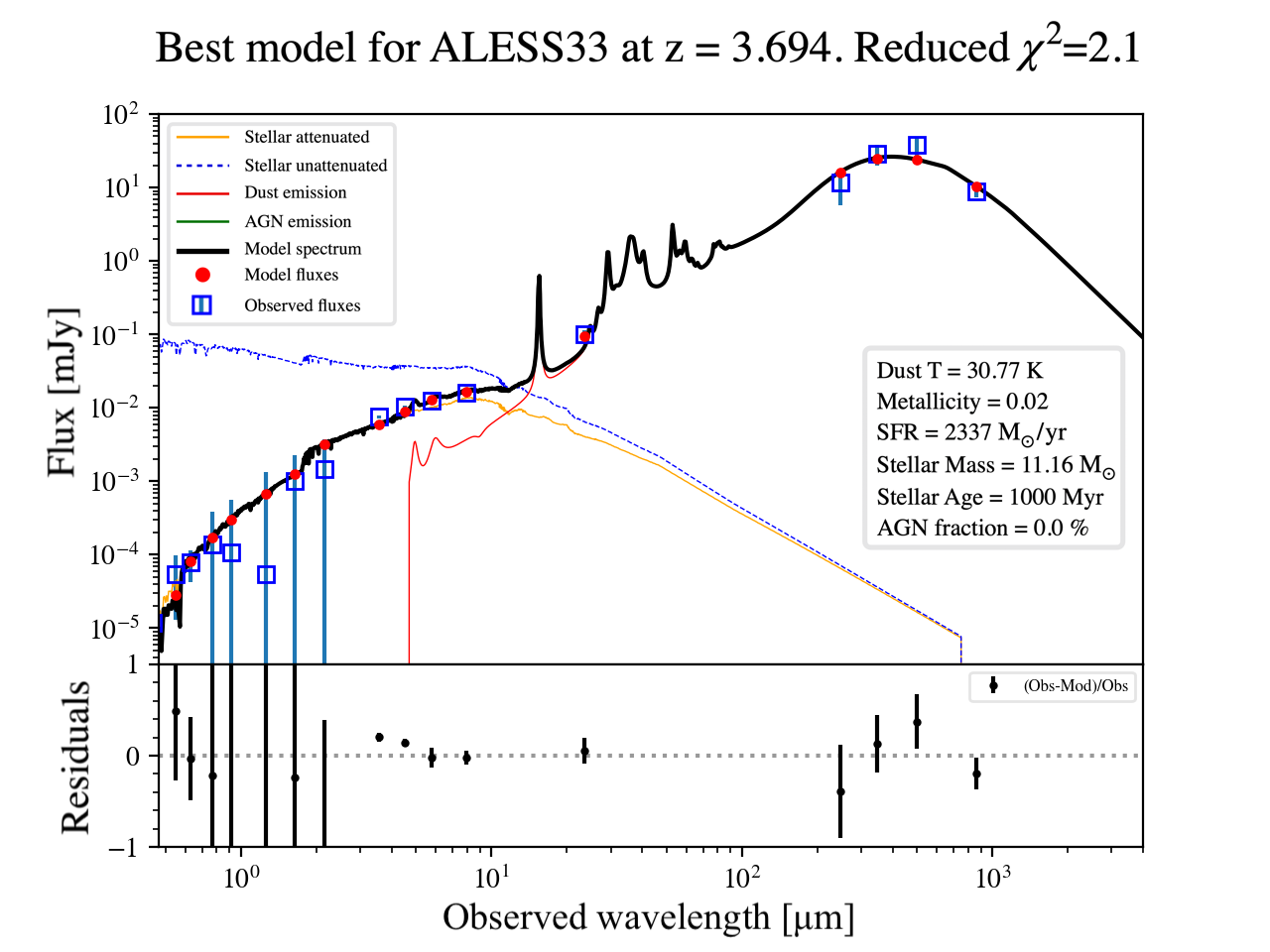}
\includegraphics[width=0.88\columnwidth]{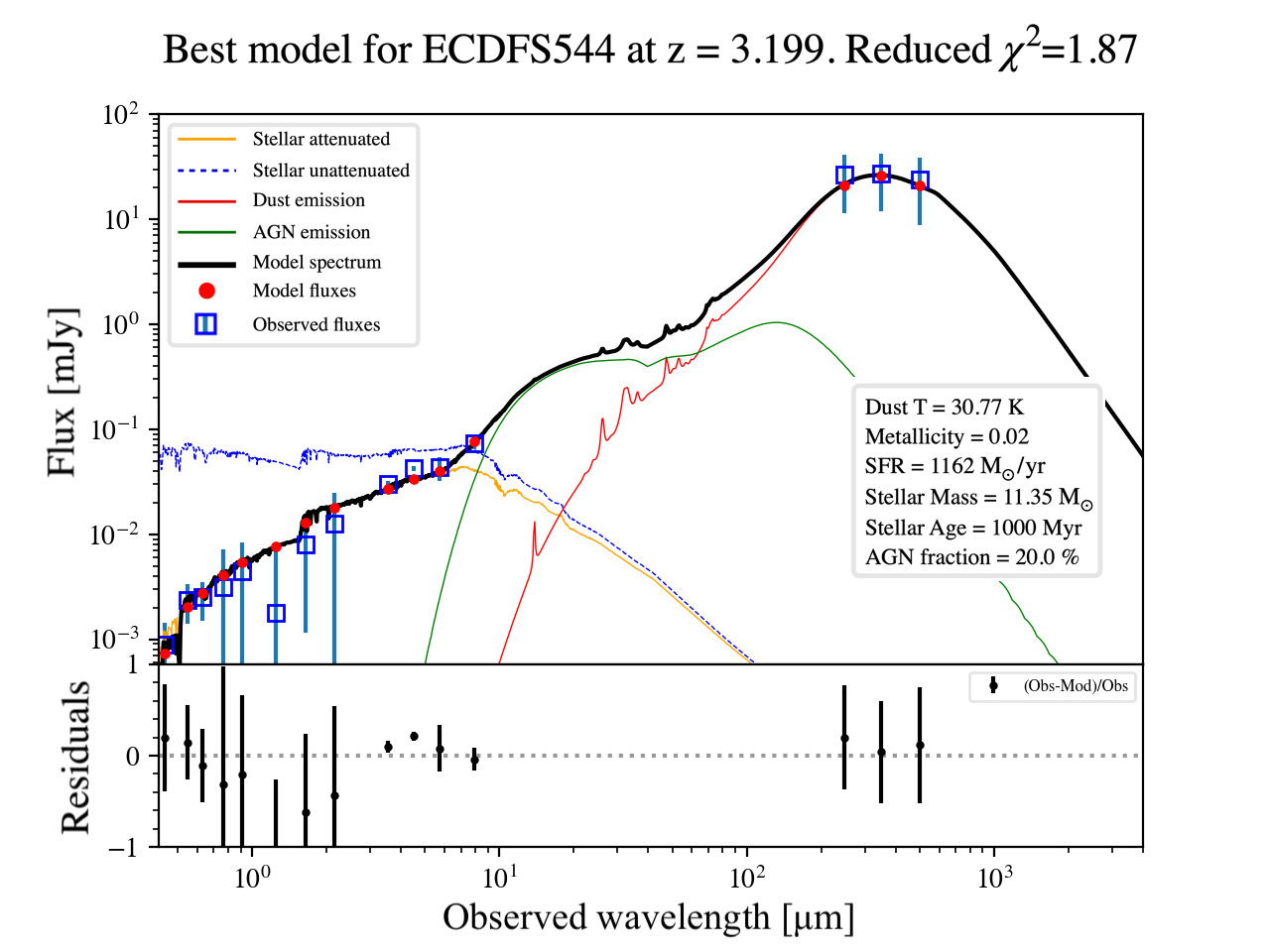}
\caption{CIGALE SED fitting results for the eight galaxies of our sample. The redshift values for CVLA1002, CVLA958, ALESS33, ECDFS544, ALESS14, and CVLA1000 were fixed to those derived from the literature, and the redshift for CVLA100 and CVLA028 was estimated from a range of values separated by $\Delta z =0.1$.}
\label{all_photometry}
\end{figure*}

The SED fitting analysis was conducted using the CIGALE software, employing consistent input models and parameters for the entire sample. The data fluxes from the Cosmos2015 catalogue were used in the analysis, while the radio information as well as any nebular emission component were not taken into consideration. For the sources that have been identified spectroscopically in the literature (i.e. CVLA1002, CVLA958, ALESS33, and ECDFS544), the redshift input was fixed to the corresponding spectroscopic value. From the remaining galaxies, ALESS14 and CVLA1000, the redshifts were fixed to the photometric values retrieved from the literature, while for CVLA100 and CVLA028, the best-fitting SED was selected by introducing $\Delta z=0.1$ intervals. The resulting fits, along with the model parameters used in this analysis, are presented in Fig. \ref{all_photometry} and Table \ref{CIGALE_models}.

To probe the significance of the AGN contribution, we applied the same fitting strategy using CIGALE without the AGN component. The updated SED fits revealed that for sources that initially required a lower AGN contribution (e.g. CVLA1000), the $\chi^2$ values changed marginally by approximately $\sim 10 \%$. On the other hand, for galaxies such as CVLA100, which presents a substantial AGN contribution, no accurate SED fit could be provided without an AGN component ($\Delta \chi^2 > 3$). Consequently, this exercise demonstrates the necessity of incorporating an AGN component in the SED fit for five out of eight sources, generally leading to improved fits as compared to those obtained without an AGN component.

\begin{table*}
\caption{Model parameters of CIGALE that were used for the SED fitting.}
\label{CIGALE_models}
\centering

\begin{tabular}{cccc}
\centering
Type       & Model name  & Reference     & Parameters used \\ \hline
\begin{tabular}[c]{@{}c@{}}Star formation\\ history\end{tabular} & sfhdelayed  & \citep[][]{sfhdelayed}      & \begin{tabular}[c]{@{}c@{}}age$\_$main = 1000, 2500, 4500, 6000, 8000 \\ tau$\_$burst = 10000.0\\ age$\_$burst = 10, 50, 80, 110\\ f$\_$burst = 0.001, 0.01, 0.03, 0.1, 0.2, 0.3\\ sfr$\_$A = 10.0,100.0,1000.0,3000.0 \end{tabular}  \\ \hline
Stellar emission       & bc03  &  \citep[][]{bc03}     & \begin{tabular}[c]{@{}c@{}}imf = 1(Chabrier)\\ metallicity = 0.0001,0.02\\separation$\_$age = 10\end{tabular}  \\ \hline
Dust attenuation       & dustatt$\_$2powerlaws & \citep[][]{dustatt_2powerlaws}      & \begin{tabular}[c]{@{}c@{}}Av$\_$BC = 0.3, 0.8, 1.2, 1.7, 2.3, 2.8, 3.3, 3.8\\ slope$\_$BC = -0.7 \\ BC$\_$to$\_$ISM$\_$factor = 0.3, 0.5, 0.8, 1.0\\ slope$\_$ISM = -0.7 \\filters = V$\_$B90, FUV\end{tabular}  \\ \hline
\begin{tabular}[c]{@{}c@{}}Lyman Continuum\\ escape\end{tabular} & lyc$\_$absorption & \citep[][]{lyc_absorption}      & \begin{tabular}[c]{@{}c@{}}$\rm f_{esc}$ = 0.0,0.5 \\ $\rm f_{dust}$ = 0.0,0.5\end{tabular}  \\ \hline
Dust emission       & dl2014  & \citep[][]{dale2014} & \begin{tabular}[c]{@{}c@{}}qpah = 0.47, 1.12, 2.5, 3.9\\ umin = 5.0, 10.0, 25.0\\ alpha = 2.0\\gamma = 0.02,0.9\end{tabular}  \\ \hline
AGN        & fritz2006  & \citep[][]{fritz2006} & \begin{tabular}[c]{@{}c@{}}r$\_$ratio=60.0\\ tau=1.0,6.0\\ beta=-0.5\\opening$\_$angle=100.0\\psy=0.001\\fracAGN$_{escape}$=0.0,0.1,0.2,0.6,0.8\end{tabular} \\ \hline
Redshift        & redshifting  & \citep[][]{redshifting} & redshift=0.01-7.5 (or fixed) \\ \hline
\end{tabular}
\end{table*}

\subsection{Sample presentation}
\subsubsection*{COSMOSVLA J100256.53+021158.4}
The COSMOS field deep multi-wavelength data provide the opportunity of examining the structure of this source, which appears to be composed of two galaxies: an optically bright quasar, and a fainter millimetre source (see the top left panel of Fig. \ref{figure3}). Spectroscopic data from the BOSS survey \citep{2016MNRAS.455.1553R} secured the detection of the Ly-$\alpha$ emission line at a redshift of $z=3.503 \pm 0.0007$. \textcolor{black}{Because the BOSS capabilities cannot resolve the two sources in study here}, this archival spectrum corresponds to both galaxies. However, considering that the main source (i.e. the quasar) is the brightest galaxy at optical wavelengths, we assume that the emission line and subsequently derived redshift correspond to this source. \\
\indent Furthermore, inspection of the 3 GHz data reveals that the quasar causes most of the radio flux density emitted from this system with a value of $f_{\rm3GHz-QSO}=0.052 \pm 0.02 \, mJy$ ($\sim 80 \, \%$ of the total), while the second source has a flux density of $f_{\rm3GHz-2^{nd} source}=0.013 \pm 0.009 \,mJy$. \textcolor{black}{When we assume} a typical radio spectral index of 0.8 for both sources\footnote{$S_{\nu} \propto \nu^{-\alpha}$, where $S_{\nu}$ corresponds to the flux density, $\nu$ is the frequency, and $\alpha$ is the radio spectral index, which is 0.8 for pure radio synchrotron emission.}, the corresponding radio flux density at 1.4 GHz is equal to $f_{\rm 1.4GHz-QSO}=0.1 \pm 0.04 \, mJy$ and $f_{\rm 1.4GHz-2^{nd} source}=0.024 \pm 0.017 \, mJy$. Therefore, if both galaxies are located at $z \sim 3.5$ and the detected radio emission was generated solely from SF, the corresponding SFR \citep[][ ]{2003ApJ...586..794B} would be $SFR_{\rm radio-QSO}=4246 \pm 1633 \, M_{\odot}/yr$ and $SFR_{\rm radio-2^{nd} source}=1061 \pm 735 \, M_{\odot}/yr$ for the quasar and SFG, respectively ($SFR_{\rm radio-total}=5307 \pm 1790 \, M_{\odot}/yr$). According to \citet{2017A&A...602A.123L}, this system of galaxies has a total stellar mass of $log(M_{\star}/M_{\odot})=11.69 \pm 0.01$ and $SFR_{\rm IR-total}$ of 1549 $M_{\odot}/yr$. The latter value is derived from the far-IR flux integrated at the position of the quasar and is rather close to our SED-fitting, best-derived SFR estimate of 1721 $M_{\odot}/yr$. Curiously, this value is similar to the value derived from the radio for the second source in the system (i.e. $SFR_{\rm radio}-2^{nd} source$). This is compatible with a system with a powerful quasar and a companion where most of the SF is taking place, and where no AGN is detected.\\
\indent It has to be noted that for most bands, we cannot separate the brightness contribution from each of the two sources because of the sub-arcsec separation and the insufficient angular resolution of telescopes such as \textit{Spitzer}. The SED fit was therefore performed using only total fluxes. A separation of the SED fitting analysis for the two sources would be non-trivial and would yield high uncertainties in the fluxes of each source. In the X-ray regime, COSMOS has been observed both with \textit{XMM-Newton} \citep[\textit{XMM}-COSMOS survey;][]{Cappelluti} and the \textit{Chandra} X-ray Observatory \citep[\textit{Chandra} COSMOS Survey and \textit{Chandra} COSMOS Legacy Survey;][]{2009ApJ..184..158E,2016ApJ...819...62C}, reaching flux limits of $f_{\rm 0.5-2 keV}\sim 1.7 \cdot 10^{-15} \, erg/s/cm^2$ and $f_{\rm 0.5-2 keV}\sim 2.7 \cdot 10^{-16} \, erg/s/cm^2$, respectively. The source CVLA1002 has been observed and detected by the \textit{Chandra} COSMOS survey with a flux of $f_{\rm 2-10keV}=6.67 \cdot 10^{-15}\, erg/s/cm^2$ \citep[see][]{2016ApJ...817...34M}.

\subsubsection*{ALESS J033152.49-280319.1}
Based on the current photometric data, this source seems to be a single galaxy, and according to the SED photometric analysis by \citet{2015ApJ...806..110D}, it has a photometric redshift of $z \sim 3.38$, an $SFR_{\rm IR}$ of 1202 $M_{\odot}/yr,$ and a stellar mass of $10^{10.96} \, M_{\odot}$. \textcolor{black}{For} our photometric analysis, which takes the possibility of an AGN component into account, the estimated best SFR value has a similar value of $SFR_{\rm IR}= 1188$ \, $M_{\odot}/yr$. Assuming that the galaxy is located at $z=3.38$ and its radio emission is solely due to star formation, the corresponding SFR from the radio emission would be equal to $SFR_{\rm radio}\sim 4213 \pm 757 \, M_{\odot}/yr$. A comparison between this value and the value derived from our SED fitting analysis or by \citet{2015ApJ...806..110D} suggests that the source hosts intense SF, and an accreting SMBH explains the clear radio excess that is observed. The X-ray energies have been observed in the CDF-S field with \textit{XMM-Newton} and the \textit{Chandra} observatories \citep[for more details, see][]{2013A&A...555A..42R, 2017ApJ..228....2L} in deep surveys, resulting in sensitivities of $6.6 \cdot 10^{-16} \, erg/s/cm^2$ and $2.7 \cdot 10^{-17} \, erg/s/cm^2$ at 2-10 keV, respectively. However, ALESS14 is not detected in any of these surveys, supporting the view that if an AGN exists, it contributes mostly to the radio emission, and at a much lower level at X-ray and optical/IR wavelengths. This was also suggested by the $\sim 20 \%$ AGN contribution to the IR emission from our CIGALE SED fit.

\subsubsection*{COSMOSVLA J100004.81+023045.2}

\begin{figure*}
\centering
\includegraphics[width=1.39\columnwidth]{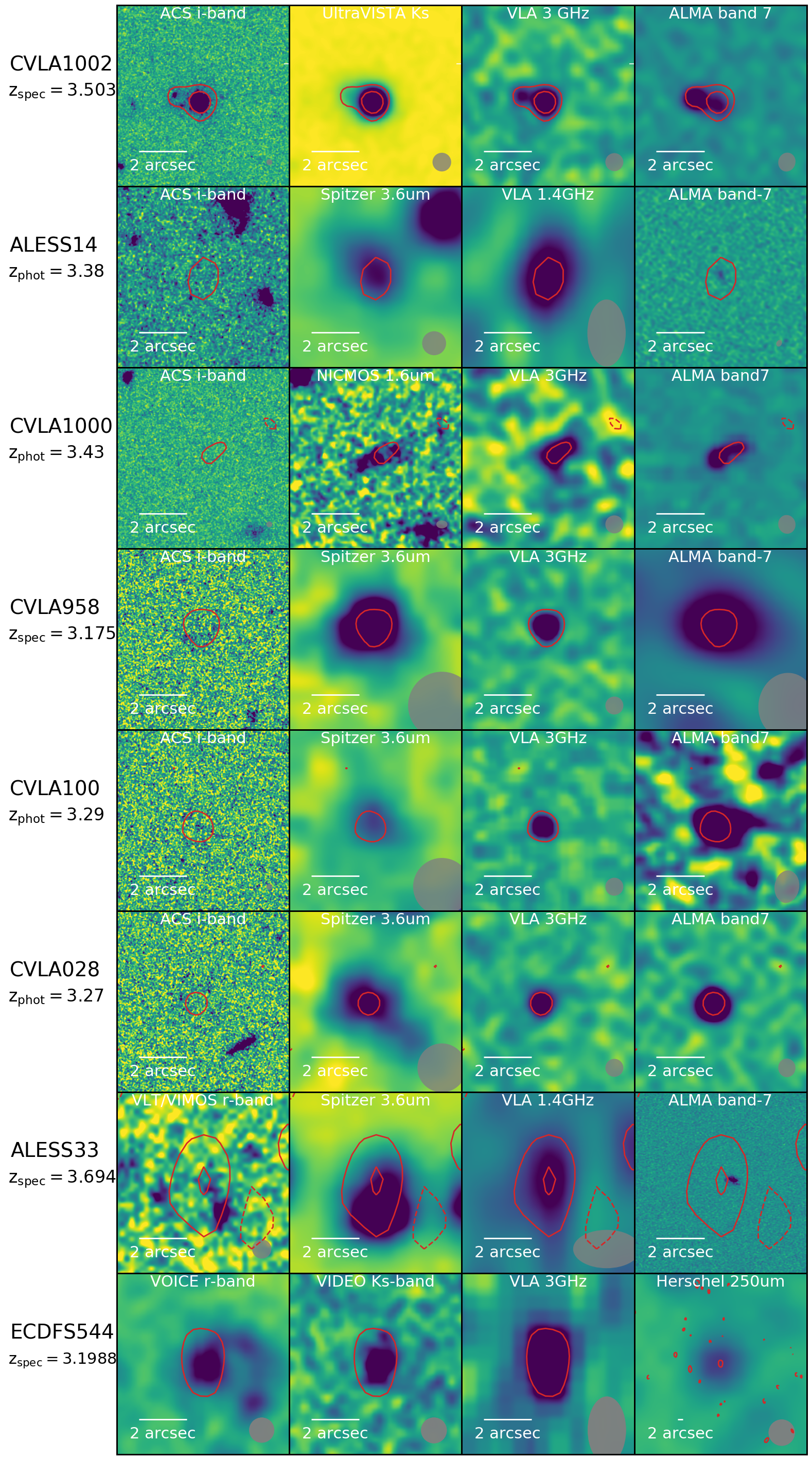}
\caption{Multi-wavelength view of the eight high-redshift galaxies of our sample, ordered according to Table \ref{radio_sources_table_2}. From left to right, we present the i- or r-band (HST, VIMOS, and VOICE data), the \textit{Spitzer} 3.6$\mu$m/VIDEO Ks-band/NICMOS 1.6$\mu$m images, the radio VLA 1.4/3 GHz and the FIR data from ALMA and \textit{Herschel}, and the red contours correspond to the VLA emission. From top to bottom, the contour levels of the radio image for each source are ($7.715\cdot 10^{-6}$, $2.093\cdot 10^{-5}$), $2.827\cdot 10^{-5}$, ($-4.710\cdot 10^{-6}$, $1.058\cdot 10^{-5}$), $5.640\cdot 10^{-6}$, ($-4.845\cdot 10^{-6}$, $5.461\cdot 10^{-6}$), ($-5.078\cdot 10^{-6}$, $1.407\cdot 10^{-5}$), ($-4.633\cdot 10^{-6}$, $1.124\cdot 10^{-5}$, $2.712\cdot 10^{-5}$), and $2.092\cdot 10^{-5}$ Jy/beam. Each panel has a size of 7.2x7.2 $arcsec^2$ , except for the \textit{Herschel} image, which for visualisation purposes has a larger extent. Additionally, the beam size (shaded light grey area) of each instrument and a 2 arcsec scale bar are added in each panel.}
\label{figure3}
\end{figure*}

According to the HELP catalogues \citep[][]{2021MNRAS.507..129S}, this source is located at a redshift of 3.43 with an $SFR_{\rm IR}$ of 1519 $M_{\odot}/yr$, while our SED fitting analysis predicts a lower $SFR_{\rm IR}$ of 690 $M_{\odot}/yr$. Based on the SED fitting results, modelling of the IR emission of CVLA1000 requires an AGN contribution. According to the Cosmos2015 and Cosmos2020 \citep[][]{2022ApJS..258...11W} catalogues, the photometric redshifts are 3.26 and 3.62, respectively. When we accept an average redshift value of $\sim 3.3$ for this source, then according to its radio flux density of $f_{\rm 1.4GHz}=0.096 \, mJy,$ the $SFR_{\rm radio}$ should be $\sim 3723 \, M_{\odot}/yr$, which in comparison with the previous estimate from the IR emission strongly suggests an active SMBH at the centre of this galaxy.\\
\indent Due to its morphology in the near-IR, radio, and mm wavelengths, this source could be classified as a group of galaxies (see the third row of Fig. \ref{figure3}), however, catalogues with large samples of galaxies (i.e. Cosmos2015 and Cosmos2020) identify this galaxy as a single source \textcolor{black}{due to the lack of resolution}. With a radial extent of $\sim 1.146$ arcsec, measured at the 1.6 $\mu m$ - NICMOS data, and a scaling factor of $7.524 \, kpc/''$ (at $z=3.3$), the actual size of this galaxy can be estimated as $\sim 8.6\, kpc$. This value is compatible with an ultra-compact dwarf galaxy \citep[e.g.][]{2021MNRAS.505.3192R}, however, the images reveal a source that resembles a disk rather than a spherical object. Inspection of all the available images cannot exclude the possibility of a merging system, considering that the NICMOS data depict a structure that could resemble a complex of galaxies. No detection in the X-ray bands is reported in the literature for this galaxy.

\subsubsection*{COSMOSVLA J095845.94+024329.2}
Source CVLA958 was recently spectroscopically observed by our team with the Institut de Radioastronomie Millimetrique (IRAM) 30m antenna (project code: 227-19; for more details regarding these observations, see Appendix \ref{appC}). Unfortunately, its mm faint nature prevented us from acquiring a secure redshift determination with the IRAM telescope. According to Cosmos2015, the photometric redshift of this galaxy is $z=3.172$, while our photometric analysis suggests a minimum of $\chi^2$ at 3.21. The latter redshift solution agrees with the only line that is detected from recent ALMA archival spectra (project code: 2019.1.01600.S). The detected emission line at $\sim 110.5 \, GHz$ has a full width at half maximum of $\sim$0.4 GHz ($\sim 70.4\, km/s$), possibly indicating a merging system or an orderly rotating spiral galaxy \citep[see][for more details]{2022ApJ...929..159C}. The recent work by \citet{2021MNRAS.501.3926B} also identified the emission line as due to the CO4-3 rotational transition at band 3 of ALMA, locating this source at $z=3.1755$. According to our photometric SED fitting analysis, this source has a stellar mass of $10^{11.07}\, M_{\odot}$, an SFR of 1177 $M_{\odot}/yr,$ and a significant AGN contribution. Considering the measured flux density at 1.4 GHz of 90 $\mu Jy$, we estimate that its $SFR_{\rm radio}$ (assuming that its radio emission comes purely from star formation) should be $\sim 3376 \, M_{\odot}/yr$. This value is significantly higher than the value determined from the SED fitting (see Table \ref{table2} for further details), supporting the presence of an AGN in this source as well. CVLA958 is not detected in any available dataset in the X-ray wavelengths.

\setlength{\tabcolsep}{4pt}
\begin{table*}
\caption{Derived and observed properties of the galaxy sample. The SFR values derived from the IR emission are a product of the SED fitting described in the text, while the radio-derived SFRs are derived by assuming that the radio emission is solely due to star formation without contribution by an AGN. The second row of the table corresponds to the units of each parameter.}
\label{table2}
\centering
\resizebox{\textwidth}{!}{%
\begin{tabular}{ccccccccc}
Name & Redshift$^{(1)}$ & Stellar Mass & $\rm SFR_{IR}$ & $\rm SFR_{radio}$ & AGN $^{(2)}$ & $\rm L_{1.4GHz}$ & $\rm L_{IR}$ $^{(3)}$ & $q_{TIR}^{(4)}$ \\
-- & --& [$M_{\odot}$] & [$M_{\odot}/yr$] & [$M_{\odot}/yr$] & [$\%$] & [$10^{24} W/Hz$] & [$10^{39} W$] & --  \\ \hline
CVLA1002 & $3.503 \pm 0.001^{(s)}$ & $11.62 \pm 0.44$  & $2496 \pm 1266$ & $5307 \pm 1790$ & $41 \pm 22$ & $8.34 \pm 1.60$ & $3.15 \pm 0.01$ & $2.00 \pm 0.08$ \\
ALESS14 & $3.380 \pm 0.25^{(p)}$  & $11.07 \pm 0.85$ & $1123 \pm 142$  & $4213 \pm 757$ & $20 \pm 1$ & $7.63 \pm 1.37$ & $2.19\pm 0.20$ & $1.88\pm 0.09$ \\
CVLA1000 & $3.430 \pm 0.32^{(p)}$  & $12.13 \pm 1.30$ & $711 \pm 40$  & $4057 \pm 792$ & $10 \pm 2$ & $7.35 \pm 1.44$ & $2.15\pm 0.25$ & $1.89\pm 0.10$ \\
CVLA958 & $3.175 \pm 0.001^{(s)}$ & $11.19 \pm 0.44$ & $1225 \pm 546$  & 3382$ \pm 641$ & $20 \pm 4$ & $6.13 \pm 1.16$ & $0.58\pm 0.02$ & $1.40\pm 0.11$ \\
CVLA100 & $3.290 \pm 0.39^{(p)}$  & $10.36 \pm 0.52$ & $528 \pm 96$  & $3391 \pm 781$ & $60 \pm 3$ & $6.14 \pm 1.41$ & $1.98\pm 0.32$ & $1.93\pm 0.10$ \\
CVLA028 & $3.270 \pm 0.24^{(p)}$  & $11.13 \pm 0.30$ & $1565 \pm 753$  & $2167 \pm 602$ & $3 \pm 7$  & $3.92 \pm 1.09$ & $2.52\pm 0.22$ & $2.23\pm 0.13$ \\
ALESS33 & $3.694\pm 0.001^{(s)}$  & $11.27 \pm 0.81$ & $978 \pm 113$  & $2140 \pm 597$ & $10 \pm 1$ & $3.88 \pm 1.08$ & $2.20\pm 0.01$ & $2.18\pm 0.12$ \\
ECDFS544 & $3.199\pm 0.001^{(s)}$ & $11.42 \pm 0.80$ & $1083 \pm 470$  & $1520 \pm 507$ & $20 \pm 3$ & $2.75 \pm 0.92$ & $2.01\pm 0.01$ & $2.29\pm 0.15$ \\
\hline
\end{tabular}%
}

\scriptsize $^{(1)}$(s): spectroscopic redshift; (p): photometric redshift (accepting the solution with the lowest $\chi^2$ or the best available in the literature), $^{(2)}$: the AGN percentage (Bayesian values) is estimated by our CIGALE fitting and corresponds to the same wavelength range in which the dust luminosity of each source was estimated, $^{(3)}$ this infrared luminosity corresponds to the wavelength range 42-122 $\mu$m (rest frame), $^{(4)}$ according to equation 5 by \citet{2022ApJ...927..204E}: $\rm q_{TIR}=log(L_{IR}/3.75 \cdot 10^{12})-logL_{1.4GHz}$
\end{table*}

\subsubsection*{COSMOSVLA J100233.16+020626.5}
The source CVLA100 seems to be a single galaxy as well. Our photometric analysis places this source at a redshift of $z_{\rm phot}\sim 3.3$. This result agrees with the estimates of other photometric catalogues \citep[see][]{Laigle16,2022ApJS..258...11W}. ALMA spectroscopic data acquired by our team (project code: 2017.1.01713.S; see Appendix \ref{appC} for more details) reveal an emission line at $\sim 107.3 \, GHz$, which corresponds to the CO(4-3) emission line with the galaxy located at $z=3.29$. For the derived physical properties of this source, our SED fitting analysis indicates a stellar mass of $10^{10.27} \, M_{\odot}$, an SFR of 566 $M_{\odot}/yr,$ and an AGN fraction of 60 $\%$. Application of equation 6 by \citet{2003ApJ...586..794B} provides us with an estimate of the $SFR_{\rm radio}$ of 2760 $M_{\odot}/yr$. The SED fitting models for this redshift suggest an SFR of $\sim 500 \, M_{\odot}/yr$, which could mean that the observed 1.4 GHz radio flux cannot be obtained solely from SF, but by a combination of star emission and AGN activity (see Table \ref{table2} for additional information). Similarly to CVLA958, this source is not detected at any high-energy band.

\subsubsection*{COSMOSVLA J100028.75+023203.5}
As mentioned in the previous section, this source (which seems to be a single galaxy) has a spectroscopic redshift of 4.76 \citep[see][]{2018ApJ...858...77H}, \textcolor{black}{however,} inspection of the DEIMOS spectrum does not reveal a secure emission line (i.e. the Ly-$\alpha$ line, according to the authors). No further spectroscopic data are available in the literature. Our own photometric SED fitting analysis suggests a redshift of 3.27, which is consistent with the redshift provided by the HELP database and \citet{2015A&A...577A..29M}. When we assume this photo-$z$ to be the true redshift of this galaxy, our SED fit analysis proposes a stellar mass of $10^{10.74} \, M_{\odot}/yr$, an SFR of 2554 $M_{\odot}/yr,$ and no AGN contribution. The derived SFR from the radio emission of this galaxy (i.e. 2166 $M_{\odot}/yr$) further supports the argument that no radio emission is added by SMBH activity. This result establishes CVLA028 as the only galaxy in our sample without an indication of an AGN by the SED fit or radio excess. Although this galaxy has been observed by both \textit{XMM-Newton} and \textit{Chandra}, the images at all bands reveal no clear detection.

\subsubsection*{ALESS J033211.34-275211.9}
This source is recognised in the literature as an X-ray AGN with faint radio emission. Photometric analysis from various deep surveys \citep[e.g.][]{2020MNRAS.492.1887G} reveals a redshift of $z\sim 3.7$, while ALMA spectroscopic data \citep[see][]{2021MNRAS.501.3926B} show two clear emission lines (namely the CO4-3 and [CI] lines at $\sim 98.22\, GHz$ and $\sim 104.85\, GHz$, respectively). This secures the galaxy at $z=3.694$. Assuming that this spectroscopic redshift and that a radio flux density of $0.043 \, mJy$ solely originate from star formation, the corresponding SFR is $\sim 1884 \, M_{\odot}/yr$. However, according to \citet{2020MNRAS.492.1887G}, following the SED fitting method using CIGALE, this galaxy is a type II AGN with a stellar mass of $\sim 10^{11.3}\, M_{\odot}$, an SFR of $\sim 31 \, M_{\odot}/yr$ (our CIGALE SED fitting analysis suggests a higher value of 1194 $M_{\odot}/yr$), and an AGN fraction of 72 $\%$ (our analysis proposes a lower value of 10 $\%$ ). The source detected in the HST image (see Fig. \ref{figure3}), which is rather close to the ALMA detection ($\sim 0.5 \, arcsec$), seems to cause the radio and IR emission \citep[see][]{2011MNRAS.415.1479W}. Therefore, this source could be a complex system of at least two galaxies possibly interacting, with the presence of a radio-quiet AGN and an SFG. Even though ALESS33 is not detected in any of the X-ray surveys, another source lies at a distance of $\sim 2$ arcsec \citep[named $\rm LBX2017-14$; see][for more details]{2013A&A...555A..42R} that seems to be an X-ray bright AGN with a flux of $f_{\rm 2-10keV}=6.26 \cdot 10^{-15} erg/s/cm^2$ and no counterparts. This X-ray source was spectroscopically identified by \citet{2013A&A...555A..42R} at $z=3.74 \pm 0.06$, meaning that it could be part of a merging system along with the sub-millimetre and radio galaxy identified with the source ALESS33.

\subsubsection*{[DSS2017] 554}
This source is present in the VIMOS spectroscopic survey presented by \citet{2017ApJ...840...78D} with a spectroscopic redshift of 3.1988, which was used in this work as a fixed value for the subsequent SED fitting analysis. This galaxy is placed at the outskirts of the ECD-S field, meaning that it was not selected in the photometric analysis by \citet{2020MNRAS.492.1887G}. Inspection of the VST Optical Imaging of the CDFS and ES1 \citep[VOICE;][]{2016heas.confE..26V} data for this source in bands r, g, and i (see the bottom row of Fig. \ref{figure3}) suggests the possibility of multiple components. These observations might signify a merging or interacting system. The predicted SFR derived solely from the radio emission is 1520 $M_{\odot}/yr$, while our SED fitting analysis suggests a value of 1162 $M_{\odot}/yr$ and an AGN contribution of $20\%$ . No clear detection has been reported regarding the possible X-ray nature of this source.

\section{Discussion}\label{discussion}
\begin{figure*}[t!]
\centering
\includegraphics[width=2\columnwidth]{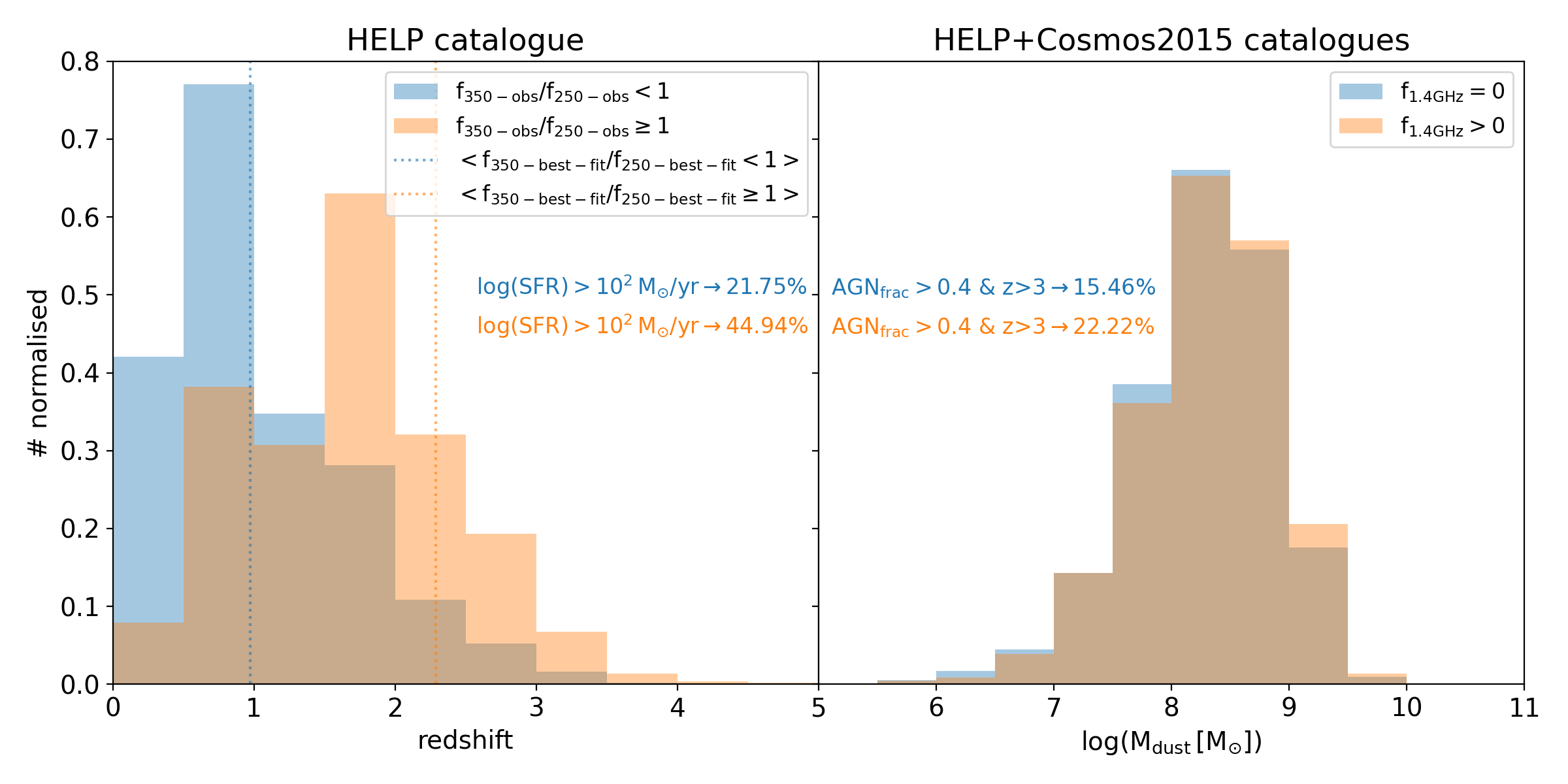}
\caption{Normalised redshift histograms of all galaxies from the HELP catalogue (panel a), separated with the criterion $f_{350 \rm \mu m}/f_{250 \rm \mu m}=1$. Flux densities were estimated using the \textit{Herschel} observational data as provided from the same catalogue. The vertically dotted coloured lines correspond to the average values of the same histograms, employing the flux densities from the CIGALE best model fit, and are coloured accordingly. The text in the panel denotes the percentage of the sources that present an SFR above $10^2 \, M_{\odot}/yr$ and are coloured according to the aforementioned criterion. Panel b illustrates the normalised histograms for the dust content of the galaxies in both the HELP and Cosmos2015 catalogues. The blue and orange histograms are derived from galaxies without and with radio VLA 1.4 GHz emission, respectively. The text of the panel demonstrates the percentage of the galaxies that present an AGN component ($AGN_{\rm frac}>40 \%$) and are placed at $z>3$.}
\label{criteria_test}
\end{figure*}
We have examined a sample of eight high-redshift dusty galaxies with an AGN component (3 <\,$z_{\rm spec,phot}$\,< 4). The selection process involved using the 1.4 GHz VLA radio catalogues from the COSMOS and ECDF-S fields in combination with the far-IR peak criterion ($f_{350 \rm \mu m}/f_{250 \rm \mu m}>1$) using the HerMES survey. Moreover, a compact morphology was considered as an additional criterion. In the following paragraphs, we assess the efficacy of these criteria in selecting these sources. Furthermore, we provide a comparison between the derived properties of our sample with those obtained in previous studies, focusing on similar redshift ranges.

\subsection{Validity of the selection criteria}
While the compactness criterion merely serves as a method for avoiding local extended galaxies, the far-IR and radio criteria have been employed extensively in the past in the study of the distant Universe. For instance, \citet{2010A&A...518L...9A} demonstrated that the redshift of a source tends to increase for galaxies that satisfy the criteria $f_{350 \rm \mu m}/f_{250 \rm \mu m}>1$ and $f_{500 \rm \mu m}/f_{350 \rm \mu m}>1$. These results are expected, considering that the far-IR emission of an SED traces the dust content of a galaxy, and more specifically, its blackbody radiation, which is shifted to higher wavelengths with increasing redshift. However, this emission is affected by the temperature of the dust as well, which consequently introduces a degeneracy that is \textcolor{black}{non-trivial} to break. Another point of attention is the sensitivity of the \textit{Herschel} surveys, which are inherently biased towards the detection of the brightest dust-rich far-IR galaxies.\\
\indent In order to evaluate this argument and selection method further, we employed the HELP catalogue, which contains far-IR observational data (i.e. 15747 \textit{Herschel} detected galaxies) along with SED fitting results from CIGALE \citep[for a detailed description of the CIGALE parameters used in this catalogue, see][]{2018A&A...620A..50M}. The galaxies from this catalogue are presented in the left panel of Fig. \ref{criteria_test}, in which the blue and orange histograms correspond to the sources that follow the criterion $f_{350 \rm \mu m}/f_{250 \rm \mu m}<1$ and $f_{350 \rm \mu m}/f_{250 \rm \mu m}>1$, respectively. As anticipated, galaxies that adhere to the far-IR criterion are situated at higher redshifts on average. Interestingly, based on the CIGALE fits, a higher percentage (i.e. $\sim 45 \%$) of the galaxies that follow the $f_{350 \rm \mu m}/f_{250 \rm \mu m}\geq 1$ criterion exhibit SFRs of $SFR>10^2 \, M_{\odot}/yr$. In contrast, for the rest of the sources, a smaller percentage ($\sim 22 \%$) demonstrates high SFR values like this.\\
\indent A similar exercise was repeated for the radio criterion by estimating the AGN fraction and dust mass, derived from the CIGALE fits, for each galaxy. \textcolor{black}{Because} the HELP catalogues do not consider the radio emission in their fits, we cross-matched this catalogue with Cosmos2015, which includes the 1.4 GHz VLA data. The results of this analysis are displayed in the right panel of Fig. \ref{criteria_test}, which shows histograms of the dust mass. The shape and mean values of these histograms indicate that the radio selection criterion does not guarantee the preferential selection of highly dusty SFGs. However, upon examining galaxies at redshifts above 3, it becomes evident that the radio selection method identifies a higher proportion of sources (i.e. $\sim 22 \pm 9 \%$) with a substantial AGN fraction (i.e. $AGN_{\rm frac}>40 \%$) compared to cases without detected radio emission ($\sim 15 \pm 4\%$).\\
\indent While this percentage difference may not be statistically significant, the combination of the two far-IR and radio selection methods still provides motivation for their application in identifying high-redshift SFGs with an AGN component. The detailed analysis of the eight sources in this sample has reinforced these findings and encourages their application in large-scale surveys, such as EMU.

\subsection{Comparison with previous work}
Overall, our sample has a mean redshift of $<z_{\rm}>=3.37 \pm 0.16$, an average stellar mass of $<log(M_{\rm \star}/M_{\odot})>= 11.27 \pm 0.47$, $<L_{\rm IR,42-122\mu m}>=2.1 \pm 0.67 \cdot 10^{39} \, W$ (total IR luminosity of $<L_{\rm IR,8-1000\mu m}>=3.9 \pm 0.84 \cdot 10^{39} \, W \sim 10^{13}\, L_{\odot}$), $<L_{1.4GHz}>=5.77\pm 1.9 \cdot  10^{24} \,W/Hz,$ and a mean star formation rate (derived from the $42-122 \, \mu m$ rest-frame IR luminosity of each source) of $<SFR_{\rm IR}> 1213 \pm 567\, M_{\odot}/yr$. This average SFR is consistent with previous results that reported an increasing SFR at higher redshift. For instance, \citet{2016MNRAS.461.1100R} and \citet{2022A&A...668A..54G} showed that \textit{Herschel}-detected samples of hyper-luminous infrared galaxies (HLIRGs; $L \sim 10^{12}-10^{13}\, L_{\odot}$) at $z>3$ have SFR values ranging between $\sim 1000-10000 \, M_{\odot}/yr$. Therefore, the inclusion of the 500 $\mu m$ emission in our selection criteria naturally provides us with the high bright tail of the SFR distribution at these redshifts. Similarly, \citet{2019A&A...624A..98W} reported that the selection of \textit{Herschel} galaxies above $z>3$ revealed dusty SFR galaxies with infrared luminosities of $\sim 10^{12}-10^{13} \, L_{\odot}$, similar to the derived values of our sample.\\
\indent Furthermore, for 87.5 $\%$ of the sources in our sample (seven out of eight), an AGN is indicated either by the radio excess or by the mid-IR signature obtained from the best SED fit. Although the morphology of these sources still remains uncertain, five out of eight (63 $\%$) might consist of multiple galaxies, with significant levels of both SF and AGN activity. In comparison, \citet{2021A&A...654A.117G} and \citet{2021A&A...648A...8W}, who analysed a sample of HLIRGs selected using data from \textit{Herschel} and the Deep Fields of the LOFAR Two-metre Sky Survey \citep[LOTSS;][]{2017A&A...598A.104S}. described an AGN contamination of 34-64 $\%$ (using six different models to fit the AGN component) for galaxies at the redshift range $3<z<4$. These values, although model dependent, seem to be lower than the AGN contribution of our sample, which can be attributed to the fact that the LOTSS survey reaches sensitivities of 20-30 $\mu Jy/beam$ at 150 MHz. When we assume a typical radio synchrotron slope, this value translates into a lower flux density value at 1.4 GHz than the COSMOS-VLA sensitivity (i.e. 7 $\mu Jy/beam$). Therefore, it seems that by selecting radio sources from a deeper radio survey, the authors revealed a population of SF galaxies that receive contributions from AGN less frequently. However, it is important to emphasise that a comparison of our results with studies involving larger sample sizes should be approached with caution, considering the limited statistics that our data provides.\\
\begin{figure*}
\centering
\includegraphics[width=2\columnwidth]{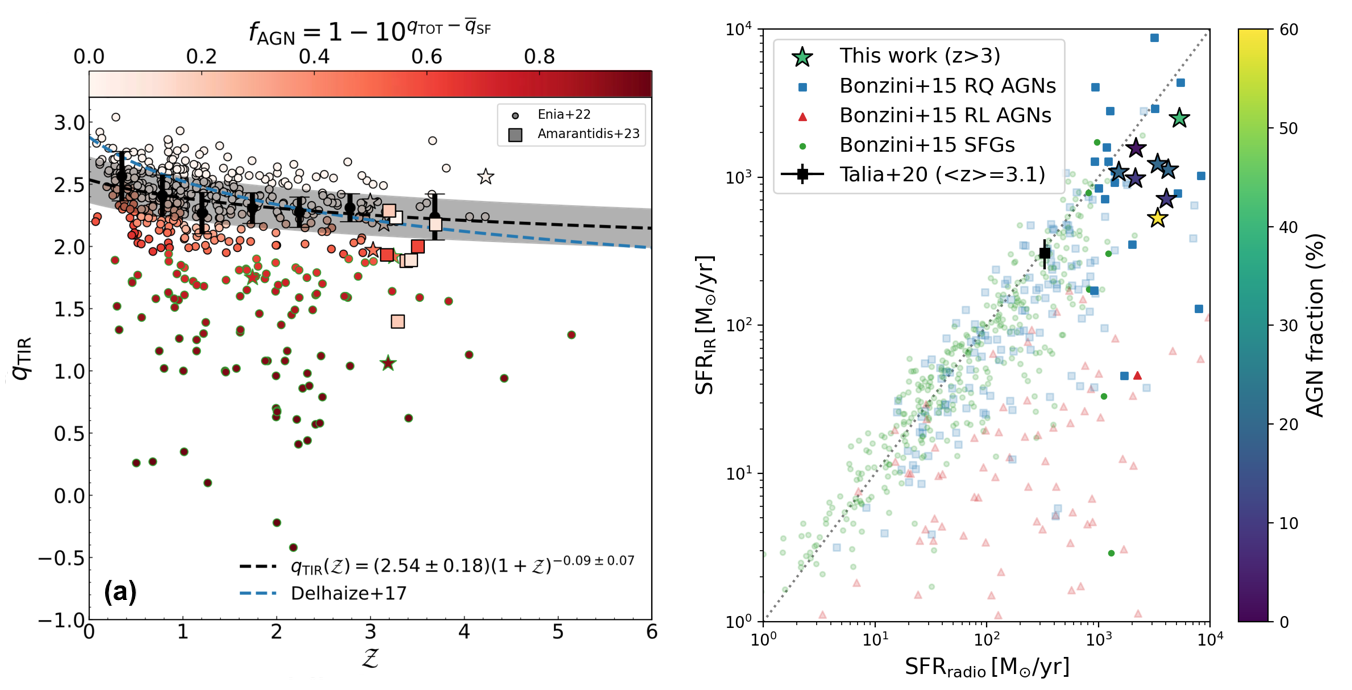}
\caption{The $q_{TIR}$ parameter over redshift for the sample explored by \citet{2022ApJ...927..204E} and for our 8 galaxies (panel a). This figure was retrieved by \citet{2022ApJ...927..204E}, published under the CC BY 4.0 international licence and was modified by adding the results from our work (square points). The dashed black line and grey shaded region correspond to the best-fit relation of their total sample and the standard deviation, respectively, while the dashed blue line shows a similar relation by \citet{2017A&A...602A...4D}. All points by \citet{2022ApJ...927..204E} are coloured according to the contribution of AGN to the radio emission defined as $f_{\rm AGN}=1-10^{q_{\rm tot}-\bar{q_{\rm SF}}}$, with $q_{\rm tot}$ and $\bar{q_{\rm SF}}$ corresponding to the $q_{\rm TIR}$ and its evolution with redshift, respectively. The star symbols depict what is defined by \citet{2022ApJ...927..204E} as H-dark sources (i.e. not detected at the \textit{Hubble}-WFC3 H-band). Panel b, presents the SFRs derived from the IR and radio luminosity, for the 8 sources of our sample, along with the results by \citet{2015MNRAS.453.1079B} (radio-quiet AGN: blue squares, radio-loud AGN: red triangles, SF galaxies: green circles) and \citet{2021ApJ...909...23T} (average values of their sample). All data points with redshift below 3 are coloured with transparency, while the ones with z>3 have normal colours.}
\label{last_figure}
\end{figure*}
\indent \textcolor{black}{Additionally, \citet{2022ApJ...927..204E} reported a negligible percentage of AGN for the entire sample ($0<z<6$)}, which was selected using deep radio 1.4 GHz observations from the GOODS-N field (sensitivity of 2.2 $\mu Jy$). This can be perceived with the parameter $\rm q_{TIR}=log(L_{IR}/3.75 \cdot 10^{12})-logL_{1.4GHz}$, which is commonly used as an indicator of the presence of AGN contribution to the radio emission. Figure \ref{last_figure} depicts the $q_{TIR}$ values for our sample and those of \citet{2022ApJ...927..204E}, demonstrating the AGN contamination in the radio emission in most of our galaxies. In more detail, 62.5 $\%$ of the sources in our sample have lower values of $q_{\rm TIR}$ than the grey region of the plot (which denotes the average redshift-dependent $q_{\rm TIR}$). These statistics do not fully match the number of 87.5 $\%$ of AGN contamination that was derived from the SED fitting and described earlier in the text. However, this discrepancy might originate from the low AGN contribution and uncertainties in the measurements of the radio and infrared fluxes of the SED of some galaxies. In contrast, the sample by \citet{2022ApJ...927..204E} for the redshift bin of $3<z<4$ indicates an AGN contamination of 37.5 $\%$ . This difference can be attributed to the fact that in their selection criteria, they did not require the far-IR emission (or better, far-IR peak spectrum) that we employed. As a result, only 38.4 $\%$ of their total sample has a 500 $\mu m$ \textit{Herschel} detection. \\
\indent Another comparison can be conducted with the $q_{TIR}$ relation derived by \citet{2021A&A...647A.123D} and presents a mass-dependent and almost redshift-independent equation (i.e. $q_{TIR}(M_{\star},z)=(2.646\pm 0.024)(1+z)^{-0.023\pm0.008}-(logM_{\star}/M_{\odot}-10)(0.148\pm0.013)$). According to this relation, all of the galaxies in our sample satisfy the criterion, placing them below the mean of the SFR population. However, none of the sources in our sample follows the relation $log(L_{1.4GHz}/SFR_{IR})=21.984 \times (1+z)^{0.013}$ that was derived earlier by \citet{2017A&A...602A...3D} and corresponds to the $3 \sigma$ deviation from the peak distribution of the ratio of $L_{\rm 1.4 GHz}$ and $SFR_{\rm IR}$ as a function of redshift.\\
\indent In the same manner, \citet{2017A&A...602A...5N}, which selected sources from the VLA 3GHz survey without a far-IR emission requirement, similarly presented the trend of a lower contamination of AGN (23 $\%$ and 47 $\%$ of the total sample and $3<z<4$ redshift bin, respectively) than in our sample. This comparison favours the inclusion of the criterion of a far-IR rising spectrum to the radio emission when AGN at high redshifts are to be identified and studied. In contrast, an interesting comparison can be performed with studies such as that by \citet{2015MNRAS.453.1079B}, in which the authors presented a sample with galactic properties similar to those of our galaxies. This sample was derived from the radio 1.4 GHz ECDF-S catalogues (6 $\mu Jy$ sensitivity) with the addition of the \textit{Herschel} detection in all sources at the redshift bin of interest (i.e. $z=3-4$). Fifteen of the 28 galaxies presented in this redshift range were categorised as radio-quiet (RQ) AGN while one as a radio-loud (RL) AGN, totalling 16 galaxies, or an AGN contamination of 73 $\%$ . This value is close to the value derived from our sample, which further supports the necessity of including the far-IR criterion to detect high-$z$ AGN. Additionally, the right panel of Fig. \ref{last_figure} shows that our galaxies fall in the RQ region of their work, even though our estimates indicate that at least two of our sources should have a significant AGN contribution.\\
\indent In this sense, our selection method might provide an increased probability for the detection of dusty galaxies at high-$z$ that present high SFRs with an additional AGN component, with weak radio emission, as part of a merging system. Additionally, our galaxies seem to be a more extreme group of interacting systems, having brighter IR and radio luminosities as well as stellar masses and SFRs as compared, for instance, to the work by \citet{2021ApJ...909...23T}. This outcome might originate from the fact that their sample, once again, did not require a far-IR \textit{Herschel} detection and was originally derived from the 3 GHz - COSMOS survey, which has a higher detection limit than the VLA-1.4 GHz-COSMOS survey we employed. Nevertheless, \citet{2017A&A...602A...5N} also explored the same field at 3 GHz and reached the same flux density sensitivity (rms of 2.3 $\mu Jy$), from which they achieved more AGN detections. The main difference lies in their selection criteria, with the latter work requiring an optical/near-IR counterpart, while \citet{2021ApJ...909...23T} excluded these sources. The selection of radio sources with a far-IR peak spectrum and an optical counterpart might therefore reveal merging systems that at high $z$, present high SFRs ($\sim 10^3 \, M_{\odot}/yr$) and an AGN counterpart in most cases.\\
\indent Nevertheless, selecting galaxies that are considered as extreme in terms of mass and luminosity will naturally limit the study to a rare sample, hence to a reduced number of selected sources. This infrequency, along with the complexity of the adopted selection  method, will unavoidably provide a less accurate estimation of the SFRD at $z>3$, which can be only considered as a lower limit \citep[following the $V_{\rm max}$ method introduced by][]{1980ApJ...235..694A}, and for our study, this is equal to $SFRD>8 \cdot 10^{-3}\, M_{\odot}/yr/Mpc^3$. However, these selection criteria might be quite powerful in the aforementioned identification and in studying the interplay between AGN and SF activity at high redshift, especially with their application to the upcoming next-generation radio surveys (e.g. the EMU survey) and current/future observational data from mm facilities \citep[e.g.][]{2023arXiv230507054B}.

\section{Conclusions}\label{conclusions}
\textcolor{black}{In this work, we present the exploration of} a sample of a high-redshift ($3<z<4$) radio galaxy population that was selected with the criterion of a rising far-IR spectrum (i.e. $f_{350\mu m}>f_{250\mu m}$) along with the detection of non-extended (<10 arcsec) radio emission at 1.4 GHz. The aim behind this selection method was to investigate the potential of combining these two criteria in identifying high-redshift dusty SFGs that are contaminated by an AGN and might be `hidden' by other selection methods. A population like this, due to its extremity and low occurrence, \textcolor{black}{might} not contribute significantly to the SFRD of the Universe, but it can provide us with further insight into the interplay between SMBHs and SF activity at high $z$.

In order to explore the morphology of our sample and its physical characteristics in depth, we took advantage of the multi-wavelength deep data available in the literature from the COSMOS and ECDF-S fields. These data sets enabled us to identify that five out of eight galaxies in our sample might consist of more than one galaxy with possible interaction/merging. It has to be noted that the low angular extent of these galaxies did not permit confirmation of this statement \textcolor{black}{due to} the spatial resolution of the current instrumentation. Nevertheless, future observations with telescopes such as the \textit{James Webb} Space Telescope (\textit{JWST}) may provide further insights into this argument. Furthermore, we determined the presence of an AGN component in seven out of eight of these sources, which in most cases represents a weak radio emission, \textcolor{black}{identifying} these sources as radio-quiet AGN. An additional highly star-forming component seems to be present in all cases, with a typical SFR of $\sim 10^3 \, M_{\odot}/yr$.

We have demonstrated that the combination of the far-IR and radio emission criteria is highly efficient in selecting merging systems at $3<z<4$, typically with the presence of a radio-quiet AGN and a highly SFG. With upcoming wide radio surveys (e.g. the Evolutionary Map of the Universe,  EMU) and next-generation telescopes such as  the SKA, these results motivate further exploration of alternative selection methods. In order to advance the study of systems similar to those in our sample, particularly during the epoch of reionisation (EoR), sources with a mm rise in flux in combination with deep radio surveys might further be studued. A possible rising mm spectrum profile, which could be explored for faint individual sources using interferometric facilities (e.g. ALMA and NOEMA) and single-dish instruments (e.g. the NIKA2 camera of the IRAM 30m antenna, or the ToLTEC of the Large Millimeter Telescope) targeting a brighter population from larger surveys could assist in the identification of the redshifted dust component of SFGs at $z>6$. Similarly, the morphology of these galaxies could be further understood with infrared facilities, including the \textit{JWST}.


\begin{acknowledgements}
This work was supported by Fundação para a Ciência e a Tecnologia (FCT) through the research grants PTDC/FIS-AST/29245/2017, 1220UID/FIS/04434/2019, UIDB/04434/2020, and UIDP/04434/2020. C.P. acknowledges support from DL 57/2016 (P2460) from the ‘Departamento de Física, Faculdade de Ciências da Universidade de Lisboa'. This paper makes use of ALMA data. ALMA is a partnership of ESO (representing its member states), NSF (USA), and NINS (Japan); together with NRC (Canada), MOST, and ASIAA (Taiwan); and KASI (Republic of Korea), in cooperation with the Republic of Chile. The Joint ALMA Observatory is operated by ESO, AUI/NRAO, and NAOJ. This research made extensive use of core PYTHON packages for Astronomy as well as TOPCAT. We thank the anonymous referee for all the valuable comments, which improved the quality of the paper.
\end{acknowledgements}

\bibliographystyle{aa}
\bibliography{ref2}

\begin{thebibliography}{102}
\expandafter\ifx\csname natexlab\endcsname\relax\def\natexlab#1{#1}\fi

\bibitem[{{Amblard} {et~al.}(2010){Amblard}, {Cooray}, {Serra}, {Temi},
  {Barton}, {Negrello}, {Auld}, {Baes}, {Baldry}, {Bamford}, {Blain}, {Bock},
  {Bonfield}, {Burgarella}, {Buttiglione}, {Cameron}, {Cava}, {Clements},
  {Croom}, {Dariush}, {de Zotti}, {Driver}, {Dunlop}, {Dunne}, {Dye}, {Eales},
  {Frayer}, {Fritz}, {Gardner}, {Gonzalez-Nuevo}, {Herranz}, {Hill}, {Hopkins},
  {Hughes}, {Ibar}, {Ivison}, {Jarvis}, {Jones}, {Kelvin}, {Lagache}, {Leeuw},
  {Liske}, {Lopez-Caniego}, {Loveday}, {Maddox}, {Micha{\l}owski}, {Norberg},
  {Parkinson}, {Peacock}, {Pearson}, {Pascale}, {Pohlen}, {Popescu},
  {Prescott}, {Robotham}, {Rigby}, {Rodighiero}, {Samui}, {Sansom}, {Scott},
  {Serjeant}, {Sharp}, {Sibthorpe}, {Smith}, {Thompson}, {Tuffs}, {Valtchanov},
  {van Kampen}, {van der Werf}, {Verma}, {Vieira}, \&
  {Vlahakis}}]{2010A&A...518L...9A}
{Amblard}, A., {Cooray}, A., {Serra}, P., {et~al.} 2010, \aap, 518, L9

\bibitem[{{Avni} \& {Bahcall}(1980)}]{1980ApJ...235..694A}
{Avni}, Y. \& {Bahcall}, J.~N. 1980, \apj, 235, 694

\bibitem[{{Bell}(2003)}]{2003ApJ...586..794B}
{Bell}, E.~F. 2003, ApJ, 586, 794

\bibitem[{{Bing} {et~al.}(2023){Bing}, {B{\'e}thermin}, {Lagache}, {Adam},
  {Ade}, {Ajeddig}, {Andr{\'e}}, {Artis}, {Aussel}, {Beelen}, {Beno{\^\i}t},
  {Berta}, {Billot}, {Bourrion}, {Calvo}, {Catalano}, {De Petris},
  {D{\'e}sert}, {Doyle}, {Driessen}, {Elbaz}, {Gkogkou}, {Gomez}, {Goupy},
  {Hanser}, {K{\'e}ruzor{\'e}}, {Kramer}, {Ladjelate}, {Liu}, {Leclercq},
  {Lestrade}, {Lustig}, {Mac{\'\i}as-P{\'e}rez}, {Maury}, {Mauskopf}, {Mayet},
  {Monfardini}, {Mu{\~n}oz-Echeverr{\'\i}a}, {Perotto}, {Pisano}, {Ponthieu},
  {Rev{\'e}ret}, {Rigby}, {Ritacco}, {Romero}, {Roussel}, {Ruppin}, {Schuster},
  {Sievers}, {Tucker}, \& {Zylka}}]{2023arXiv230507054B}
{Bing}, L., {B{\'e}thermin}, M., {Lagache}, G., {et~al.} 2023, arXiv e-prints,
  arXiv:2305.07054

\bibitem[{{Birkin} {et~al.}(2021){Birkin}, {Weiss}, {Wardlow}, {Smail},
  {Swinbank}, {Dudzevi{\v{c}}i{\={u}}t{\.{e}}}, {An}, {Ao}, {Chapman}, {Chen},
  {da Cunha}, {Dannerbauer}, {Gullberg}, {Hodge}, {Ikarashi}, {Ivison},
  {Matsuda}, {Stach}, {Walter}, {Wang}, \& {van der
  Werf}}]{2021MNRAS.501.3926B}
{Birkin}, J.~E., {Weiss}, A., {Wardlow}, J.~L., {et~al.} 2021, \mnras, 501,
  3926

\bibitem[{{Bonzini} {et~al.}(2015){Bonzini}, {Mainieri}, {Padovani},
  {Andreani}, {Berta}, {Bethermin}, {Lutz}, {Rodighiero}, {Rosario}, {Tozzi},
  \& {Vattakunnel}}]{2015MNRAS.453.1079B}
{Bonzini}, M., {Mainieri}, V., {Padovani}, P., {et~al.} 2015, \mnras, 453, 1079

\bibitem[{{Bouwens} {et~al.}(2015){Bouwens}, {Illingworth}, {Oesch}, {Caruana},
  {Holwerda}, {Smit}, \& {Wilkins}}]{2015ApJ...811..140B}
{Bouwens}, R.~J., {Illingworth}, G.~D., {Oesch}, P.~A., {et~al.} 2015, \apj,
  811, 140

\bibitem[{{Bruzual} \& {Charlot}(2003)}]{bc03}
{Bruzual}, G. \& {Charlot}, S. 2003, MNRAS, 344, 1000

\bibitem[{{Cappelluti} {et~al.}(2009){Cappelluti}, {Brusa}, {Hasinger},
  {Comastri}, {Zamorani}, {Finoguenov}, {Gilli}, {Puccetti}, {Miyaji},
  {Salvato}, {Vignali}, {Aldcroft}, {B{\"o}hringer}, {Brunner}, {Civano},
  {Elvis}, {Fiore}, {Fruscione}, {Griffiths}, {Guzzo}, {Iovino}, {Koekemoer},
  {Mainieri}, {Scoville}, {Shopbell}, {Silverman}, \& {Urry}}]{Cappelluti}
{Cappelluti}, N., {Brusa}, M., {Hasinger}, G., {et~al.} 2009, A\&A, 497, 635

\bibitem[{{Carter} {et~al.}(2012){Carter}, {Lazareff}, {Maier}, {Chenu},
  {Fontana}, {Bortolotti}, {Boucher}, {Navarrini}, {Blanchet}, {Greve}, {John},
  {Kramer}, {Morel}, {Navarro}, {Pe{\~n}alver}, {Schuster}, \&
  {Thum}}]{2012A&A...538A..89C}
{Carter}, M., {Lazareff}, B., {Maier}, D., {et~al.} 2012, A\&A, 538, A89

\bibitem[{{CASA Team} {et~al.}(2022){CASA Team}, {Bean}, {Bhatnagar}, {Castro},
  {Donovan Meyer}, {Emonts}, {Garcia}, {Garwood}, {Golap}, {Gonzalez Villalba},
  {Harris}, {Hayashi}, {Hoskins}, {Hsieh}, {Jagannathan}, {Kawasaki},
  {Keimpema}, {Kettenis}, {Lopez}, {Marvil}, {Masters}, {McNichols},
  {Mehringer}, {Miel}, {Moellenbrock}, {Montesino}, {Nakazato}, {Ott}, {Petry},
  {Pokorny}, {Raba}, {Rau}, {Schiebel}, {Schweighart}, {Sekhar}, {Shimada},
  {Small}, {Steeb}, {Sugimoto}, {Suoranta}, {Tsutsumi}, {van Bemmel},
  {Verkouter}, {Wells}, {Xiong}, {Szomoru}, {Griffith}, {Glendenning}, \&
  {Kern}}]{2022PASP..134k4501C}
{CASA Team}, {Bean}, B., {Bhatnagar}, S., {et~al.} 2022, \pasp, 134, 114501

\bibitem[{{Casey} {et~al.}(2021){Casey}, {Zavala}, {Manning}, {Aravena},
  {B{\'e}thermin}, {Caputi}, {Champagne}, {Clements}, {Drew}, {Finkelstein},
  {Fujimoto}, {Hayward}, {Dekel}, {Kokorev}, {Lagos}, {Long}, {Magdis}, {Man},
  {Mitsuhashi}, {Popping}, {Spilker}, {Staguhn}, {Talia}, {Toft}, {Treister},
  {Weaver}, \& {Yun}}]{2021ApJ...923..215C}
{Casey}, C.~M., {Zavala}, J.~A., {Manning}, S.~M., {et~al.} 2021, \apj, 923,
  215

\bibitem[{{Charlot} \& {Fall}(2000)}]{dustatt_2powerlaws}
{Charlot}, S. \& {Fall}, S.~M. 2000, ApJ, 539, 718

\bibitem[{{Chen} {et~al.}(2022){Chen}, {Liao}, {Smail}, {Swinbank}, {Ao},
  {Bunker}, {Chapman}, {Hatsukade}, {Ivison}, {Lee}, {Serjeant}, {Umehata},
  {Wang}, \& {Zhao}}]{2022ApJ...929..159C}
{Chen}, C.-C., {Liao}, C.-L., {Smail}, I., {et~al.} 2022, \apj, 929, 159

\bibitem[{{Chenu} {et~al.}(2016){Chenu}, {Navarrini}, {Bortolotti}, {Butin},
  {Fontana}, {Mahieu}, {Maier}, {Mattiocco}, {Serres}, {Berton}, {Garnier},
  {Moutote}, {Parioleau}, {Pissard}, \& {Reverdy}}]{2016ITTST...6..223C}
{Chenu}, J.-Y., {Navarrini}, A., {Bortolotti}, Y., {et~al.} 2016, IEEE
  Transactions on Terahertz Science and Technology, 6, 223

\bibitem[{{Ciesla} {et~al.}(2017){Ciesla}, {Elbaz}, \& {Fensch}}]{sfhdelayed}
{Ciesla}, L., {Elbaz}, D., \& {Fensch}, J. 2017, A\&A, 608, A41

\bibitem[{{Civano} {et~al.}(2016){Civano}, {Marchesi}, {Comastri}, {Urry},
  {Elvis}, {Cappelluti}, {Puccetti}, {Brusa}, {Zamorani}, {Hasinger},
  {Aldcroft}, {Alexand er}, {Allevato}, {Brunner}, {Capak}, {Finoguenov},
  {Fiore}, {Fruscione}, {Gilli}, {Glotfelty}, {Griffiths}, {Hao}, {Harrison},
  {Jahnke}, {Kartaltepe}, {Karim}, {LaMassa}, {Lanzuisi}, {Miyaji}, {Ranalli},
  {Salvato}, {Sargent}, {Scoville}, {Schawinski}, {Schinnerer}, {Silverman},
  {Smolcic}, {Stern}, {Toft}, {Trakhtenbrot}, {Treister}, \&
  {Vignali}}]{2016ApJ...819...62C}
{Civano}, F., {Marchesi}, S., {Comastri}, A., {et~al.} 2016, ApJ, 819, 62

\bibitem[{{Coppin} {et~al.}(2012){Coppin}, {Danielson}, {Geach}, {Hodge},
  {Swinbank}, {Wardlow}, {Bertoldi}, {Biggs}, {Brandt}, {Caselli}, {Chapman},
  {Dannerbauer}, {Dunlop}, {Greve}, {Hamann}, {Ivison}, {Karim}, {Knudsen},
  {Menten}, {Schinnerer}, {Smail}, {Spaans}, {Walter}, {Webb}, \& {van der
  Werf}}]{2012MNRAS.427..520C}
{Coppin}, K.~E.~K., {Danielson}, A.~L.~R., {Geach}, J.~E., {et~al.} 2012,
  MNRAS, 427, 520

\bibitem[{{da Cunha} {et~al.}(2015){da Cunha}, {Walter}, {Smail}, {Swinbank},
  {Simpson}, {Decarli}, {Hodge}, {Weiss}, {van der Werf}, {Bertoldi},
  {Chapman}, {Cox}, {Danielson}, {Dannerbauer}, {Greve}, {Ivison}, {Karim}, \&
  {Thomson}}]{2015ApJ...806..110D}
{da Cunha}, E., {Walter}, F., {Smail}, I.~R., {et~al.} 2015, \apj, 806, 110

\bibitem[{{Dale} {et~al.}(2014){Dale}, {Helou}, {Magdis}, {Armus},
  {D{\'\i}az-Santos}, \& {Shi}}]{dale2014}
{Dale}, D.~A., {Helou}, G., {Magdis}, G.~E., {et~al.} 2014, ApJ, 784, 83

\bibitem[{{Danielson} {et~al.}(2017){Danielson}, {Swinbank}, {Smail},
  {Simpson}, {Casey}, {Chapman}, {da Cunha}, {Hodge}, {Walter}, {Wardlow},
  {Alexander}, {Brandt}, {de Breuck}, {Coppin}, {Dannerbauer}, {Dickinson},
  {Edge}, {Gawiser}, {Ivison}, {Karim}, {Kovacs}, {Lutz}, {Menten},
  {Schinnerer}, {Wei{\ss}}, \& {van der Werf}}]{2017ApJ...840...78D}
{Danielson}, A.~L.~R., {Swinbank}, A.~M., {Smail}, I., {et~al.} 2017, ApJ, 840,
  78

\bibitem[{{Delhaize} {et~al.}(2017){Delhaize}, {Smol{\v{c}}i{\'c}},
  {Delvecchio}, {Novak}, {Sargent}, {Baran}, {Magnelli}, {Zamorani},
  {Schinnerer}, {Murphy}, {Aravena}, {Berta}, {Bondi}, {Capak}, {Carilli},
  {Ciliegi}, {Civano}, {Ilbert}, {Karim}, {Laigle}, {Le F{\`e}vre}, {Marchesi},
  {McCracken}, {Salvato}, {Seymour}, \& {Tasca}}]{2017A&A...602A...4D}
{Delhaize}, J., {Smol{\v{c}}i{\'c}}, V., {Delvecchio}, I., {et~al.} 2017, \aap,
  602, A4

\bibitem[{{Delvecchio} {et~al.}(2021){Delvecchio}, {Daddi}, {Sargent},
  {Jarvis}, {Elbaz}, {Jin}, {Liu}, {Whittam}, {Algera}, {Carraro}, {D'Eugenio},
  {Delhaize}, {Kalita}, {Leslie}, {Moln{\'a}r}, {Novak}, {Prandoni},
  {Smol{\v{c}}i{\'c}}, {Ao}, {Aravena}, {Bournaud}, {Collier},
  {Randriamampandry}, {Randriamanakoto}, {Rodighiero}, {Schober}, {White}, \&
  {Zamorani}}]{2021A&A...647A.123D}
{Delvecchio}, I., {Daddi}, E., {Sargent}, M.~T., {et~al.} 2021, \aap, 647, A123

\bibitem[{{Delvecchio} {et~al.}(2017){Delvecchio}, {Smol{\v{c}}i{\'c}},
  {Zamorani}, {Lagos}, {Berta}, {Delhaize}, {Baran}, {Alexander}, {Rosario},
  {Gonzalez-Perez}, {Ilbert}, {Lacey}, {Le F{\`e}vre}, {Miettinen}, {Aravena},
  {Bondi}, {Carilli}, {Ciliegi}, {Mooley}, {Novak}, {Schinnerer}, {Capak},
  {Civano}, {Fanidakis}, {Herrera Ruiz}, {Karim}, {Laigle}, {Marchesi},
  {McCracken}, {Middleberg}, {Salvato}, \& {Tasca}}]{2017A&A...602A...3D}
{Delvecchio}, I., {Smol{\v{c}}i{\'c}}, V., {Zamorani}, G., {et~al.} 2017, \aap,
  602, A3

\bibitem[{{Donevski} {et~al.}(2018){Donevski}, {Buat}, {Boone}, {Pappalardo},
  {Bethermin}, {Schreiber}, {Mazyed}, {Alvarez-Marquez}, \&
  {Duivenvoorden}}]{2018A&A...614A..33D}
{Donevski}, D., {Buat}, V., {Boone}, F., {et~al.} 2018, A\&A, 614, A33

\bibitem[{{Dowell} {et~al.}(2014){Dowell}, {Conley}, {Glenn}, {Arumugam},
  {Asboth}, {Aussel}, {Bertoldi}, {B{\'e}thermin}, {Bock}, {Boselli}, {Bridge},
  {Buat}, {Burgarella}, {Cabrera-Lavers}, {Casey}, {Chapman}, {Clements},
  {Conversi}, {Cooray}, {Dannerbauer}, {De Bernardis}, {Ellsworth-Bowers},
  {Farrah}, {Franceschini}, {Griffin}, {Gurwell}, {Halpern}, {Hatziminaoglou},
  {Heinis}, {Ibar}, {Ivison}, {Laporte}, {Marchetti}, {Mart{\'\i}nez-Navajas},
  {Marsden}, {Morrison}, {Nguyen}, {O'Halloran}, {Oliver}, {Omont}, {Page},
  {Papageorgiou}, {Pearson}, {Petitpas}, {P{\'e}rez-Fournon}, {Pohlen},
  {Riechers}, {Rigopoulou}, {Roseboom}, {Rowan-Robinson}, {Sayers}, {Schulz},
  {Scott}, {Seymour}, {Shupe}, {Smith}, {Streblyanska}, {Symeonidis},
  {Vaccari}, {Valtchanov}, {Vieira}, {Viero}, {Wang}, {Wardlow}, {Xu}, \&
  {Zemcov}}]{2014ApJ...780...75D}
{Dowell}, C.~D., {Conley}, A., {Glenn}, J., {et~al.} 2014, ApJ, 780, 75

\bibitem[{{Duncan} {et~al.}(2019){Duncan}, {Conselice}, {Mundy}, {Bell},
  {Donley}, {Galametz}, {Guo}, {Grogin}, {Hathi}, {Kartaltepe}, {Kocevski},
  {Koekemoer}, {P{\'e}rez-Gonz{\'a}lez}, {Mantha}, {Snyder}, \&
  {Stefanon}}]{2019ApJ...876..110D}
{Duncan}, K., {Conselice}, C.~J., {Mundy}, C., {et~al.} 2019, \apj, 876, 110

\bibitem[{{Dunlop} {et~al.}(2017){Dunlop}, {McLure}, {Biggs}, {Geach},
  {Micha{\l}owski}, {Ivison}, {Rujopakarn}, {van Kampen}, {Kirkpatrick},
  {Pope}, {Scott}, {Swinbank}, {Targett}, {Aretxaga}, {Austermann}, {Best},
  {Bruce}, {Chapin}, {Charlot}, {Cirasuolo}, {Coppin}, {Ellis}, {Finkelstein},
  {Hayward}, {Hughes}, {Ibar}, {Jagannathan}, {Khochfar}, {Koprowski},
  {Narayanan}, {Nyland}, {Papovich}, {Peacock}, {Rieke}, {Robertson},
  {Vernstrom}, {Werf}, {Wilson}, \& {Yun}}]{2017MNRAS.466..861D}
{Dunlop}, J.~S., {McLure}, R.~J., {Biggs}, A.~D., {et~al.} 2017, \mnras, 466,
  861

\bibitem[{{Elvis} {et~al.}(2009){Elvis}, {Civano}, {Vignali}, {Puccetti},
  {Fiore}, {Cappelluti}, {Aldcroft}, {Fruscione}, {Zamorani}, {Comastri},
  {Brusa}, {Gilli}, {Miyaji}, {Damiani}, {Koekemoer}, {Finoguenov}, {Brunner},
  {Urry}, {Silverman}, {Mainieri}, {Hasinger}, {Griffiths}, {Carollo}, {Hao},
  {Guzzo}, {Blain}, {Calzetti}, {Carilli}, {Capak}, {Ettori}, {Fabbiano},
  {Impey}, {Lilly}, {Mobasher}, {Rich}, {Salvato}, {Sand ers}, {Schinnerer},
  {Scoville}, {Shopbell}, {Taylor}, {Taniguchi}, \&
  {Volonteri}}]{2009ApJ..184..158E}
{Elvis}, M., {Civano}, F., {Vignali}, C., {et~al.} 2009, ApJ, 184, 158

\bibitem[{{Enia} {et~al.}(2022){Enia}, {Talia}, {Pozzi}, {Cimatti},
  {Delvecchio}, {Zamorani}, {D'Amato}, {Bisigello}, {Gruppioni}, {Rodighiero},
  {Calura}, {Dallacasa}, {Giulietti}, {Barchiesi}, {Behiri}, \&
  {Romano}}]{2022ApJ...927..204E}
{Enia}, A., {Talia}, M., {Pozzi}, F., {et~al.} 2022, \apj, 927, 204

\bibitem[{{Faisst} {et~al.}(2020){Faisst}, {Schaerer}, {Lemaux}, {Oesch},
  {Fudamoto}, {Cassata}, {B{\'e}thermin}, {Capak}, {Le F{\`e}vre}, {Silverman},
  {Yan}, {Ginolfi}, {Koekemoer}, {Morselli}, {Amor{\'\i}n}, {Bardelli},
  {Boquien}, {Brammer}, {Cimatti}, {Dessauges-Zavadsky}, {Fujimoto},
  {Gruppioni}, {Hathi}, {Hemmati}, {Ibar}, {Jones}, {Khusanova}, {Loiacono},
  {Pozzi}, {Talia}, {Tasca}, {Riechers}, {Rodighiero}, {Romano}, {Scoville},
  {Toft}, {Vallini}, {Vergani}, {Zamorani}, \& {Zucca}}]{2020ApJS..247...61F}
{Faisst}, A.~L., {Schaerer}, D., {Lemaux}, B.~C., {et~al.} 2020, \apjs, 247, 61

\bibitem[{{Fritz} {et~al.}(2006){Fritz}, {Franceschini}, \&
  {Hatziminaoglou}}]{fritz2006}
{Fritz}, J., {Franceschini}, A., \& {Hatziminaoglou}, E. 2006, MNRAS, 366, 767

\bibitem[{{Gao} {et~al.}(2021){Gao}, {Wang}, {Efstathiou}, {Ma{\l}ek}, {Best},
  {Bonato}, {Farrah}, {Kondapally}, {McCheyne}, \&
  {R{\"o}ttgering}}]{2021A&A...654A.117G}
{Gao}, F., {Wang}, L., {Efstathiou}, A., {et~al.} 2021, \aap, 654, A117

\bibitem[{{Gao} {et~al.}(2022){Gao}, {Wang}, {Ramos Padilla}, {Clements},
  {Farrah}, \& {Huang}}]{2022A&A...668A..54G}
{Gao}, F., {Wang}, L., {Ramos Padilla}, A.~F., {et~al.} 2022, \aap, 668, A54

\bibitem[{{Griffin} {et~al.}(2010){Griffin}, {Abergel}, {Abreu}, {Ade},
  {Andr{\'e}}, {Augueres}, {Babbedge}, {Bae}, {Baillie}, {Baluteau}, {Barlow},
  {Bendo}, {Benielli}, {Bock}, {Bonhomme}, {Brisbin}, {Brockley-Blatt},
  {Caldwell}, {Cara}, {Castro-Rodriguez}, {Cerulli}, {Chanial}, {Chen},
  {Clark}, {Clements}, {Clerc}, {Coker}, {Communal}, {Conversi}, {Cox},
  {Crumb}, {Cunningham}, {Daly}, {Davis}, {de Antoni}, {Delderfield}, {Devin},
  {di Giorgio}, {Didschuns}, {Dohlen}, {Donati}, {Dowell}, {Dowell}, {Duband},
  {Dumaye}, {Emery}, {Ferlet}, {Ferrand}, {Fontignie}, {Fox}, {Franceschini},
  {Frerking}, {Fulton}, {Garcia}, {Gastaud}, {Gear}, {Glenn}, {Goizel},
  {Griffin}, {Grundy}, {Guest}, {Guillemet}, {Hargrave}, {Harwit}, {Hastings},
  {Hatziminaoglou}, {Herman}, {Hinde}, {Hristov}, {Huang}, {Imhof}, {Isaak},
  {Israelsson}, {Ivison}, {Jennings}, {Kiernan}, {King}, {Lange}, {Latter},
  {Laurent}, {Laurent}, {Leeks}, {Lellouch}, {Levenson}, {Li}, {Li},
  {Lilienthal}, {Lim}, {Liu}, {Lu}, {Madden}, {Mainetti}, {Marliani}, {McKay},
  {Mercier}, {Molinari}, {Morris}, {Moseley}, {Mulder}, {Mur}, {Naylor},
  {Nguyen}, {O'Halloran}, {Oliver}, {Olofsson}, {Olofsson}, {Orfei}, {Page},
  {Pain}, {Panuzzo}, {Papageorgiou}, {Parks}, {Parr-Burman}, {Pearce},
  {Pearson}, {P{\'e}rez-Fournon}, {Pinsard}, {Pisano}, {Podosek}, {Pohlen},
  {Polehampton}, {Pouliquen}, {Rigopoulou}, {Rizzo}, {Roseboom}, {Roussel},
  {Rowan-Robinson}, {Rownd}, {Saraceno}, {Sauvage}, {Savage}, {Savini},
  {Sawyer}, {Scharmberg}, {Schmitt}, {Schneider}, {Schulz}, {Schwartz},
  {Shafer}, {Shupe}, {Sibthorpe}, {Sidher}, {Smith}, {Smith}, {Smith},
  {Spencer}, {Stobie}, {Sudiwala}, {Sukhatme}, {Surace}, {Stevens}, {Swinyard},
  {Trichas}, {Tourette}, {Triou}, {Tseng}, {Tucker}, {Turner}, {Vaccari},
  {Valtchanov}, {Vigroux}, {Virique}, {Voellmer}, {Walker}, {Ward}, {Waskett},
  {Weilert}, {Wesson}, {White}, {Whitehouse}, {Wilson}, {Winter}, {Woodcraft},
  {Wright}, {Xu}, {Zavagno}, {Zemcov}, {Zhang}, \&
  {Zonca}}]{2010A&A...518L...3G}
{Griffin}, M.~J., {Abergel}, A., {Abreu}, A., {et~al.} 2010, A\&A, 518, L3

\bibitem[{{Gruppioni} {et~al.}(2020){Gruppioni}, {B{\'e}thermin}, {Loiacono},
  {Le F{\`e}vre}, {Capak}, {Cassata}, {Faisst}, {Schaerer}, {Silverman}, {Yan},
  {Bardelli}, {Boquien}, {Carraro}, {Cimatti}, {Dessauges-Zavadsky}, {Ginolfi},
  {Fujimoto}, {Hathi}, {Jones}, {Khusanova}, {Koekemoer}, {Lagache}, {Lemaux},
  {Oesch}, {Pozzi}, {Riechers}, {Rodighiero}, {Romano}, {Talia}, {Vallini},
  {Vergani}, {Zamorani}, \& {Zucca}}]{2020A&A...643A...8G}
{Gruppioni}, C., {B{\'e}thermin}, M., {Loiacono}, F., {et~al.} 2020, \aap, 643,
  A8

\bibitem[{{Gruppioni} {et~al.}(2013){Gruppioni}, {Pozzi}, {Rodighiero},
  {Delvecchio}, {Berta}, {Pozzetti}, {Zamorani}, {Andreani}, {Cimatti},
  {Ilbert}, {Le Floc'h}, {Lutz}, {Magnelli}, {Marchetti}, {Monaco}, {Nordon},
  {Oliver}, {Popesso}, {Riguccini}, {Roseboom}, {Rosario}, {Sargent},
  {Vaccari}, {Altieri}, {Aussel}, {Bongiovanni}, {Cepa}, {Daddi},
  {Dom{\'\i}nguez-S{\'a}nchez}, {Elbaz}, {F{\"o}rster Schreiber}, {Genzel},
  {Iribarrem}, {Magliocchetti}, {Maiolino}, {Poglitsch}, {P{\'e}rez
  Garc{\'\i}a}, {Sanchez-Portal}, {Sturm}, {Tacconi}, {Valtchanov}, {Amblard},
  {Arumugam}, {Bethermin}, {Bock}, {Boselli}, {Buat}, {Burgarella},
  {Castro-Rodr{\'\i}guez}, {Cava}, {Chanial}, {Clements}, {Conley}, {Cooray},
  {Dowell}, {Dwek}, {Eales}, {Franceschini}, {Glenn}, {Griffin},
  {Hatziminaoglou}, {Ibar}, {Isaak}, {Ivison}, {Lagache}, {Levenson}, {Lu},
  {Madden}, {Maffei}, {Mainetti}, {Nguyen}, {O'Halloran}, {Page}, {Panuzzo},
  {Papageorgiou}, {Pearson}, {P{\'e}rez-Fournon}, {Pohlen}, {Rigopoulou},
  {Rowan-Robinson}, {Schulz}, {Scott}, {Seymour}, {Shupe}, {Smith}, {Stevens},
  {Symeonidis}, {Trichas}, {Tugwell}, {Vigroux}, {Wang}, {Wright}, {Xu},
  {Zemcov}, {Bardelli}, {Carollo}, {Contini}, {Le F{\'e}vre}, {Lilly},
  {Mainieri}, {Renzini}, {Scodeggio}, \& {Zucca}}]{2013MNRAS.432...23G}
{Gruppioni}, C., {Pozzi}, F., {Rodighiero}, G., {et~al.} 2013, \mnras, 432, 23

\bibitem[{{Guo} {et~al.}(2020){Guo}, {Gu}, {Ding}, {Contini}, \&
  {Chen}}]{2020MNRAS.492.1887G}
{Guo}, X., {Gu}, Q., {Ding}, N., {Contini}, E., \& {Chen}, Y. 2020, \mnras,
  492, 1887

\bibitem[{{Harikane} {et~al.}(2023){Harikane}, {Ouchi}, {Oguri}, {Ono},
  {Nakajima}, {Isobe}, {Umeda}, {Mawatari}, \& {Zhang}}]{2023ApJS..265....5H}
{Harikane}, Y., {Ouchi}, M., {Oguri}, M., {et~al.} 2023, \apjs, 265, 5

\bibitem[{{Hasinger} {et~al.}(2018){Hasinger}, {Capak}, {Salvato}, {Barger},
  {Cowie}, {Faisst}, {Hemmati}, {Kakazu}, {Kartaltepe}, {Masters}, {Mobasher},
  {Nayyeri}, {Sanders}, {Scoville}, {Suh}, {Steinhardt}, \&
  {Yang}}]{2018ApJ...858...77H}
{Hasinger}, G., {Capak}, P., {Salvato}, M., {et~al.} 2018, ApJ, 858, 77

\bibitem[{{Holland} {et~al.}(2013){Holland}, {Bintley}, {Chapin},
  {Chrysostomou}, {Davis}, {Dempsey}, {Duncan}, {Fich}, {Friberg}, {Halpern},
  {Irwin}, {Jenness}, {Kelly}, {MacIntosh}, {Robson}, {Scott}, {Ade},
  {Atad-Ettedgui}, {Berry}, {Craig}, {Gao}, {Gibb}, {Hilton}, {Hollister},
  {Kycia}, {Lunney}, {McGregor}, {Montgomery}, {Parkes}, {Tilanus}, {Ullom},
  {Walther}, {Walton}, {Woodcraft}, {Amiri}, {Atkinson}, {Burger}, {Chuter},
  {Coulson}, {Doriese}, {Dunare}, {Economou}, {Niemack}, {Parsons},
  {Reintsema}, {Sibthorpe}, {Smail}, {Sudiwala}, \&
  {Thomas}}]{2013MNRAS.430.2513H}
{Holland}, W.~S., {Bintley}, D., {Chapin}, E.~L., {et~al.} 2013, \mnras, 430,
  2513

\bibitem[{{Inoue}(2011)}]{lyc_absorption}
{Inoue}, A.~K. 2011, MNRAS, 415, 2920

\bibitem[{{Ishigaki} {et~al.}(2018){Ishigaki}, {Kawamata}, {Ouchi}, {Oguri},
  {Shimasaku}, \& {Ono}}]{2018ApJ...854...73I}
{Ishigaki}, M., {Kawamata}, R., {Ouchi}, M., {et~al.} 2018, \apj, 854, 73

\bibitem[{{Iwasawa} {et~al.}(2012){Iwasawa}, {Gilli}, {Vignali}, {Comastri},
  {Brandt}, {Ranalli}, {Vito}, {Cappelluti}, {Carrera}, {Falocco},
  {Georgantopoulos}, {Mainieri}, \& {Paolillo}}]{2012A&A...546A..84I}
{Iwasawa}, K., {Gilli}, R., {Vignali}, C., {et~al.} 2012, APP, 546, A84

\bibitem[{{Kistler} {et~al.}(2009){Kistler}, {Y{\"u}ksel}, {Beacom}, {Hopkins},
  \& {Wyithe}}]{2009ApJ...705L.104K}
{Kistler}, M.~D., {Y{\"u}ksel}, H., {Beacom}, J.~F., {Hopkins}, A.~M., \&
  {Wyithe}, J. S.~B. 2009, \apjl, 705, L104

\bibitem[{{Klein} {et~al.}(2012){Klein}, {Hochg{\"u}rtel}, {Kr{\"a}mer},
  {Bell}, {Meyer}, \& {G{\"u}sten}}]{2012A&A...542L...3K}
{Klein}, B., {Hochg{\"u}rtel}, S., {Kr{\"a}mer}, I., {et~al.} 2012, A\&A, 542,
  L3

\bibitem[{{Klein} {et~al.}(2008){Klein}, {Kr{\"a}mer}, {Hochg{\"u}rtel},
  {G{\"u}sten}, {Bell}, {Meyer}, \& {Chetik}}]{2008stt..conf..192K}
{Klein}, B., {Kr{\"a}mer}, I., {Hochg{\"u}rtel}, S., {et~al.} 2008, in
  Ninteenth International Symposium on Space Terahertz Technology, ed.
  W.~{Wild}, 192

\bibitem[{{Klein} {et~al.}(2009){Klein}, {Kr{\"a}mer}, {Hochg{\"u}rtel},
  {G{\"u}sten}, {Bell}, {Meyer}, \& {Chetik}}]{2009stt..conf..199K}
{Klein}, B., {Kr{\"a}mer}, I., {Hochg{\"u}rtel}, S., {et~al.} 2009, in
  Twentieth International Symposium on Space Terahertz Technology, ed.
  E.~{Bryerton}, A.~{Kerr}, \& A.~{Lichtenberger}, 199

\bibitem[{{Lagache}(2018)}]{2018IAUS..333..228L}
{Lagache}, G. 2018, in Peering towards Cosmic Dawn, ed. V.~{Jeli{\'c}} \&
  T.~{van der Hulst}, Vol. 333, 228--233

\bibitem[{{Laigle} {et~al.}(2016){Laigle}, {McCracken}, {Ilbert}, {Hsieh},
  {Davidzon}, {Capak}, {Hasinger}, {Silverman}, {Pichon}, {Coupon}, {Aussel},
  {Le Borgne}, {Caputi}, {Cassata}, {Chang}, {Civano}, {Dunlop}, {Fynbo},
  {Kartaltepe}, {Koekemoer}, {Le F{\`e}vre}, {Le Floc'h}, {Leauthaud}, {Lilly},
  {Lin}, {Marchesi}, {Milvang-Jensen}, {Salvato}, {Sanders}, {Scoville},
  {Smolcic}, {Stockmann}, {Taniguchi}, {Tasca}, {Toft}, {Vaccari}, \&
  {Zabl}}]{Laigle16}
{Laigle}, C., {McCracken}, H.~J., {Ilbert}, O., {et~al.} 2016, ApJ, 224, 24

\bibitem[{{Lanzuisi} {et~al.}(2017){Lanzuisi}, {Delvecchio}, {Berta}, {Brusa},
  {Comastri}, {Gilli}, {Gruppioni}, {Marchesi}, {Perna}, {Pozzi}, {Salvato},
  {Symeonidis}, {Vignali}, {Vito}, {Volonteri}, \&
  {Zamorani}}]{2017A&A...602A.123L}
{Lanzuisi}, G., {Delvecchio}, I., {Berta}, S., {et~al.} 2017, \aap, 602, A123

\bibitem[{{Lehmer} {et~al.}(2005){Lehmer}, {Brandt}, {Alexander}, {Bauer},
  {Schneider}, {Tozzi}, {Bergeron}, {Garmire}, {Giacconi}, {Gilli}, {Hasinger},
  {Hornschemeier}, {Koekemoer}, {Mainieri}, {Miyaji}, {Nonino}, {Rosati},
  {Silverman}, {Szokoly}, \& {Vignali}}]{2005ApJS..161...21L}
{Lehmer}, B.~D., {Brandt}, W.~N., {Alexander}, D.~M., {et~al.} 2005, APJS, 161,
  21

\bibitem[{{Luo} {et~al.}(2017){Luo}, {Brandt}, {Xue}, {Lehmer}, {Alexander},
  {Bauer}, {Vito}, {Yang}, {Basu-Zych}, {Comastri}, {Gilli}, {Gu},
  {Hornschemeier}, {Koekemoer}, {Liu}, {Mainieri}, {Paolillo}, {Ranalli},
  {Rosati}, {Schneider}, {Shemmer}, {Smail}, {Sun}, {Tozzi}, {Vignali}, \&
  {Wang}}]{2017ApJ..228....2L}
{Luo}, B., {Brandt}, W.~N., {Xue}, Y.~Q., {et~al.} 2017, ApJ, 228, 2

\bibitem[{{Madau} \& {Fragos}(2017)}]{2017ApJ...840...39M}
{Madau}, P. \& {Fragos}, T. 2017, \apj, 840, 39

\bibitem[{{Magnelli} {et~al.}(2019){Magnelli}, {Karim}, {Staguhn},
  {Kov{\'a}cs}, {Jim{\'e}nez-Andrade}, {Casey}, {Zavala}, {Schinnerer},
  {Sargent}, {Aravena}, {Bertoldi}, {Capak}, {Riechers}, \&
  {Benford}}]{2019ApJ...877...45M}
{Magnelli}, B., {Karim}, A., {Staguhn}, J., {et~al.} 2019, ApJ, 877, 45

\bibitem[{{Malefahlo} {et~al.}(2022){Malefahlo}, {Jarvis}, {Santos}, {White},
  {Adams}, \& {Bowler}}]{2022MNRAS.509.4291M}
{Malefahlo}, E.~D., {Jarvis}, M.~J., {Santos}, M.~G., {et~al.} 2022, \mnras,
  509, 4291

\bibitem[{{Ma{\l}ek} {et~al.}(2018){Ma{\l}ek}, {Buat}, {Roehlly}, {Burgarella},
  {Hurley}, {Shirley}, {Duncan}, {Efstathiou}, {Papadopoulos}, {Vaccari},
  {Farrah}, {Marchetti}, \& {Oliver}}]{2018A&A...620A..50M}
{Ma{\l}ek}, K., {Buat}, V., {Roehlly}, Y., {et~al.} 2018, \aap, 620, A50

\bibitem[{{Marchesi} {et~al.}(2016){Marchesi}, {Civano}, {Elvis}, {Salvato},
  {Brusa}, {Comastri}, {Gilli}, {Hasinger}, {Lanzuisi}, {Miyaji}, {Treister},
  {Urry}, {Vignali}, {Zamorani}, {Allevato}, {Cappelluti}, {Cardamone},
  {Finoguenov}, {Griffiths}, {Karim}, {Laigle}, {LaMassa}, {Jahnke}, {Ranalli},
  {Schawinski}, {Schinnerer}, {Silverman}, {Smolcic}, {Suh}, \&
  {Trakhtenbrot}}]{2016ApJ...817...34M}
{Marchesi}, S., {Civano}, F., {Elvis}, M., {et~al.} 2016, \apj, 817, 34

\bibitem[{{McLeod} {et~al.}(2016){McLeod}, {McLure}, \&
  {Dunlop}}]{2016MNRAS.459.3812M}
{McLeod}, D.~J., {McLure}, R.~J., \& {Dunlop}, J.~S. 2016, \mnras, 459, 3812

\bibitem[{{Meiksin}(2006)}]{redshifting}
{Meiksin}, A. 2006, MNRAS, 365, 807

\bibitem[{{Miettinen} {et~al.}(2015){Miettinen}, {Smol{\v{c}}i{\'c}}, {Novak},
  {Aravena}, {Karim}, {Masters}, {Riechers}, {Bussmann}, {McCracken}, {Ilbert},
  {Bertoldi}, {Capak}, {Feruglio}, {Halliday}, {Kartaltepe}, {Navarrete},
  {Salvato}, {Sand ers}, {Schinnerer}, \& {Sheth}}]{2015A&A...577A..29M}
{Miettinen}, O., {Smol{\v{c}}i{\'c}}, V., {Novak}, M., {et~al.} 2015, A\&A,
  577, A29

\bibitem[{{Miller} {et~al.}(2013){Miller}, {Bonzini}, {Fomalont}, {Kellermann},
  {Mainieri}, {Padovani}, {Rosati}, {Tozzi}, \&
  {Vattakunnel}}]{2013ApJS..205...13M}
{Miller}, N.~A., {Bonzini}, M., {Fomalont}, E.~B., {et~al.} 2013, APJS, 205, 13

\bibitem[{{Muzzin} {et~al.}(2013){Muzzin}, {Marchesini}, {Stefanon}, {Franx},
  {Milvang-Jensen}, {Dunlop}, {Fynbo}, {Brammer}, {Labb{\'e}}, \& {van
  Dokkum}}]{2013ApJS..206....8M}
{Muzzin}, A., {Marchesini}, D., {Stefanon}, M., {et~al.} 2013, \apjs, 206, 8

\bibitem[{{Nakajima} {et~al.}(2023){Nakajima}, {Ouchi}, {Isobe}, {Harikane},
  {Zhang}, {Ono}, {Umeda}, \& {Oguri}}]{2023arXiv230112825N}
{Nakajima}, K., {Ouchi}, M., {Isobe}, Y., {et~al.} 2023, arXiv e-prints,
  arXiv:2301.12825

\bibitem[{{Nguyen} {et~al.}(2010){Nguyen}, {Schulz}, {Levenson}, {Amblard},
  {Arumugam}, {Aussel}, {Babbedge}, {Blain}, {Bock}, {Boselli}, {Buat},
  {Castro-Rodriguez}, {Cava}, {Chanial}, {Chapin}, {Clements}, {Conley},
  {Conversi}, {Cooray}, {Dowell}, {Dwek}, {Eales}, {Elbaz}, {Fox},
  {Franceschini}, {Gear}, {Glenn}, {Griffin}, {Halpern}, {Hatziminaoglou},
  {Ibar}, {Isaak}, {Ivison}, {Lagache}, {Lu}, {Madden}, {Maffei}, {Mainetti},
  {Marchetti}, {Marsden}, {Marshall}, {O'Halloran}, {Oliver}, {Omont}, {Page},
  {Panuzzo}, {Papageorgiou}, {Pearson}, {Perez Fournon}, {Pohlen}, {Rangwala},
  {Rigopoulou}, {Rizzo}, {Roseboom}, {Rowan-Robinson}, {Scott}, {Seymour},
  {Shupe}, {Smith}, {Stevens}, {Symeonidis}, {Trichas}, {Tugwell}, {Vaccari},
  {Valtchanov}, {Vigroux}, {Wang}, {Ward}, {Wiebe}, {Wright}, {Xu}, \&
  {Zemcov}}]{2010A&A...518L...5N}
{Nguyen}, H.~T., {Schulz}, B., {Levenson}, L., {et~al.} 2010, \aap, 518, L5

\bibitem[{{Noll} {et~al.}(2009){Noll}, {Burgarella}, {Giovannoli}, {Buat},
  {Marcillac}, \& {Mu{\~n}oz-Mateos}}]{cigale1}
{Noll}, S., {Burgarella}, D., {Giovannoli}, E., {et~al.} 2009, A\&A, 507, 1793

\bibitem[{{Norris}(2009)}]{norris}
{Norris}, R.~P. 2009, in Panoramic Radio Astronomy: Wide-field 1-2 GHz Research
  on Galaxy Evolution, 33

\bibitem[{{Norris} {et~al.}(2021){Norris}, {Marvil}, {Collier}, {Kapi{\'n}ska},
  {O'Brien}, {Rudnick}, {Andernach}, {Asorey}, {Brown}, {Br{\"u}ggen},
  {Crawford}, {English}, {Rahman}, {Filipovi{\'c}}, {Gordon}, {G{\"u}rkan},
  {Hale}, {Hopkins}, {Huynh}, {HyeongHan}, {James Jee}, {Koribalski}, {Lenc},
  {Luken}, {Parkinson}, {Prandoni}, {Raja}, {Reiprich}, {Riseley}, {Shabala},
  {Sheil}, {Vernstrom}, {Whiting}, {Allison}, {Anderson}, {Ball}, {Bell},
  {Bunton}, {Galvin}, {Gupta}, {Hotan}, {Jacka}, {Macgregor}, {Mahony}, {Maio},
  {Moss}, {Pandey-Pommier}, \& {Voronkov}}]{2021PASA...38...46N}
{Norris}, R.~P., {Marvil}, J., {Collier}, J.~D., {et~al.} 2021, \pasa, 38, e046

\bibitem[{{Novak} {et~al.}(2017){Novak}, {Smol{\v{c}}i{\'c}}, {Delhaize},
  {Delvecchio}, {Zamorani}, {Baran}, {Bondi}, {Capak}, {Carilli}, {Ciliegi},
  {Civano}, {Ilbert}, {Karim}, {Laigle}, {Le F{\`e}vre}, {Marchesi},
  {McCracken}, {Miettinen}, {Salvato}, {Sargent}, {Schinnerer}, \&
  {Tasca}}]{2017A&A...602A...5N}
{Novak}, M., {Smol{\v{c}}i{\'c}}, V., {Delhaize}, J., {et~al.} 2017, \aap, 602,
  A5

\bibitem[{{Oesch} {et~al.}(2018){Oesch}, {Bouwens}, {Illingworth}, {Labb{\'e}},
  \& {Stefanon}}]{2018ApJ...855..105O}
{Oesch}, P.~A., {Bouwens}, R.~J., {Illingworth}, G.~D., {Labb{\'e}}, I., \&
  {Stefanon}, M. 2018, \apj, 855, 105

\bibitem[{{Oliver} {et~al.}(2012){Oliver}, {Bock}, {Altieri}, {Amblard},
  {Arumugam}, {Aussel}, {Babbedge}, {Beelen}, {B{\'e}thermin}, {Blain},
  {Boselli}, {Bridge}, {Brisbin}, {Buat}, {Burgarella},
  {Castro-Rodr{\'\i}guez}, {Cava}, {Chanial}, {Cirasuolo}, {Clements},
  {Conley}, {Conversi}, {Cooray}, {Dowell}, {Dubois}, {Dwek}, {Dye}, {Eales},
  {Elbaz}, {Farrah}, {Feltre}, {Ferrero}, {Fiolet}, {Fox}, {Franceschini},
  {Gear}, {Giovannoli}, {Glenn}, {Gong}, {Gonz{\'a}lez Solares}, {Griffin},
  {Halpern}, {Harwit}, {Hatziminaoglou}, {Heinis}, {Hurley}, {Hwang}, {Hyde},
  {Ibar}, {Ilbert}, {Isaak}, {Ivison}, {Lagache}, {Le Floc'h}, {Levenson},
  {Faro}, {Lu}, {Madden}, {Maffei}, {Magdis}, {Mainetti}, {Marchetti},
  {Marsden}, {Marshall}, {Mortier}, {Nguyen}, {O'Halloran}, {Omont}, {Page},
  {Panuzzo}, {Papageorgiou}, {Patel}, {Pearson}, {P{\'e}rez-Fournon}, {Pohlen},
  {Rawlings}, {Raymond}, {Rigopoulou}, {Riguccini}, {Rizzo}, {Rodighiero},
  {Roseboom}, {Rowan-Robinson}, {S{\'a}nchez Portal}, {Schulz}, {Scott},
  {Seymour}, {Shupe}, {Smith}, {Stevens}, {Symeonidis}, {Trichas}, {Tugwell},
  {Vaccari}, {Valtchanov}, {Vieira}, {Viero}, {Vigroux}, {Wang}, {Ward},
  {Wardlow}, {Wright}, {Xu}, \& {Zemcov}}]{hermes}
{Oliver}, S.~J., {Bock}, J., {Altieri}, B., {et~al.} 2012, MNRAS, 424, 1614

\bibitem[{{Pantoni} {et~al.}(2021){Pantoni}, {Massardi}, {Lapi}, {Donevski},
  {D'Amato}, {Giulietti}, {Pozzi}, {Talia}, {Vignali}, {Cimatti}, {Silva},
  {Bressan}, \& {Ronconi}}]{2021MNRAS.507.3998P}
{Pantoni}, L., {Massardi}, M., {Lapi}, A., {et~al.} 2021, \mnras, 507, 3998

\bibitem[{{P{\^a}ris} {et~al.}(2014){P{\^a}ris}, {Petitjean}, {Aubourg},
  {Ross}, {Myers}, {Streblyanska}, {Bailey}, {Hall}, {Strauss}, {Anderson},
  {Bizyaev}, {Borde}, {Brinkmann}, {Bovy}, {Brandt}, {Brewington},
  {Brownstein}, {Cook}, {Ebelke}, {Fan}, {Filiz Ak}, {Finley}, {Font-Ribera},
  {Ge}, {Hamann}, {Ho}, {Jiang}, {Kinemuchi}, {Malanushenko}, {Malanushenko},
  {Marchante}, {McGreer}, {McMahon}, {Miralda-Escud{\'e}}, {Muna},
  {Noterdaeme}, {Oravetz}, {Palanque-Delabrouille}, {Pan}, {Perez-Fournon},
  {Pieri}, {Riffel}, {Schlegel}, {Schneider}, {Simmons}, {Viel}, {Weaver},
  {Wood-Vasey}, {Y{\`e}che}, \& {York}}]{2014A&A...563A..54P}
{P{\^a}ris}, I., {Petitjean}, P., {Aubourg}, {\'E}., {et~al.} 2014, AAP, 563,
  A54

\bibitem[{{P{\'e}rez-Gonz{\'a}lez} {et~al.}(2023){P{\'e}rez-Gonz{\'a}lez},
  {Barro}, {Annunziatella}, {Costantin}, {Garc{\'\i}a-Argum{\'a}nez},
  {McGrath}, {M{\'e}rida}, {Zavala}, {Haro}, {Bagley}, {Backhaus}, {Behroozi},
  {Bell}, {Bisigello}, {Buat}, {Calabr{\`o}}, {Casey}, {Cleri}, {Coogan},
  {Cooper}, {Cooray}, {Dekel}, {Dickinson}, {Elbaz}, {Ferguson}, {Finkelstein},
  {Fontana}, {Franco}, {Gardner}, {Giavalisco}, {G{\'o}mez-Guijarro},
  {Grazian}, {Grogin}, {Guo}, {Huertas-Company}, {Jogee}, {Kartaltepe},
  {Kewley}, {Kirkpatrick}, {Kocevski}, {Koekemoer}, {Long}, {Lotz}, {Lucas},
  {Papovich}, {Pirzkal}, {Ravindranath}, {Somerville}, {Tacchella}, {Trump},
  {Wang}, {Wilkins}, {Wuyts}, {Yang}, \& {Aaron Yung}}]{2023ApJ...946L..16P}
{P{\'e}rez-Gonz{\'a}lez}, P.~G., {Barro}, G., {Annunziatella}, M., {et~al.}
  2023, \apjl, 946, L16

\bibitem[{{Pilbratt} {et~al.}(2010){Pilbratt}, {Riedinger}, {Passvogel},
  {Crone}, {Doyle}, {Gageur}, {Heras}, {Jewell}, {Metcalfe}, {Ott}, \&
  {Schmidt}}]{2010A&A...518L...1P}
{Pilbratt}, G.~L., {Riedinger}, J.~R., {Passvogel}, T., {et~al.} 2010, A\&A,
  518, L1

\bibitem[{{Pillepich} {et~al.}(2018){Pillepich}, {Springel}, {Nelson}, {Genel},
  {Naiman}, {Pakmor}, {Hernquist}, {Torrey}, {Vogelsberger}, {Weinberger}, \&
  {Marinacci}}]{2018MNRAS.473.4077P}
{Pillepich}, A., {Springel}, V., {Nelson}, D., {et~al.} 2018, \mnras, 473, 4077

\bibitem[{{Poglitsch} {et~al.}(2010){Poglitsch}, {Waelkens}, {Geis},
  {Feuchtgruber}, {Vandenbussche}, {Rodriguez}, {Krause}, {Renotte}, {van
  Hoof}, {Saraceno}, {Cepa}, {Kerschbaum}, {Agn{\`e}se}, {Ali}, {Altieri},
  {Andreani}, {Augueres}, {Balog}, {Barl}, {Bauer}, {Belbachir}, {Benedettini},
  {Billot}, {Boulade}, {Bischof}, {Blommaert}, {Callut}, {Cara}, {Cerulli},
  {Cesarsky}, {Contursi}, {Creten}, {De Meester}, {Doublier}, {Doumayrou},
  {Duband }, {Exter}, {Genzel}, {Gillis}, {Gr{\"o}zinger}, {Henning},
  {Herreros}, {Huygen}, {Inguscio}, {Jakob}, {Jamar}, {Jean}, {de Jong},
  {Katterloher}, {Kiss}, {Klaas}, {Lemke}, {Lutz}, {Madden}, {Marquet},
  {Martignac}, {Mazy}, {Merken}, {Montfort}, {Morbidelli}, {M{\"u}ller},
  {Nielbock}, {Okumura}, {Orfei}, {Ottensamer}, {Pezzuto}, {Popesso},
  {Putzeys}, {Regibo}, {Reveret}, {Royer}, {Sauvage}, {Schreiber}, {Stegmaier},
  {Schmitt}, {Schubert}, {Sturm}, {Thiel}, {Tofani}, {Vavrek}, {Wetzstein},
  {Wieprecht}, \& {Wiezorrek}}]{2010A&A...518L...2P}
{Poglitsch}, A., {Waelkens}, C., {Geis}, N., {et~al.} 2010, A\&A, 518, L2

\bibitem[{{Ranalli} {et~al.}(2013){Ranalli}, {Comastri}, {Vignali}, {Carrera},
  {Cappelluti}, {Gilli}, {Puccetti}, {Brand t}, {Brunner}, {Brusa},
  {Georgantopoulos}, {Iwasawa}, \& {Mainieri}}]{2013A&A...555A..42R}
{Ranalli}, P., {Comastri}, A., {Vignali}, C., {et~al.} 2013, A\&A, 555, A42

\bibitem[{{Reid} {et~al.}(2016){Reid}, {Ho}, {Padmanabhan}, {Percival},
  {Tinker}, {Tojeiro}, {White}, {Eisenstein}, {Maraston}, {Ross},
  {S{\'a}nchez}, {Schlegel}, {Sheldon}, {Strauss}, {Thomas}, {Wake}, {Beutler},
  {Bizyaev}, {Bolton}, {Brownstein}, {Chuang}, {Dawson}, {Harding}, {Kitaura},
  {Leauthaud}, {Masters}, {McBride}, {More}, {Olmstead}, {Oravetz}, {Nuza},
  {Pan}, {Parejko}, {Pforr}, {Prada}, {Rodr{\'\i}guez-Torres},
  {Salazar-Albornoz}, {Samushia}, {Schneider}, {Sc{\'o}ccola}, {Simmons}, \&
  {Vargas-Magana}}]{2016MNRAS.455.1553R}
{Reid}, B., {Ho}, S., {Padmanabhan}, N., {et~al.} 2016, \mnras, 455, 1553

\bibitem[{{Rodighiero} {et~al.}(2023){Rodighiero}, {Bisigello}, {Iani},
  {Marasco}, {Grazian}, {Sinigaglia}, {Cassata}, \&
  {Gruppioni}}]{2023MNRAS.518L..19R}
{Rodighiero}, G., {Bisigello}, L., {Iani}, E., {et~al.} 2023, \mnras, 518, L19

\bibitem[{{Rodriguez} {et~al.}(2021){Rodriguez}, {Montero-Dorta}, {Angulo},
  {Artale}, \& {Merch{\'a}n}}]{2021MNRAS.505.3192R}
{Rodriguez}, F., {Montero-Dorta}, A.~D., {Angulo}, R.~E., {Artale}, M.~C., \&
  {Merch{\'a}n}, M. 2021, \mnras, 505, 3192

\bibitem[{{Rowan-Robinson} {et~al.}(2016){Rowan-Robinson}, {Oliver}, {Wang},
  {Farrah}, {Clements}, {Gruppioni}, {Marchetti}, {Rigopoulou}, \&
  {Vaccari}}]{2016MNRAS.461.1100R}
{Rowan-Robinson}, M., {Oliver}, S., {Wang}, L., {et~al.} 2016, \mnras, 461,
  1100

\bibitem[{{Schinnerer} {et~al.}(2010){Schinnerer}, {Sargent}, {Bondi},
  {Smol{\v{c}}i{\'c}}, {Datta}, {Carilli}, {Bertoldi}, {Blain}, {Ciliegi},
  {Koekemoer}, \& {Scoville}}]{2010ApJS..188..384S}
{Schinnerer}, E., {Sargent}, M.~T., {Bondi}, M., {et~al.} 2010, APJS, 188, 384

\bibitem[{{Scoville} {et~al.}(2007){Scoville}, {Aussel}, {Brusa}, {Capak},
  {Carollo}, {Elvis}, {Giavalisco}, {Guzzo}, {Hasinger}, {Impey}, {Kneib},
  {LeFevre}, {Lilly}, {Mobasher}, {Renzini}, {Rich}, {Sanders}, {Schinnerer},
  {Schminovich}, {Shopbell}, {Taniguchi}, \& {Tyson}}]{cosmos_paper}
{Scoville}, N., {Aussel}, H., {Brusa}, M., {et~al.} 2007, ApJ, 172, 1

\bibitem[{{Serra} {et~al.}(2011){Serra}, {Amblard}, {Temi}, {Burgarella},
  {Giovannoli}, {Buat}, {Noll}, \& {Im}}]{cigale2}
{Serra}, P., {Amblard}, A., {Temi}, P., {et~al.} 2011, ApJ, 740, 22

\bibitem[{{Shimwell} {et~al.}(2017){Shimwell}, {R{\"o}ttgering}, {Best},
  {Williams}, {Dijkema}, {de Gasperin}, {Hardcastle}, {Heald}, {Hoang},
  {Horneffer}, {Intema}, {Mahony}, {Mandal}, {Mechev}, {Morabito}, {Oonk},
  {Rafferty}, {Retana-Montenegro}, {Sabater}, {Tasse}, {van Weeren},
  {Br{\"u}ggen}, {Brunetti}, {Chy{\.z}y}, {Conway}, {Haverkorn}, {Jackson},
  {Jarvis}, {McKean}, {Miley}, {Morganti}, {White}, {Wise}, {van Bemmel},
  {Beck}, {Brienza}, {Bonafede}, {Calistro Rivera}, {Cassano}, {Clarke},
  {Cseh}, {Deller}, {Drabent}, {van Driel}, {Engels}, {Falcke}, {Ferrari},
  {Fr{\"o}hlich}, {Garrett}, {Harwood}, {Heesen}, {Hoeft}, {Horellou},
  {Israel}, {Kapi{\'n}ska}, {Kunert-Bajraszewska}, {McKay}, {Mohan},
  {Orr{\'u}}, {Pizzo}, {Prandoni}, {Schwarz}, {Shulevski}, {Sipior}, {Smith},
  {Sridhar}, {Steinmetz}, {Stroe}, {Varenius}, {van der Werf}, {Zensus}, \&
  {Zwart}}]{2017A&A...598A.104S}
{Shimwell}, T.~W., {R{\"o}ttgering}, H.~J.~A., {Best}, P.~N., {et~al.} 2017,
  A\&A, 598, A104

\bibitem[{{Shirley} {et~al.}(2021){Shirley}, {Duncan}, {Campos Varillas},
  {Hurley}, {Ma{\l}ek}, {Roehlly}, {Smith}, {Aussel}, {Bakx}, {Buat},
  {Burgarella}, {Christopher}, {Duivenvoorden}, {Eales}, {Efstathiou},
  {Gonz{\'a}lez Solares}, {Griffin}, {Jarvis}, {Faro}, {Marchetti}, {McCheyne},
  {Papadopoulos}, {Penner}, {Pons}, {Prescott}, {Rigby}, {Rottgering},
  {Saxena}, {Scudder}, {Vaccari}, {Wang}, \& {Oliver}}]{2021MNRAS.507..129S}
{Shirley}, R., {Duncan}, K., {Campos Varillas}, M.~C., {et~al.} 2021, \mnras,
  507, 129

\bibitem[{{Smol{\v c}i{\'c}} {et~al.}(2017){Smol{\v c}i{\'c}}, {Novak},
  {Delvecchio}, {Ceraj}, {Bondi}, {Delhaize}, {Marchesi}, {Murphy},
  {Schinnerer}, {Vardoulaki}, \& {Zamorani}}]{smolcic17}
{Smol{\v c}i{\'c}}, V., {Novak}, M., {Delvecchio}, I., {et~al.} 2017, A\&A,
  602, A6

\bibitem[{{Suh} {et~al.}(2015){Suh}, {Hasinger}, {Steinhardt}, {Silverman}, \&
  {Schramm}}]{2015ApJ...815..129S}
{Suh}, H., {Hasinger}, G., {Steinhardt}, C., {Silverman}, J.~D., \& {Schramm},
  M. 2015, ApJ, 815, 129

\bibitem[{{Tacconi} {et~al.}(2020){Tacconi}, {Genzel}, \&
  {Sternberg}}]{2020ARA&A..58..157T}
{Tacconi}, L.~J., {Genzel}, R., \& {Sternberg}, A. 2020, \araa, 58, 157

\bibitem[{{Talia} {et~al.}(2021){Talia}, {Cimatti}, {Giulietti}, {Zamorani},
  {Bethermin}, {Faisst}, {Le F{\`e}vre}, \&
  {Smol{\c{c}}i{\'c}}}]{2021ApJ...909...23T}
{Talia}, M., {Cimatti}, A., {Giulietti}, M., {et~al.} 2021, \apj, 909, 23

\bibitem[{{Vaccari} {et~al.}(2016){Vaccari}, {Covone}, {Radovich}, {Grado},
  {Limatola}, {Botticella}, {Cappellaro}, {Paolillo}, {Pignata}, {De Cicco},
  {Falocco}, {Marchetti}, {Brescia}, {Cavuoti}, {Longo}, {Capaccioli},
  {Napolitano}, \& {Schipani}}]{2016heas.confE..26V}
{Vaccari}, M., {Covone}, G., {Radovich}, M., {et~al.} 2016, in The 4th Annual
  Conference on High Energy Astrophysics in Southern Africa (HEASA 2016), 26

\bibitem[{{Viero} {et~al.}(2013){Viero}, {Moncelsi}, {Quadri}, {Arumugam},
  {Assef}, {B{\'e}thermin}, {Bock}, {Bridge}, {Casey}, {Conley}, {Cooray},
  {Farrah}, {Glenn}, {Heinis}, {Ibar}, {Ikarashi}, {Ivison}, {Kohno},
  {Marsden}, {Oliver}, {Roseboom}, {Schulz}, {Scott}, {Serra}, {Vaccari},
  {Vieira}, {Wang}, {Wardlow}, {Wilson}, {Yun}, \&
  {Zemcov}}]{2013ApJ...779...32V}
{Viero}, M.~P., {Moncelsi}, L., {Quadri}, R.~F., {et~al.} 2013, \apj, 779, 32

\bibitem[{{Wang} {et~al.}(2021){Wang}, {Gao}, {Best}, {Duncan}, {Hardcastle},
  {Kondapally}, {Ma{\l}ek}, {McCheyne}, {Sabater}, {Shimwell}, {Tasse},
  {Bonato}, {Bondi}, {Cochrane}, {Farrah}, {G{\"u}rkan}, {Haskell}, {Pearson},
  {Prandoni}, {R{\"o}ttgering}, {Smith}, {Vaccari}, \&
  {Williams}}]{2021A&A...648A...8W}
{Wang}, L., {Gao}, F., {Best}, P.~N., {et~al.} 2021, \aap, 648, A8

\bibitem[{{Wang} {et~al.}(2019){Wang}, {Pearson}, {Cowley}, {Trayford},
  {B{\'e}thermin}, {Gruppioni}, {Hurley}, \&
  {Micha{\l}owski}}]{2019A&A...624A..98W}
{Wang}, L., {Pearson}, W.~J., {Cowley}, W., {et~al.} 2019, \aap, 624, A98

\bibitem[{{Wardlow} {et~al.}(2011){Wardlow}, {Smail}, {Coppin}, {Alexander},
  {Brandt}, {Danielson}, {Luo}, {Swinbank}, {Walter}, {Wei{\ss}}, {Xue},
  {Zibetti}, {Bertoldi}, {Biggs}, {Chapman}, {Dannerbauer}, {Dunlop},
  {Gawiser}, {Ivison}, {Knudsen}, {Kov{\'a}cs}, {Lacey}, {Menten}, {Padilla},
  {Rix}, \& {van der Werf}}]{2011MNRAS.415.1479W}
{Wardlow}, J.~L., {Smail}, I., {Coppin}, K.~E.~K., {et~al.} 2011, MNRAS, 415,
  1479

\bibitem[{{Weaver} {et~al.}(2022){Weaver}, {Kauffmann}, {Ilbert}, {McCracken},
  {Moneti}, {Toft}, {Brammer}, {Shuntov}, {Davidzon}, {Hsieh}, {Laigle},
  {Anastasiou}, {Jespersen}, {Vinther}, {Capak}, {Casey}, {McPartland},
  {Milvang-Jensen}, {Mobasher}, {Sanders}, {Zalesky}, {Arnouts}, {Aussel},
  {Dunlop}, {Faisst}, {Franx}, {Furtak}, {Fynbo}, {Gould}, {Greve}, {Gwyn},
  {Kartaltepe}, {Kashino}, {Koekemoer}, {Kokorev}, {Le F{\`e}vre}, {Lilly},
  {Masters}, {Magdis}, {Mehta}, {Peng}, {Riechers}, {Salvato}, {Sawicki},
  {Scarlata}, {Scoville}, {Shirley}, {Silverman}, {Sneppen}, {Smolc̆i{\'c}},
  {Steinhardt}, {Stern}, {Tanaka}, {Taniguchi}, {Teplitz}, {Vaccari}, {Wang},
  \& {Zamorani}}]{2022ApJS..258...11W}
{Weaver}, J.~R., {Kauffmann}, O.~B., {Ilbert}, O., {et~al.} 2022, \apjs, 258,
  11

\bibitem[{{Wenger} {et~al.}(2000){Wenger}, {Ochsenbein}, {Egret}, {Dubois},
  {Bonnarel}, {Borde}, {Genova}, {Jasniewicz}, {Lalo{\"e}}, {Lesteven}, \&
  {Monier}}]{2000A&AS..143....9W}
{Wenger}, M., {Ochsenbein}, F., {Egret}, D., {et~al.} 2000, A\&As, 143, 9

\bibitem[{{Wilson} {et~al.}(2019){Wilson}, {Dong}, {di Francesco}, {Fissel},
  {Johnstone}, {Kirk}, {Matthews}, {McNamara}, {Rosolowsky}, {Rupen},
  {Sadavoy}, {Scott}, \& {van der Marel}}]{2019clrp.2020...19W}
{Wilson}, C., {Dong}, R., {di Francesco}, J., {et~al.} 2019, in Canadian Long
  Range Plan for Astronony and Astrophysics White Papers, Vol. 2020, 19

\bibitem[{{Yamaguchi} {et~al.}(2019){Yamaguchi}, {Kohno}, {Hatsukade}, {Wang},
  {Yoshimura}, {Ao}, {Caputi}, {Dunlop}, {Egami}, {Espada}, {Fujimoto},
  {Hayatsu}, {Ivison}, {Kodama}, {Kusakabe}, {Nagao}, {Ouchi}, {Rujopakarn},
  {Tadaki}, {Tamura}, {Ueda}, {Umehata}, {Wang}, \&
  {Yun}}]{2019ApJ...878...73Y}
{Yamaguchi}, Y., {Kohno}, K., {Hatsukade}, B., {et~al.} 2019, \apj, 878, 73

\bibitem[{{Yan} {et~al.}(2022){Yan}, {Ling}, \& {Ma}}]{2022MNRAS.516.5471Y}
{Yan}, H., {Ling}, C., \& {Ma}, Z. 2022, \mnras, 516, 5471

\bibitem[{{Y{\"u}ksel} {et~al.}(2008){Y{\"u}ksel}, {Kistler}, {Beacom}, \&
  {Hopkins}}]{2008ApJ...683L...5Y}
{Y{\"u}ksel}, H., {Kistler}, M.~D., {Beacom}, J.~F., \& {Hopkins}, A.~M. 2008,
  \apjl, 683, L5

\end{thebibliography}

%

\begin{appendix} 

\section{Comparison of the retrieved $z$ with photometric catalogues}\label{appA}
\begin{figure*}
\centering
\includegraphics[width=\columnwidth]{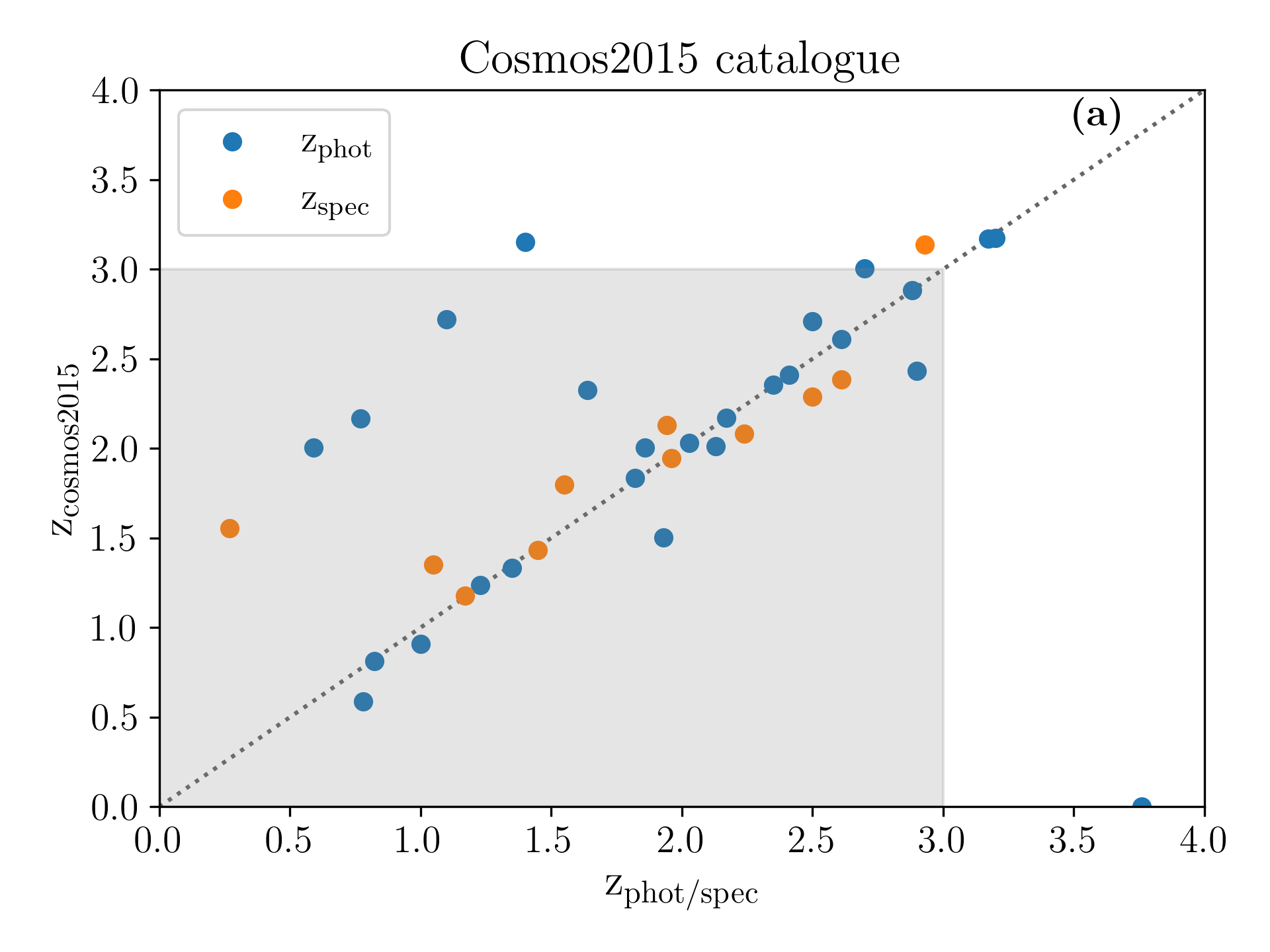}
\includegraphics[width=\columnwidth]{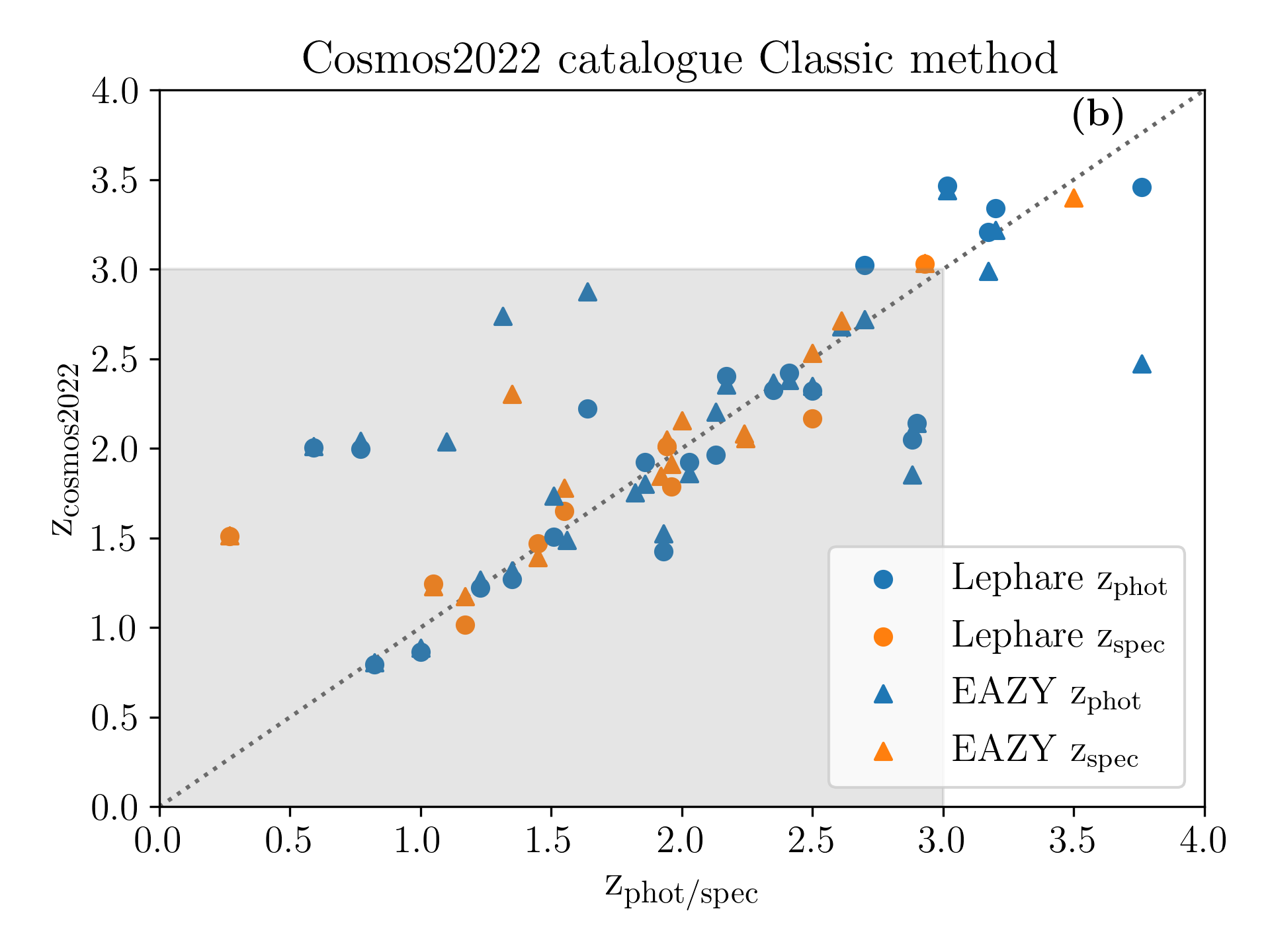}

\includegraphics[width=\columnwidth]{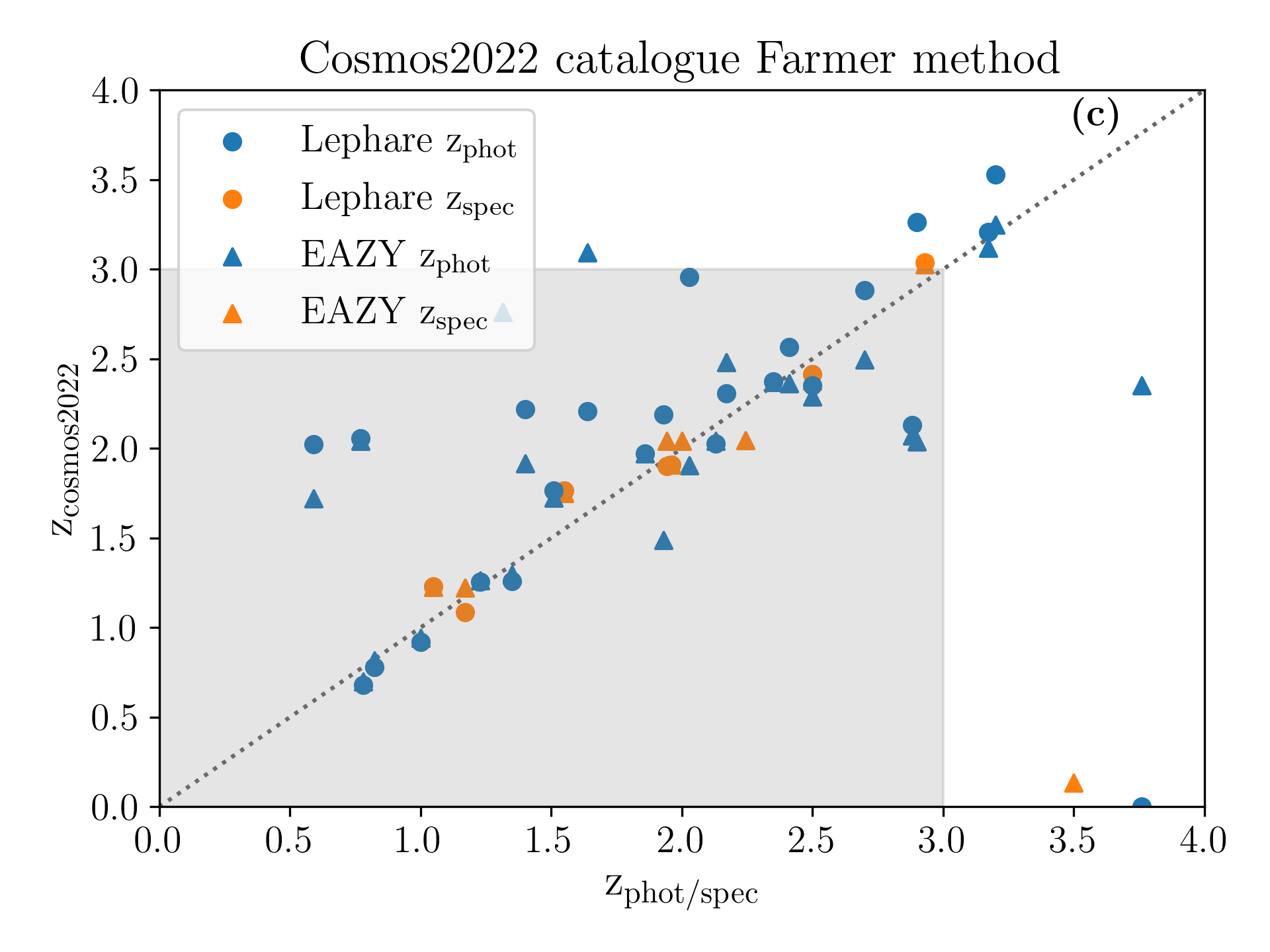}
\includegraphics[width=\columnwidth]{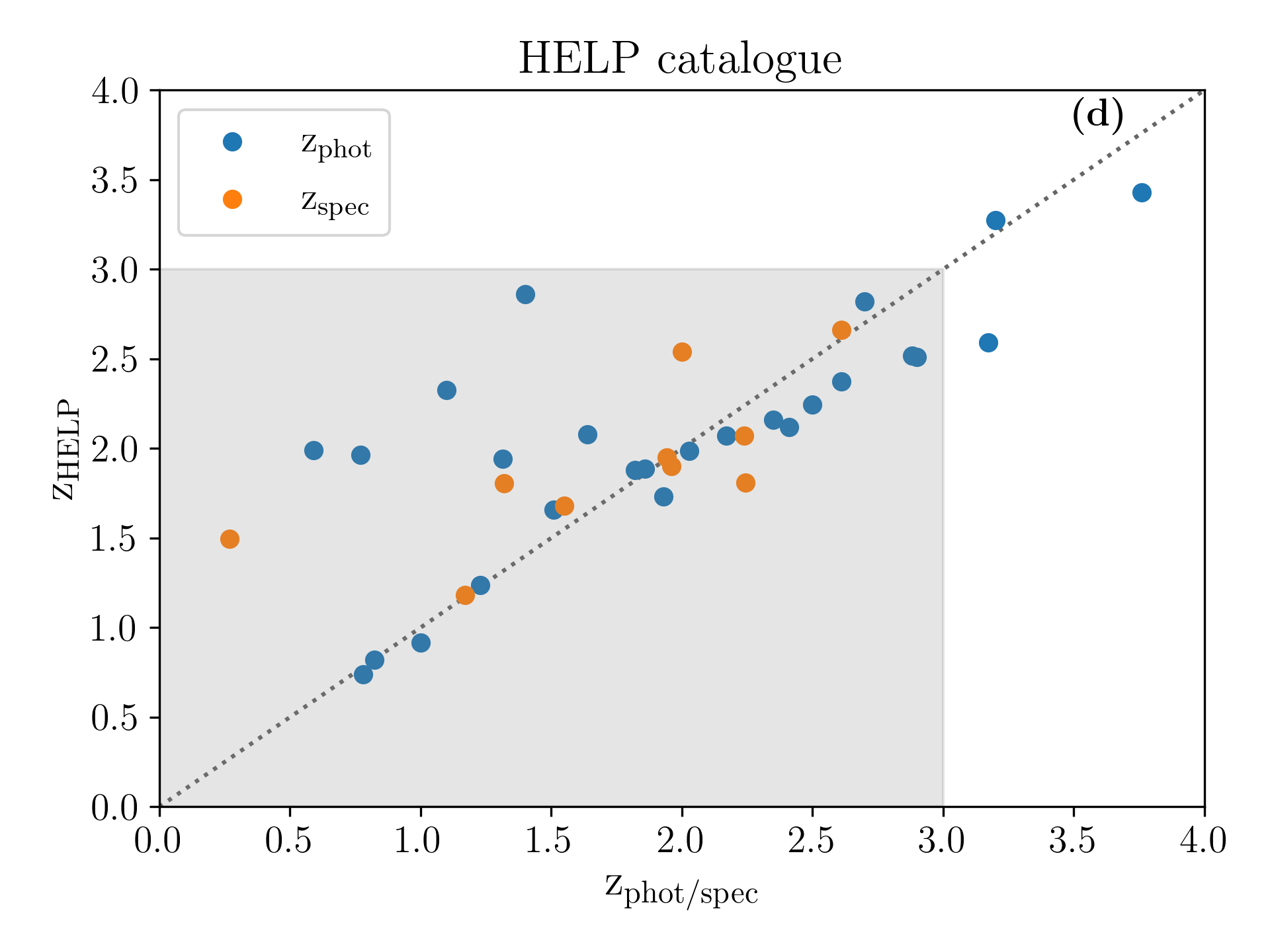}
\caption{Comparison between the Cosmos2015 (a), Cosmos2020 (b and c), and HELP (d) catalogues and our redshift values for our sample. The blue points in each plot correspond to the photometric $z$ values from each catalogue, and the orange points depict the spectroscopic values from the literature. Notably, the Cosmos2020 catalogue yields two distinct sets of results because two different SED fitting codes were used: Lephare and EAZY, denoted by circles and triangles, respectively.}
\label{z_comparison}
\end{figure*}

In this appendix, we present a comparison between the redshift values estimated by various photometric catalogues and our analysis, as described above. In more detail, the catalogues consist of the Cosmos2015 \citep[][]{Laigle16}, the Cosmos2020 \citep[][employing all possible redshift estimations, namely the classic and Farmer methods using the SED fitting codes Lephare and EAZY]{2022ApJS..258...11W} and the HELP catalogues \citep[][]{2021MNRAS.507..129S}. The redshift values derived from these catalogues are compared with the values that our team acquired from the literature and from the IRAM/ALMA observations of the aforementioned two targets (i.e. CVLA958 and CVLA100; see Sect. \ref{results}). For the cross match between catalogues, a radius of 1 arcsec was employed in order to ensure that each source from the catalogues corresponded to the correct source from our sample. The results of these comparisons are presented in Fig. \ref{z_comparison}.

In our findings, certain sources exhibit lower redshift values in the catalogues than in the present study. These differences can be attributed to the low S/N of the photometric points that were used in the SED fitting. Furthermore, the authors of these catalogues may not have individually examined each source due to the extensive volume of data. For example, the source with coordinates (Ra, Dec)=(150.02005417, 2.51254722) is depicted in the Cosmos2020 comparison plots (see Fig. \ref{z_comparison}) with a redshift of $z=2.48$, derived from the EAZY SED fit ($\chi^2$ of $\sim 15$). However, our work adopts a redshift value of 3.46 obtained from the Lephare SED fit solution for the same source. This choice was made due to the lower $\chi^2$ values obtained from both the literature Lephare analysis ($\chi^2 = 3.76$) and our own CIGALE fits ($\chi^2 = 2.24$).

Additionally, this figure shows that according to Cosmos2015, two galaxies that were identified by our search as $z<3$ sources have a higher photometric redshift value. In more detail, the sources are identified as UVISTADR1 J100131.33+020325.2 and UVISTADR1 J100114.75+023516.3 and have a $z_{\rm phot,cosmos2015}$ of 3.152 and 3.007, respectively. However, the more detailed photometric analysis by \citet{2013ApJS..206....8M} proposed lower values for both sources of $\sim 1.8$ and $\sim 2.7$, respectively. Similarly, for the cosmos2020-Farmer method, a source named UVISTADR1 J095834.49+015504.9 was estimated to be at $z_{\rm phot,cosmos2020,EAZY}\sim 3.1$, however, \citet{2013ApJS..206....8M} reported a lower value of $\sim 2.5$. The latter redshift results can be further supported by the redshift estimate by Lephare cosmos2020 of $z_{\rm phot,cosmos2020,Lephare}\sim 2.2$. These discrepancies between various photometric catalogues can mainly be attributed to the faint nature of these sources and to the subsequent low fit quality of their SED. Finally, according to the HELP catalogue, no further source could be added to our $z>3$ sample.

\section{Spectroscopic follow-up of CVLA958 and CVLA100}\label{appC}

\textcolor{black}{Because of the uncertainty of} the high-redshift nature of CVLA958 and CVLA100, these sources were originally included in a follow-up spectroscopic study by our team with the aim to understand their nature and potentially employ this knowledge to detect galaxies at higher redshifts. In this appendix, we present the retrieved spectroscopic data for these two galaxies, bearing in mind the redshift solution of $z \sim 3$ that the current photometric analysis indicated. The original idea of the spectra acquired with the Institute Radio Astronomie Millimétrique (IRAM) 30m antenna and the ALMA telescopes was the detection of two possible CO emission lines that would constrain the redshift of each galaxy.

\subsection{IRAM observations}
The IRAM\footnote{\href{https://www.iram-institute.org}{https://www.iram-institute.org}} institute is composed of two facilities, namely the interferometer NOrthern Extended Millimeter Array \citep[NOEMA;][]{2016ITTST...6..223C} and a single-dish telescope with a diameter of 30m, located in the Sierra Nevada, Spain. The adopted strategy to retrieve the millimetre spectrum of our galaxies was to scan the 3 mm band with the spectral line receiver Eight MIxer Receiver \citep[EMIR;][]{2012A&A...538A..89C} of the 30m antenna, along with the Fast Fourier Transform Spectrometers \citep[FTS200;][]{2012A&A...542L...3K} and Wideband Line Multiple Autocorrelator (WILMA\footnote{\href{https://www.iram.fr/IRAMFR/TA/backend/veleta/wilma/index.htm}{https://www.iram.fr/IRAMFR/TA/backend/veleta/wilma/index.htm}}) backends (in parallel) at the wavelength range of 73-117 GHz (band E0/E090). Given the possible AGN nature of the targeted sources, the CO spectral line energy distribution will likely peak at mid-J transitions (e.g. for the frequency range of band E0, the CO J=4-3 line could be at $z\sim 2.7-5$ and the CO J=5-4 line at $z\sim 3.7-6.5$). \textcolor{black}{The expected redshift span for this candidate is} $z \geq 3$, and therefore, the 3 mm spectral range provides an excellent trade-off of weather versus line detection to search for ideally two emission lines required to determine the source redshift. Nevertheless, in the following analysis, only the data from the FTS units were investigated since they provide a higher spectral resolution than WILMA, which is necessary to resolve faint emission lines.

The project code is 227-19, and the observations were initially scheduled for 19-24 December 2019. However, \textcolor{black}{considering that} the weather conditions were not appropriate to observe for the full length of our original allocated time, additional observing time was provided during 21-23 March 2020, 12-14 May 2020, and finally, 14-17 July 2020. Out of a total of 103 hours provided for this source, only $\sim 8$ hours of observations generated spectra with \textcolor{black}{acceptable} quality to be considered in the assembly of the final spectrum. The rest of the observing time was lost due to instrumentation malfunctions and weather conditions (strong wind and high water vapour).

\begin{figure*}
\includegraphics[width=0.5\textwidth]{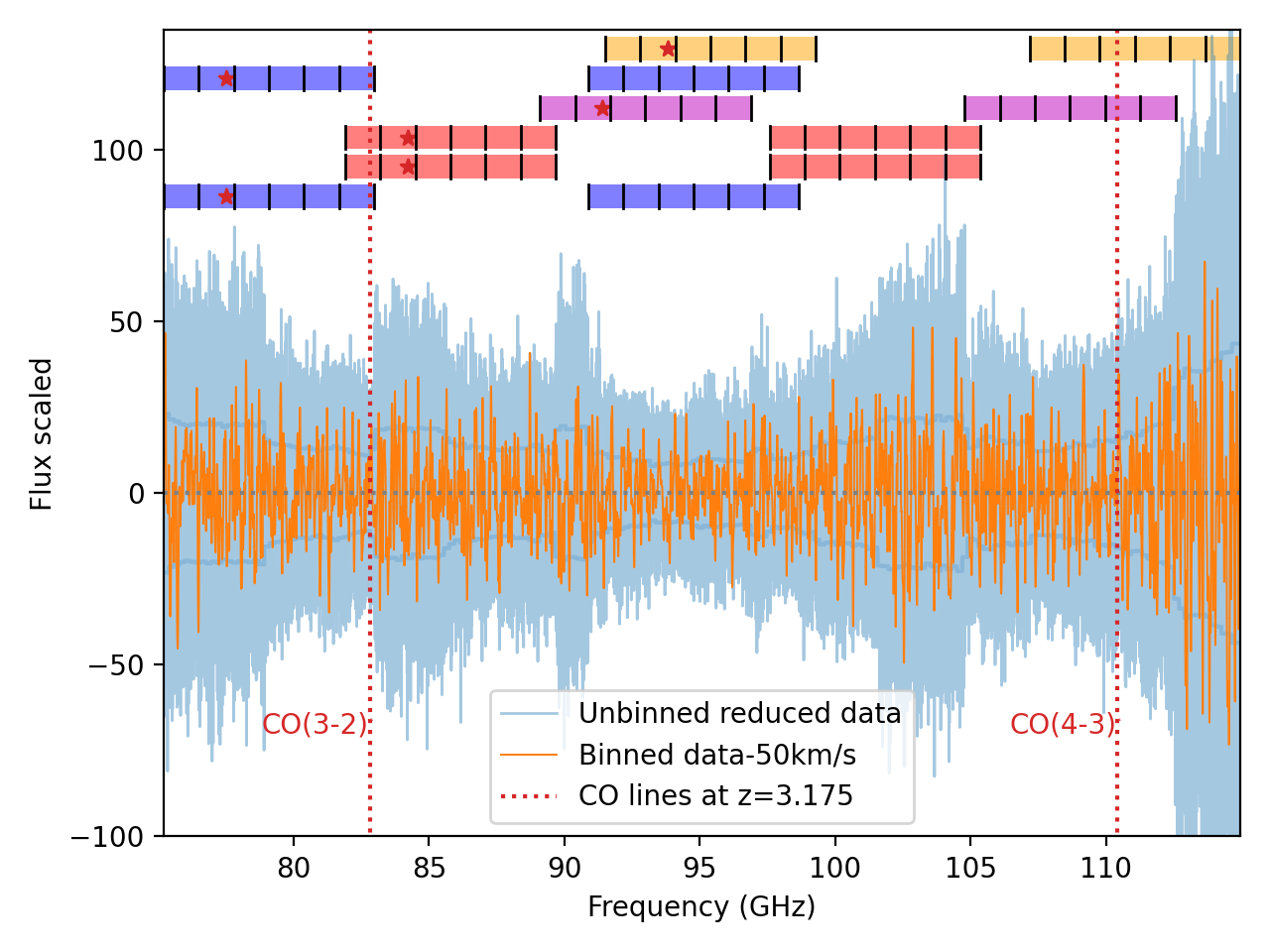}
\includegraphics[width=0.55\textwidth]{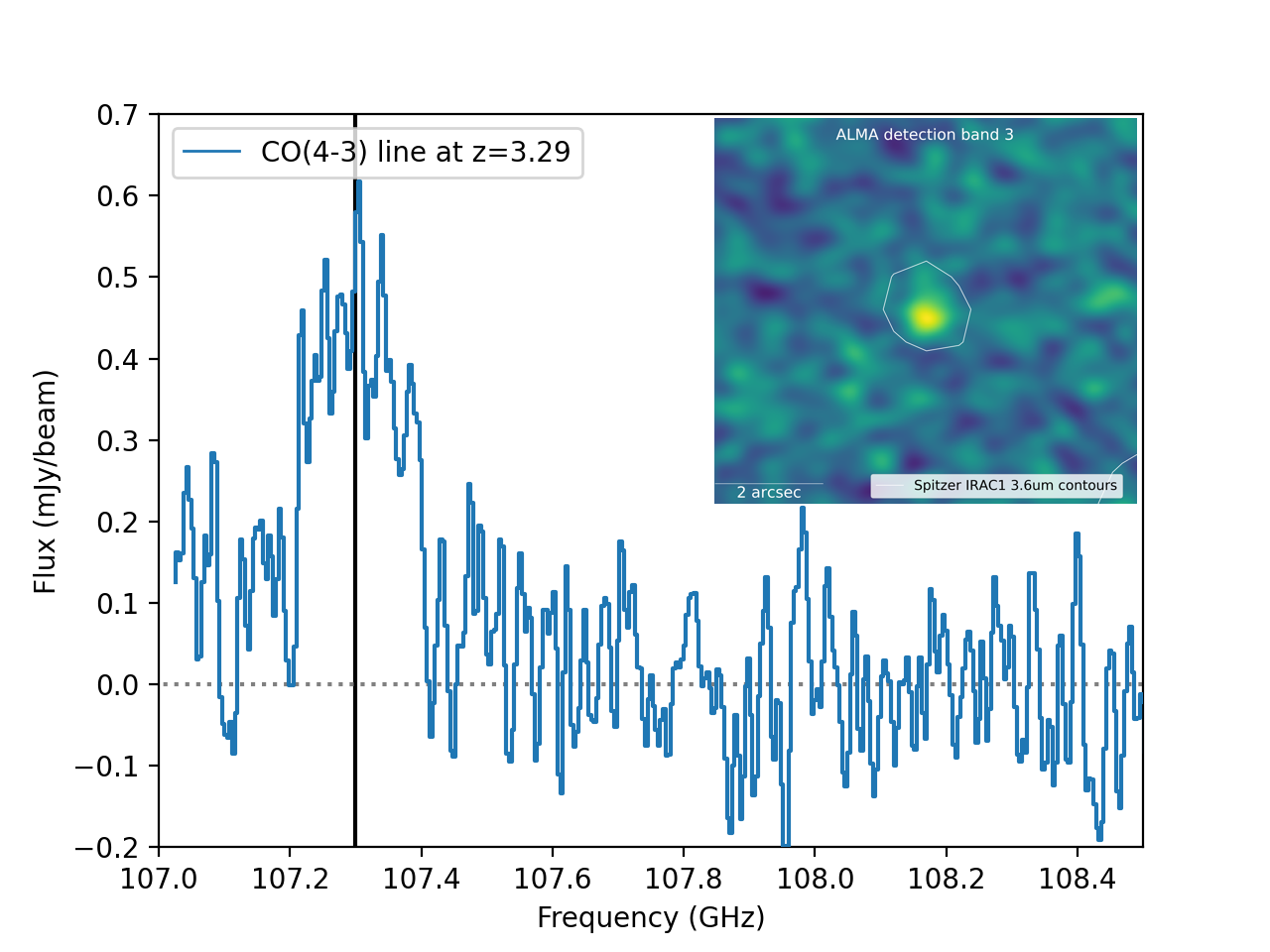}
\caption{Reduced millimetre spectra of CVLA958 and CVLA100 observed by the IRAM 30m and ALMA telescopes. The IRAM spectrum for CVLA958 is shown, with its unbinned reduced spectrum presented in blue, and the orange line corresponds to the Gaussian filtered spectrum using a width of $50\, km/s$ (left panel). The CO lines for the redshift solution of $z=3.1755$ provided by \citet{2021MNRAS.501.3926B} are depicted with the dashed red lines. The applied FTS setups are presented at the top of the plot and are coloured according to the adopted configuration and observing date. The ALMA spectrum for CVLA100 at band 3 is shown in the right panel. The vertical black line corresponds to the possible CO(4-3) emission line at $z=3.29$, located at $\nu = 107.3\, GHz$. The inset cutout illustrates the velocity-integrated map of the line emission, and the \textit{Spitzer} 3.6 $\mu m$ contour (white isocontour) is overlaid.}
\label{total_spectrum_cvla958}
\end{figure*}

The final spectrum of CVLA958 was constructed using only data from the last three days of observations (i.e. 21, 22, and 23 December 2019). We further inspected the histograms of the opacity of the clouds ($\tau$), rms random noise temperature ($T_{\rm rms}$), and receiver and system temperature ($T_{\rm rec}$ and $T_{\rm sys}$, respectively) from these observations in order to exclude possible outliers in the scans. After removing these outliers and correcting for the platforming issue \citep[e.g.][]{2008stt..conf..192K,2009stt..conf..199K}, we generated the final spectrum for the 3 mm band, depicted in the left panel of Fig. \ref{total_spectrum_cvla958}. In this plot, the total spectrum (blue line) and its filtered version for a Gaussian filter of $50 \, km/s$ (orange line) are presented, and the colours are denoted in the legend. The full applied FTS configuration is also shown, as is the location of the CO lines, accepting the recent spectroscopic redshift value of 3.1755 by \citet{2021MNRAS.501.3926B} (dashed red lines). The figure clearly shows that no secure line can be attributed to any feature because the S/N does not permit this determination.\\
\indent The high-redshift candidate CVLA100 was also observed by our team with the IRAM 30m antenna as part of the project 198-16 and during the time period of 23-27 March 2017. In these observations, we also scanned the E0 band of EMIR (frequency range of 73–117 GHz) using the FTS200 backend. Additional time was allocated for our source, and observations were conducted remotely on $26^{\rm }$  May 2107, aiming at the 2 mm spectral window. Similarly, with the acquired spectrum of CVLA958, no secure emission line could be detected for CVLA100. In order to further test possible lines that are not apparent with visual inspection, we developed a simple convolution algorithm to search for emission from CO lines at a given redshift. Essentially, the approach of this code is to sum the observed spectra at the positions where a CO detection was expected at a given redshift, and subsequently vary the redshift in steps of 0.01. Application of this analysis, presents no emission line detection, considering that the resulting values for specific redshifts ranging from $z=0$ to $z=6$ do not deviate significantly from zero.

\subsection{ALMA observations}
\textcolor{black}{No emission was detected in the IRAM spectrum for} CVLA100, our team decided to secure additional observing time for the same candidate using the ALMA telescope. More specifically, the source was observed with multiple tunings in bands 3 (84-116 GHz) and 4 (125-163 GHz) in various configurations (longest baselines from 300 m to 1400 m), under the project code 2017.1.01713.S. With acceptable weather conditions (water vapour of $\leq 3\, mm$ at all times; wind at $\sim 3 \, m/s$),  our observations achieved a mean angular
resolution of $\sim$1 arcsec and a mean continuum sensitivity of $\sim$0.2
mJy after 4.1 hours at band 3 and 1.4 hours of integration time at band 4. By analysing the spectrum, extracted using the CASA software \citep[][]{2022PASP..134k4501C}, we detected a feature at 107.3 GHz which, considering the numerous CO emission lines, assisted in constraining the redshift of the source. One plausible solution indicates that this feature corresponds to the redshifted CO(4-3) emission line at $z=3.3$. This proposed redshift possibility closely agrees with the suggested solution of $z=3.15$ obtained from our photometric analysis and the referenced literature catalogues (i.e. HELP, Cosmos2015, and cosmos2020).

\end{appendix}
\end{document}